\documentclass[prl,twocolumn,aps, secnumarabic,balancelastpage,amsmath,amssymb,superscriptaddress]{revtex4-1}
\usepackage{graphicx}
\usepackage{bm,amsmath, amssymb}
\usepackage{float}
\usepackage{comment}
\usepackage{tcolorbox}
\usepackage{todonotes}
\usepackage{makecell}
\usepackage{hyperref}
\usepackage{hhline}
\usepackage[utf8]{inputenc}

\newcommand{\ket}[1]{|#1\rangle}

\begin{document}
\title{Imaging the local charge environment of nitrogen-vacancy centers in diamond}
%\title{Imaging single charges in diamond via dark-state spectroscopy of the nitrogen-vacancy center}

%\author{T. Mittiga,$^{1,*}$ S, Hsieh,$^{1,2*}$ C. Zu,$^{1,*}$ B. Kobrin,$^{1,2}$ F. Machado,$^{1}$ P. Bhattacharyya,$^{1,2}$ N. Rui,$^{1}$ A. Jarmola,$^{1,3}$ S. Choi,$^{1}$ D. Budker,$^{1,4}$ N. Y. Yao$^{1,2}$}
%\affiliation{$^{1}$Department of Physics, University of California, Berkeley, CA 94720, USA}
%\affiliation{$^{2}$Materials Science Division, Lawrence Berkeley National Laboratory, Berkeley, CA 94720, USA}
%\affiliation{$^{3}$ODMR Technologies Inc., El Cerrito, CA 94530, USA}
%\affiliation{$^{4}$Helmholtz Institut, Johannes Gutenberg-Universitat Mainz, 55099 Mainz, Germany}

% fix author list:
\author{T. Mittiga}
\thanks{These authors contributed equally to this work}
\affiliation{Department of Physics, University of California, Berkeley, CA 94720, USA}

\author{S. Hsieh}
\thanks{These authors contributed equally to this work}
\affiliation{Department of Physics, University of California, Berkeley, CA 94720, USA}
\affiliation{Materials Science Division, Lawrence Berkeley National Laboratory, Berkeley, CA 94720, USA}

\author{C. Zu}
\thanks{These authors contributed equally to this work}
\affiliation{Department of Physics, University of California, Berkeley, CA 94720, USA}

\author{B. Kobrin}
\affiliation{Department of Physics, University of California, Berkeley, CA 94720, USA}
\affiliation{Materials Science Division, Lawrence Berkeley National Laboratory, Berkeley, CA 94720, USA}

\author{F. Machado}
\affiliation{Department of Physics, University of California, Berkeley, CA 94720, USA}

\author{P.~Bhattacharyya}
\affiliation{Department of Physics, University of California, Berkeley, CA 94720, USA}
\affiliation{Materials Science Division, Lawrence Berkeley National Laboratory, Berkeley, CA 94720, USA}

\author{N. Rui}
\affiliation{Department of Physics, University of California, Berkeley, CA 94720, USA}

\author{A. Jarmola}
\affiliation{Department of Physics, University of California, Berkeley, CA 94720, USA}
\affiliation{U.S. Army Research Laboratory, Adelphi, MD 20783, USA}

\author{S. Choi}
\affiliation{Department of Physics, University of California, Berkeley, CA 94720, USA}

\author{D. Budker}
\affiliation{Helmholtz Institut, Johannes Gutenberg-Universit{\"a}t Mainz, 55099 Mainz, Germany}
\affiliation{Department of Physics, University of California, Berkeley, CA 94720, USA}

\author{N. Y. Yao}
\affiliation{Department of Physics, University of California, Berkeley, CA 94720, USA}
\affiliation{Materials Science Division, Lawrence Berkeley National Laboratory, Berkeley, CA 94720, USA}

%Figure\ref{fig1}a shows a typical optically detected magnetic resonance (ODMR) spectrum for an NV$^-$ ensemble in a Type-Ib diamond sample (S1) treated by electron irradiation and annealing~\cite{supp}.
%%The spectrum exhibits two resonances with tails unlike that of a single NV center (Fig.~\ref{fig1}a inset).
%%The broadened resonance linewidth can be explained by dipolar interactions between NV$^{-}$ centers and abundant substitutional nitrogen defects (P1) bath in type-Ib samples~\cite{} {{\color{red} is this sentence precise?}.
%Unlike the NV$^-$ spectrum at large magnetic field, one observes two asymmetric resonances with a sharp dip in the middle which are not captured by either two Lorentzian or Gaussian profiles.

\begin{abstract}
Characterizing the local \emph{internal} environment surrounding solid-state spin defects is crucial to harnessing them as nanoscale sensors of \emph{external} fields.
This is especially germane to the case of defect ensembles which can exhibit a complex interplay between interactions, internal fields and lattice strain.
Working with the nitrogen-vacancy (NV) center in diamond, we demonstrate that local electric fields dominate the magnetic resonance behavior of NV ensembles at low magnetic field.
We introduce a simple microscopic model that quantitatively captures the observed spectra for samples with NV concentrations spanning over two orders of magnitude.
Motivated by this understanding, we propose and implement a novel method for the nanoscale localization of individual charges within the diamond lattice; our approach relies upon the fact that the charge induces an NV dark state which depends on the electric field orientation.
\end{abstract}

\maketitle

A tremendous amount of recent effort has focused on the creation and control of nanoscale defects in the solid-state \cite{Aharonovich:2016hm, Schirhagl:2014}.
The spectral properties of these defects often depend sensitively on their environment.
On the one hand, this sensitivity naturally suggests their use as nanoscale quantum sensors of \emph{external} signals.
On the other hand, accurately quantifying these signals requires the careful characterization of \emph{internal} local fields.
Here, we focus on a particular defect, the negatively charged nitrogen-vacancy (NV) color center in diamond \cite{doherty2013nitrogen,Schirhagl:2014}.
The electronic spin associated with the NV center is sensitive to a broad range of external signals, from magnetic and electric fields to pressure, temperature and gyroscopic precession \cite{maze2008nanoscale,mamin2013nanoscale,toyli2012measurement,Acosta:2010go,epstein2005anisotropic,Dolde:2011,Dolde:2014hc,doherty2014electronic,Ledbetter:2012kh,Ajoy:2012is}.
Isolated single NVs have been used to explore phenomena in biology \cite{le2013optical,Schirhagl:2014,Kucsko:2013,toyli2013fluorescence,McGuinness2011}, materials science \cite{Laraoui:2015ii,pelliccione2016scanned,du2017control,dovzhenko2016imaging,gross2017real}, and fundamental physics \cite{waldherr2011violation,bernien2013heralded,Hensen:2015dw}.
%

%alternatively discussing sqrt N enhancement instead? or as well?
More recently, many-body correlations have emerged as a powerful resource for enhancing the sensitivity of interacting spin ensembles \cite{wasilewski2010quantum,simmons2010magnetic,jones2009magnetic,cappellaro2009quantum,2018arXiv180100042C}.
To this end, a number of studies have explored and leveraged the properties of high-density NV systems \cite{Acosta:2009gu,Steinert:2010kk,maertz2010vector,Acosta:2010go,Stanwix:2010ko,Pham:2011dc,Jarmola:2012co,BarGill:2013dq,Jarmola:2015gc}.
The local environment in such systems is substantially more complex than that of isolated NVs; this arises from a competition between multiple effects, including: lattice strain, paramagnetic impurities, charge dynamics, and NV-NV dipolar interactions.
While the presence of an applied external magnetic field can suppress some of these effects, it significantly limits the scope of sensing applications such as zero-field nuclear magnetic resonance spectroscopy \cite{weitekamp1983zero,Thayer:2002hc}. %\footnote{}
Thus, characterizing and understanding the spectral properties of NV ensembles at zero field is crucial to utilizing these systems as quantum sensors.

\begin{figure}[h!]
  \centering
  \includegraphics[width=2.75in]{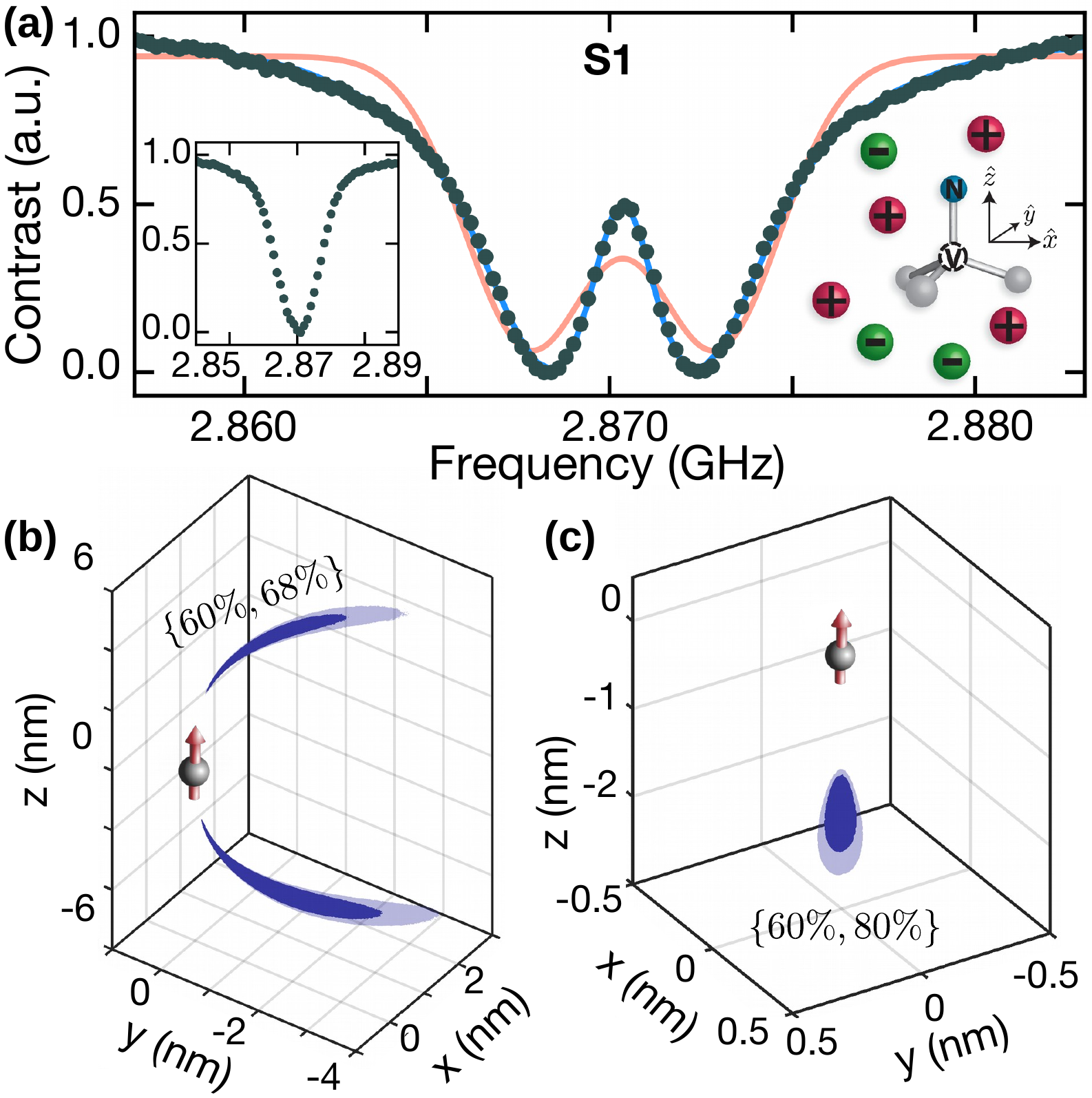}
  \caption{ Typical optically-detected magnetic resonance (ODMR) spectrum of an electron-irradiated and annealed Type-Ib diamond sample (S1) at zero magnetic field.
    %Error bars account for one standard deviation statistical error associated with photon counting.
    The spectrum exhibits heavy tails which cannot be reproduced by either a double Lorentzian or Gaussian (orange fit) profile.
    The blue theory curve is obtained via our microscopic charge model.
    (Left inset) A typical zero-field spectrum for a single NV center shows only a single resonance. (Right inset) Schematic depicting an equal density of positive (e.g.~N$^{+}$) and negative (e.g.~NV) charges, which together, create a random local electric field at each NV center's position.
    (b) Nanoscale localization ($\sim$5~nm) of a single positive charge via dark-state spectroscopy of an isolated  NV center.
   The shaded regions indicate the probable location of the charge with darker indicating a higher likelihood. Percentages shown correspond to the confidence intervals of the dark/light region, respectively.
    (c) Analogous localization of a more proximal charge defect ($\sim$2~nm) for a different NV center.
    }
        \label{fig1}
\end{figure}

In this Letter, we present three main results. First, we demonstrate that the characteristic splitting of the NV's magnetic resonance spectrum (Fig.~\ref{fig1}a), observed in ensemble NV experiments \cite{Gruber:1997,Kucsko:2013,Barson:2017,Dolde:2011,Igarashi:2012,2017arXiv170607935F,Zhu:2014,PhysRevB.87.224106,PhysRevLett.104.070801,Kubo:2010,Lai:2009,Bourgeois:2015ke,Rondin:2014kd,PhysRevB.93.024305,Simanovskaia:2013,Matsuzaki:2016,PhysRevA.95.053417,Levchenko:2015,Steele:2017cm}, originates from its local electric environment; this contrasts with the conventional picture that strain dominates the zero-field properties of these systems.
Second, we introduce a charge-based model (Fig.~\ref{fig1}a, right inset) that quantitatively reproduces the observed ODMR spectra for samples spanning two orders of magnitude in NV density.
Third, our model suggests the ability to directly \emph{image} the position of individual charges inside the diamond lattice.
To this end, we propose and implement a novel method that localizes such charges to nanometer-size volumes (Fig.~\ref{fig1}b,c).
The essence of our approach is to leverage the interplay between the polarization of the applied microwave field and the orientation of the local electric field.

%%%%%%%%%%%%%%%%%%
%
% NV Introduction + Strain discussion
%
%%%%%%%%%%%%%%%%%%

\begin{figure}[t]
  \centering
  \includegraphics[width = 3.45in]{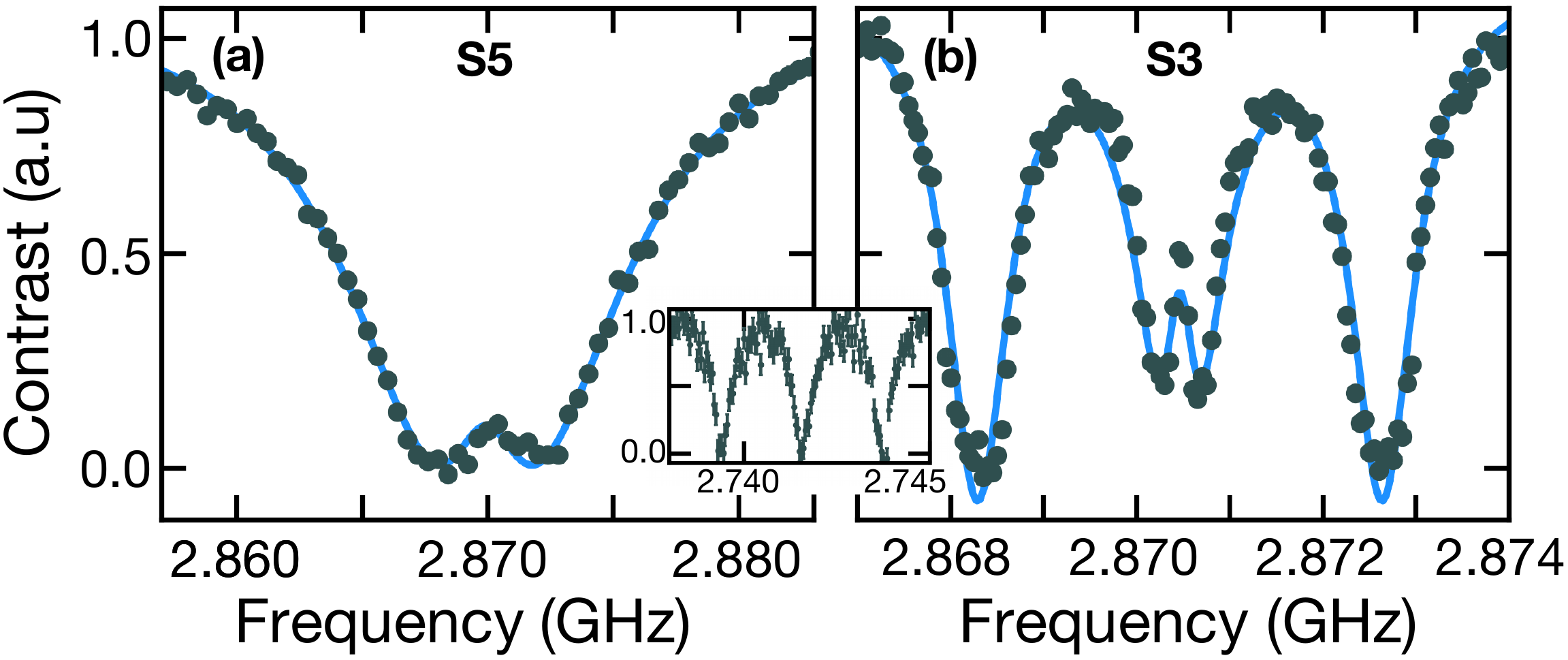}
  \caption{
    ODMR spectra at zero magnetic field for (a) a Type-Ib untreated diamond sample (S5) and (b) a Type-IIa electron-irradiated and annealed sample (S3).
    %
    %Error bars account for one standard deviation statistical error associated with photon counting.
    %
    The spectra portray the two qualitative regimes one expects based upon the average electric field strength as shown schematically in the right panel of Fig.~\ref{fig3}d.
The blue theory curve is obtained via our microscopic charge model.
(inset) The spectrum for S3 at a magnetic field $\approx 45$ G exhibits three identical hyperfine resonances.
  }
  \label{fig2}
\end{figure}

%\emph{NV$^{-}$ spectrum at zero magnetic field}---
%The magnetic spectrum of an isolated NV$^{-}$ center has been well understood.

\begin{figure}[t]
  \centering
 \includegraphics[width=3.4in]{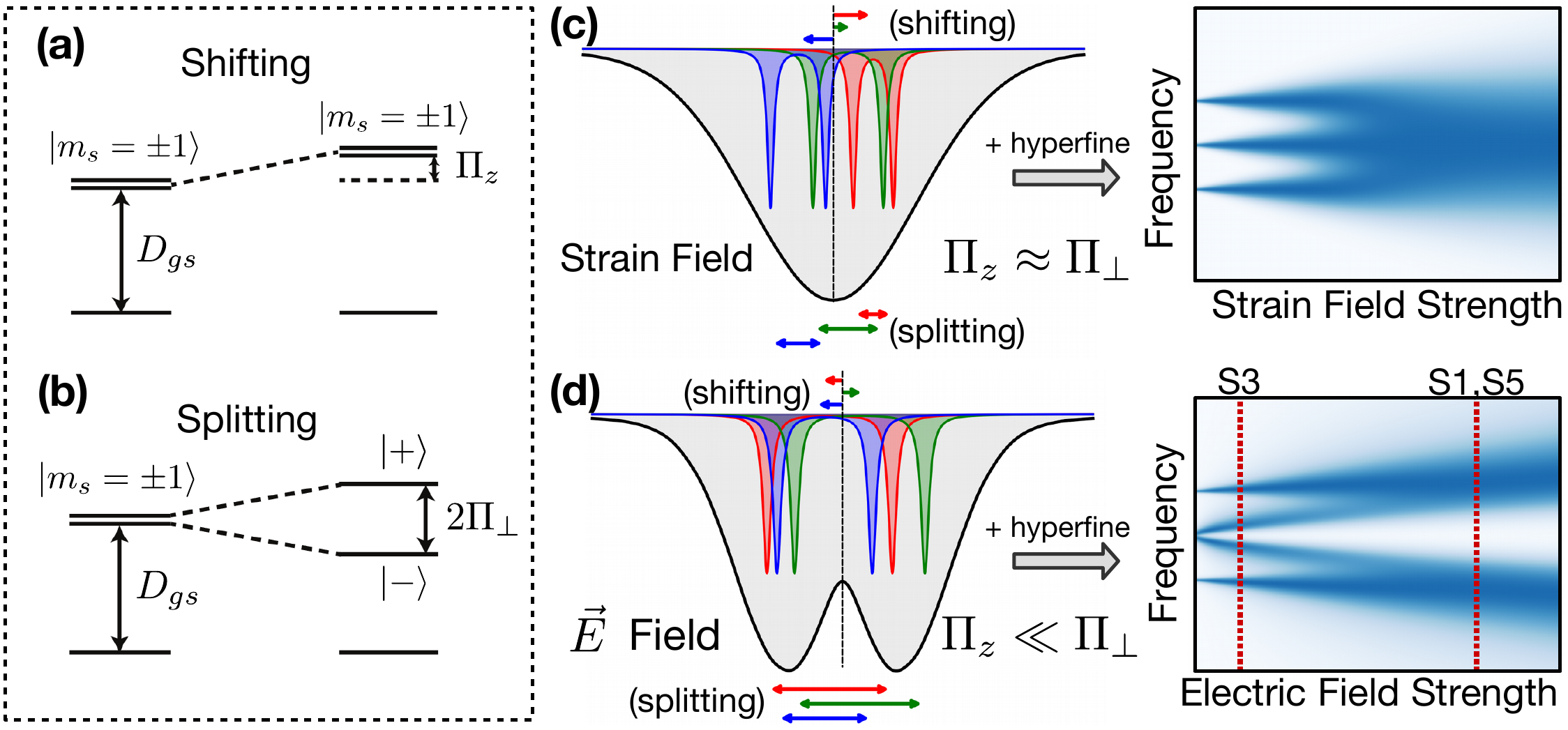}
        \caption{
          Both strain and electric fields lead to (a) shifting $\Pi_z$ and (b) splitting $2\Pi_\perp$ of the $\ket{m_s=\pm 1}$ manifold.
          (c) When averaged over an ensemble of NV centers, random local strain fields lead to a single broad spectral feature (at large strain).
           (d) In contrast, random local electric fields lead to two distinct spectral regimes:
          at small electric fields, the center hyperfine resonance splits, leading to a total of four resolvable features (S3);
          at large electric field, one obtains the characteristic split resonance seen in typical high density NV ensembles (S1, S5).
          %Shaded gray lineshape is a simulated disorder average over randomized strain and electric field given measured susceptibility parameters \cite{VANOORT1990,Barson:2017}.
          }
        \label{fig3}
\end{figure}

\emph{Magnetic spectra of NV ensembles}---The NV center has a spin triplet ground state ($\ket{m_s=\pm1,0}$), which can be initialized and read out via optical excitation and coherently manipulated using microwave fields~\cite{Maze:2011gw}.
In the absence of any external perturbations, the $\ket{m_s = \pm 1}$ states are  degenerate and separated from $\ket{m_s=0}$ by $D_{\textrm{gs}} = (2\pi) \times 2.87$ GHz (Fig.~\ref{fig3}a).
This leads to the usual expectation of a single resonance peak at $D_{\textrm{gs}}$, consistent with  experimental observations of isolated NVs (Fig.~\ref{fig1}a, inset).
However, for high-density NV ensembles, one observes a qualitatively distinct spectrum, consisting of a pair of resonances centered at $D_{\textrm{gs}}$  (Fig.~\ref{fig1}a, sample S1).
%caption should be clear about type of sample
%
This spectrum poses a number of puzzles: First, the line-shape of each resonance is  asymmetric and cannot be captured by either a Gaussian or Lorentzian profile. Second, the central feature between the resonances is sharper than the inhomogenous linewidth. Third, despite the presence of a strong \emph{splitting}, there exists almost no \emph{shift} of the NV's overall spectrum.

These generic features are present in diamond samples with NV and P1 (nitrogen impurity) densities spanning over two orders of magnitude.
Fig.~\ref{fig2} demonstrates this ubiquity. In particular, it depicts the spectrum for two other samples: one with a significantly lower NV concentration (Fig.~\ref{fig2}a, sample S5) and a second with significantly lower concentrations for both NVs and P1s (Fig.~\ref{fig2}b, sample S3).
In this latter case, the P1 density is low enough that the hyperfine interaction between the NV's electronic spin and its host $^{14}$N nuclear spin can be resolved. Normally, this hyperfine splitting would simply result in three identical resonances split from one another by $A_{zz} = (2\pi)\times 2.16$ MHz \cite{Smeltzer:2011fb} (Fig.~\ref{fig2}, inset).
However, as shown in Fig.~\ref{fig2}b, one finds that the central hyperfine resonance is split in direct analogy to the prior spectra.

% These observations have been previously attributed to diamond lattice strain.
% While strain can lift the degeneracy, the nature of the couplings should also lead to a shift
% In particular these effects are given by susceptibilities which are comparable to one another
% As a result when ensemble averaged we should see a

The most distinct of the aforementioned features -- a split central resonance -- has typically been attributed to the presence of lattice strain \cite{Barson:2017,Dolde:2011,Igarashi:2012,2017arXiv170607935F,Zhu:2014,PhysRevB.87.224106,PhysRevLett.104.070801,Kubo:2010,Lai:2009,Bourgeois:2015ke,Rondin:2014kd,PhysRevB.93.024305,Simanovskaia:2013,Matsuzaki:2016,PhysRevA.95.053417,Levchenko:2015,Steele:2017cm}.
%because it lift the degeneracies between the two spin states $\ket{m_s=\pm1}$~\cite{}.
%, whose amount may vary depending on the crystallographic orientation of NV$^{-}$ centers or the (potentially random) strength  of local strain field in diamond.
%of giving rise to a broad spectrum in an ensemble measurement.
Such strain can indeed lead to a coupling between the $\ket{m_s=\pm1}$ states, and thus split their energy levels.
However, a more careful analysis reveals an important inconsistency.
In particular, given the measured strain susceptibility parameters \cite{Barson:2017}, for each individual NV, any strain-induced splitting should be accompanied by a comparable shift of the overall spectrum (Fig.~\ref{fig3}).
Ensemble averaging then naturally leads to a spectrum that exhibits \emph{only} a single broadened resonance (Fig.~\ref{fig3}c).
%
% BRYCE TENSOR
%In fact, $\Pi_\perp \approx \Pi_\parallel$ is a direct consequence of both couplings originating from the same physical mechanism (distortion of ground state orbital) \cite{your_paper}.

%%%%%%%%%%%%%%%%%%
%
%  Charge Model
%
%%%%%%%%%%%%%%%%%%

\emph{Microscopic charge model}---In contrast, we demonstrate that all of the observed features  can be quantitatively explained via a microscopic model based upon randomly positioned charges inside the diamond lattice.
The physical intuition underlying this model is simple: each (negatively charged) NV center plays the role of an electron acceptor, and charge neutrality implies that there must be a corresponding positively charged electron donor (typically thought to be N$^{+}$, a positively charged P1 center).

Such charges produce an electric field that also (like strain) couples the $\ket{m_s=\pm1}$ states, leading to the splitting of the resulting eigenstates.
Crucially, however, the NV's susceptibility to transverse electric fields (which cause splitting)  is  $\sim$50 times larger than its susceptibility to axial electric fields (which cause shifting) \cite{VANOORT1990, Kobrin_18xx}.
This implies that even upon ensemble averaging, the electric-field-induced splitting remains prominent (Fig.~\ref{fig3}d).

%which splits and shifts the $\ket{m_s=\pm 1}$ manifold into two dressed eigenstates $\ket{\pm}$.

Qualitative picture in hand, let us now introduce the details of our microscopic model.
In particular, we consider each NV to be surrounded by an equal density, $\rho_c$, of positive and negative charges~\footnote{We assume that the charges are independently positioned in three dimensions}.
These charges generate a local electric field at the position of the NV center and couple to its spin via the Hamiltonian:
\begin{align} \label{eq:Hamil}
H=&\left(D_{gs}+\Pi_z\right) S_z^2 + (\delta B_z +A_{zz} I_z ) S_z +  \notag\\
& \Pi_x(S_y^2-S_x^2) + \Pi_y(S_xS_y+S_yS_x).
\end{align}
Here, $\hat{z}$ is the NV-axis, $\hat{x}$ is defined such that one of the carbon-vacancy bonds lies in the x-z plane (Fig.~\ref{fig1}a, right inset),
$\vec{S}$ are the electronic spin-1 operators of the NV, $\vec{I}$ are the nuclear spin-1 operators of the host $^{14}$N \footnote{We note that the the hyperfine interaction in the Hamiltonian is obtained under the secular approximation.}, and
 $\delta B_z$ represents a random local magnetic field (for example,~generated by nearby paramagnetic impurities). %
Note that we absorb the gyromagnetic ratio into $\delta B_z$.
The two terms $ \Pi_{\{x,y\}} = d_\perp E_{\{x,y\}}$ and $ \Pi_z = d_\| E_z$ characterize the NV's coupling to the electric field, $\vec{E}$, with susceptibilities $\left\{d_\parallel,d_\perp\right\} = \left\{0.35,17\right\}$ Hz cm/V~\cite{VANOORT1990}. %Cite
%
%$\Pi_\perp = \sqrt{\Pi_x^2 + \Pi_y^2}$ characterizes the electric field induced splitting of $\ket{\pm}$.
%

%In order to faithfully reproduce the spectra we further include an extra inhomogeneous Lorentzian broadening ? of each resonance. The origin of this broadening can be varied, arising from interactions, power broadening, among others.
%some intel- ligent discussion about ? should come here, including nuclear spin bath, lattice strain, power broadening, etc.

\begin{table}
\begin{tabular}{c|c|c|c|c}
  \hline \hline
  Sample &  \makecell{$\rho_c$ \\(ppm)} & \makecell{$\rho_{\text{NV}}$ \\(ppm)} & \makecell{$\rho_s$\\(ppm)} & \makecell{$\Gamma$\\ (MHz)}\\ \hline
  \makecell{Ib treated  (S1)} &  1.35(5)  & 1-10 &  70(5) & 1.16(2)\\
  \makecell{Ib treated  (S2) } &  1.7(1)  & 1-10 &  100(5)  & 0.78(3)\\
  \makecell{IIa treated  (S3)}&  0.06(2) & 0.01-0.1 & 12(3) & 0.26(2) \\ \hline
  \makecell{Ib untreated  (S4)} & 3.6(4) & 0.001-0.01 & 90(20) & 1.0(1) \\
  \makecell{Ib untreated  (S5)} & 0.9(2) & 0.001-0.01 & 130(30) & 3.3(1)\\
  \makecell{IIa untreated (S6)}&  0.05(1) & 0.001-0.01 & 16(2)& 0.08(3) \\
  \hline \hline
\end{tabular}
\caption{Summary of the measured and extracted parameters for each diamond sample.
 $\rho_c$ and $\Gamma$ are directly extracted from our microscopic model, while  $\rho_s$ is independently measured at high magnetic fields and $\rho_\textrm{NV}$ is estimated from fluorescence counts \cite{supp}.
 %
  %In samples S1 and S2, $\delta B$ is determined from the dipolar coupling to P1 spin bath, whose concentration has been independently measured.
  %In sample S3, $\delta B$ is a fitting parameter that characterizes interactions with the spin bath as well as the remnant global magnetic field $\le 0.1$ G.
}
\label{tab:Tab1}
\end{table}
\begin{figure}
  \centering
  \includegraphics[width = 3.4in]{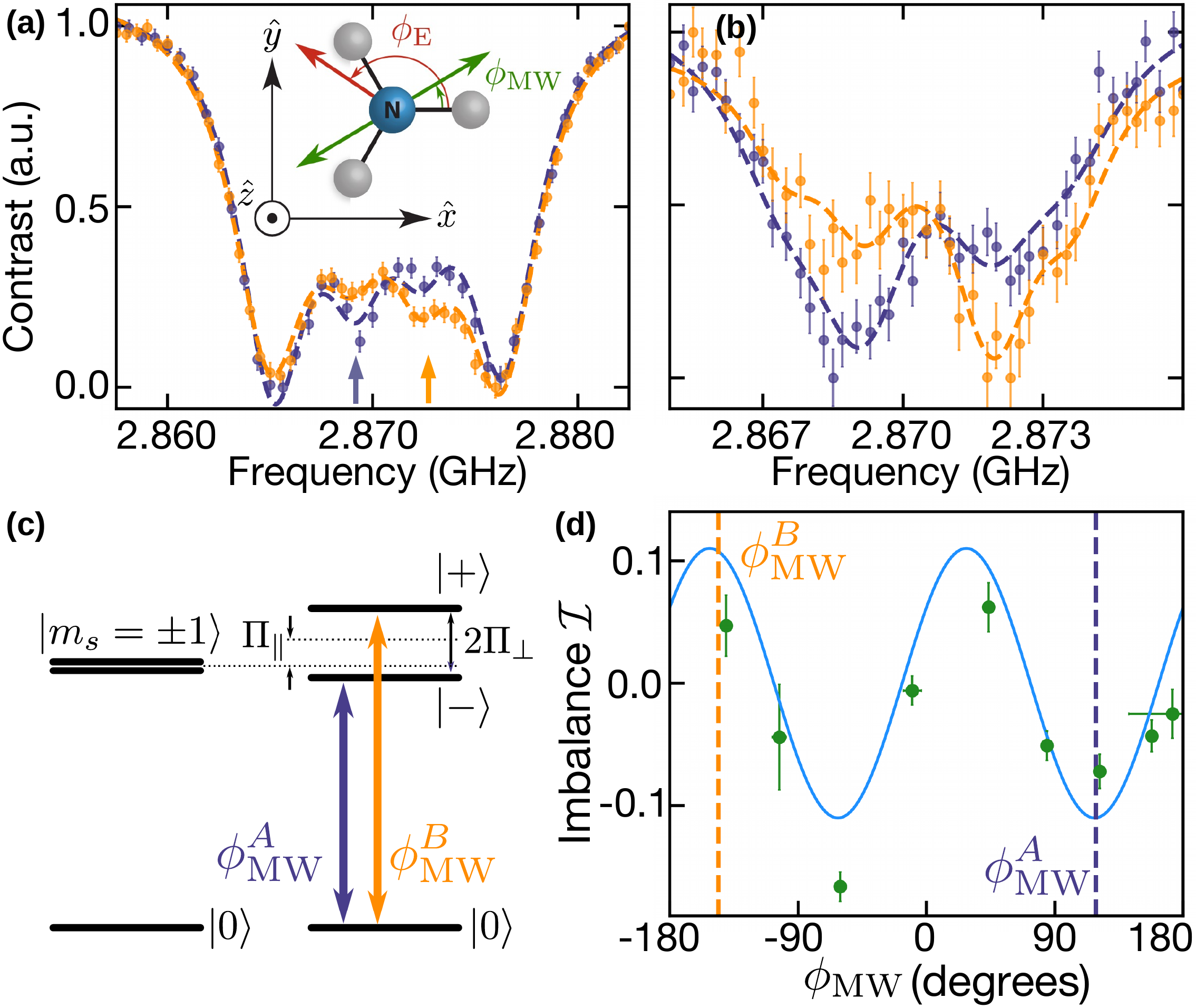}
  \caption{
    Charge localization via dark-state spectroscopy. (a) Single NV ODMR spectra (untreated Type-Ib diamond) for two different microwave polarizations, $\phi_{\textrm{MW}}$, depicting the reversal of the split-peak imbalance. The data correspond to the localized charge shown in Fig.~\ref{fig1}b. (inset) Top view through the NV-axis ($\hat{z}$), where $\phi_E$ and $\phi_{\textrm{MW}}$ are defined with respect to $\hat{x}$ (along a carbon-vacancy bond).
     (b) Analogous split-peak imbalance data corresponding to the localized charge shown in Fig.~1c.
    (c) By changing the microwave polarization, $\phi_{\textrm{MW}}$, one can directly control the coupling strength between the $\ket{0}$ and $\ket{\pm}$ states.
     (d) Measuring the change in the imbalance as a function of $\phi_{\textrm{MW}}$ allows one to extract the orientation of the electric field.  Dashed lines indicate the polarizations plotted in (a).
  }
  \label{fig4}
\end{figure}

In order to obtain the spectra for a single NV, we sample $\vec{E}$ and $\delta B_z$ from their random distributions and then diagonalize the Hamiltonian.
Moreover, to account for the natural linewidth of each resonance, we include an additional Lorentzian broadening with full-width-half-maximum, $\Gamma$ \cite{supp}.
Averaging over this procedure yields the ensemble spectrum.
The distribution of $\vec{E}$ is determined by the random positioning of the aforementioned charges.
The distribution of $\delta B_z$ is determined by the local magnetic environment, which depends sensitively on the concentration of spin defects (Table~\ref{tab:Tab1}).

In samples S1 and S5 (Type-Ib diamond), $\delta B_z$ is dominated by the dipolar interaction with a high-density P1 spin bath, whose concentration, $\rho_{\text{s}}$, is independently characterized \cite{supp}.
Meanwhile, in sample S3 (Type-IIa diamond), the P1 density is over two orders of magnitude smaller, leading to a $\delta B_z$ that is dominated by interactions with $^{13}$C nuclei (with a natural abundance of $1.1\%$); despite this difference in microscopic origin, one can also characterize the effect of this nuclear spin bath using an effective density, $\rho_s$ \cite{supp}.
For each sample, using this independently characterized $\rho_\textrm{s}$, we then fit the experimental spectrum by varying $\rho_\textrm{c}$ and $\Gamma$ .
We find excellent agreement for all three samples (Fig.~\ref{fig1}, \ref{fig2}) despite their vastly different defect concentrations (Table~\ref{tab:Tab1}).

A few remarks are in order. First, the presence of local electric fields suppresses the effect of magnetic noise when $\delta B_z \ll \Pi_\perp = \sqrt{\Pi_x^2+ \Pi_y^2}$. This is precisely the origin for both the sharpness of the inner central feature seen in Fig.~\ref{fig1}a, as well as the narrowness of the inner hyperfine resonances seen in Fig.~\ref{fig2}b.
Second, in samples where the electric field dominates, the long-range, power-law nature of the electric field leads to a particularly heavy tailed spectrum \cite{supp}.
Third, the extracted charge density, $\rho_\textrm{c}$, is consistent with the estimated NV density, $\rho_\textrm{NV}$, for all ``treated'' (electron-irradiated and annealed) samples (S1-S3).
This agrees with our previous physical intuition: NVs behave as electron acceptors while P1s behave as electron donors.
Interestingly, this simple picture does not directly translate to ``untreated'' samples (S4-S6) where the observed charge density is significantly larger than $\rho_\textrm{NV}$ (Table~\ref{tab:Tab1}); one possible explanation is that such samples harbor a higher density of non-NV charged defects (e.g.~vacancy complexes \cite{DAVIES:1977kt}).
%

%%%%%%%%%%%%%%%%
%
% Experimental Data
%
%%%%%%%%%%%%%%%%

%%%%%%%%%%%%%%%%%%
%
%  Discussion
%
%%%%%%%%%%%%%%%%%%

%- This model also provides insight into the origin of the sharp inner feature.
%% Moreover, it also elucidates the origin of the asymmetric sharp inner feature, which originates from the competition between the electric field and the local magnetic noise.
%% In the absence of electric field, the magnetic noise $B_z$ leads to a \textit{symmetric} lineshape arising from the symmetric distribution of $B_z$.
%% However, in the presence of an electric field, the energy splitting is given by $2\sqrt{(gB_z)^2 + \Pi_\perp^2}$, where $g$ is the Land\'{e} g-factor of NV$^-$ electron.
%% As a result, for the local charge configuration of each NV$^-$, there is a minimum splitting of $2\Pi_\perp$ and the presence of $B_z$ can only increase it, leading to an \emph{asymmetric lineshape}.
%% Because most configurations have a non-zero minimum splitting $2\Pi_\perp$, the ensemble averaging preserves this sharp inner feature.
%% These previous results suggest that the density of the charge defects is related to the density of NV$^-$.

%%%%%%%%%%%%%%%%%%
%
%  Single NV Localization Theory
%
%%%%%%%%%%%%%%%%%%

\textit{Nanoscale imaging of a single charge}---Our microscopic model suggests that in samples where one can resolve single NV centers, it should be possible to directly probe the \emph{local} charge environment.
However, one expects a key difference in contrast to ensemble measurements: for a single NV, the electric field has a definite orientation with respect to the NV axes (Fig.~\ref{fig4}a diagram).

Crucially, this orientation (namely, the angle, $\phi_E$, in the NV's transverse plane) dictates the way in which the electric field mixes the original $\ket{m_s= \pm1}$ states into bright and dark states: %, imparting a corresponding phase between the mixed states.
\begin{equation}
  \ket{\pm} = \frac{1}{\sqrt{2}} \left(\ket{m_s= +1}\mp e^{-i\phi_E} \ket{m_s = -1}\right).
\end{equation}
Applying a linearly polarized microwave field will then drive transitions between the $\ket{m_s=0}$ state and the $\ket{\pm}$ states.
However, the relative strength of the two transitions depends on both $\phi_E$ and the polarization of the microwave field, $\phi_{\textrm{MW}}$ (Fig.~\ref{fig4}c).
Thus, one generally expects the measured amplitudes of the corresponding resonances to be different.
These expectations are indeed borne out by the data (Fig.~\ref{fig4}a,b) \footnote{We measure the ODMR spectra of 68 single NV centers in an untreated Type-Ib sample, and find four that exhibit a significant electric-field-induced splitting with amplitude difference at zero magnetic field \cite{supp}.}.
We note that this observed imbalance in the inner hyperfine resonances for a \emph{single} NV is naturally averaged out in an ensemble measurement.

Our detailed understanding of this spectroscopy for a single NV suggests a novel method to extract the full vector electric field and to localize the position of the corresponding charge.
In particular, by measuring the imbalance as a function of $\phi_{\textrm{MW}}$, one can extract the electric field orientation, $\phi_E$. More specifically, we define the imbalance, $\mathcal{I} \equiv \frac {A_+ - A_-}{A_+ + A_-} $, where  $A_{\pm}$ are the amplitudes of the $\ket{m_s=0} \leftrightarrow \ket{\pm}$ resonances and derive \cite{supp}:
\begin{equation} \label{eq:imbalance}
  \mathcal{I} \sim -\cos(2\phi_{MW}+\phi_E).
\end{equation}
Thus, $\phi_E = 124(5)^\circ$ can be extracted as the phase offset in Fig.~4d.
In combination with the observed splitting and shifting of the inner resonances, $\Pi_z = 30(50)$ kHz, $\Pi_\perp = 650(10)$ kHz, one can fully reconstruct the local electric field vector \cite{supp}.
%\begin{align}
%  \bm{E}_{local} &= \big\{1.8(3), -3.3(4), 0.0(4)\big\} ~\mathrm{MV/m}.
%\end{align}
We do not observe any changes to this field over the course of the experiment (months) and find that it varies for different NV centers. This suggests that it originates from a stationary local charge environment.
Moreover, charge neutrality and a low defect density suggest that the electric field is generated by a single positive charge, which we can then localize to within a nanoscale volume (Fig. \ref{fig1}b,c).

\emph{Summary and outlook}---While it is abundantly asserted in the literature that the zero-field spectral features of NV ensembles owe to lattice strain, here, we demonstrate that such spectra are in fact dominated by the effect of local electric fields.
Using a microscopic charge model, we quantitatively capture the magnetic resonance spectra of NV ensembles for defect concentrations spanning two orders of magnitude.
Moreover, we introduce a method to image the spatial location of individual charges near a single NV center with nanoscale precision.

These results open the door to a number of intriguing future directions. First, although we observe charge densities that are consistent with the NV density in all treated samples (and thus consistent with a picture for charge neutrality), we find a deviation from this understanding for untreated samples which exhibit an anomalously large charge density.
Further study is necessary to reveal the precise nature of these additional charges.
Second, our results provide an improved understanding of NV ensembles at low magnetic fields;
%this is of particular relevance to the sensing of electric fields and lattice strain, as well as to studies of magnetically sensitive quantum materials
this is of particular relevance to the sensing of electric fields, lattice strain and gyroscopic precession, as well as to studies of magnetically sensitive quantum materials.
 Third, the charge-induced suppression of $\delta B_z$ suggests the possibility of enhancing the NV's resilience to magnetic noise.
 Finally, understanding the local charge environment of single NV centers could provide insights into the optical spectral diffusion observed at low temperatures \cite{Chu:2014es,Jelezko:2002kx}.

%Charge dynamics: Understand the current defects
%Charge Dynamics: controlling the NV enviroment
%Different Densities: Understand the current defects
%Different Densities:  controlling the NV enviroment

%What about other defects eg SiV? What characterizes their zero field

%  Future direction
%%  - more systematic study as function of density
%  - ways to control charge environment;study nanoscale charge dynamics
%  - apply characterization techniques to other defects

%Because the local environment is dominated by charge, controlling the charge dynamics can pave the new to even greater control over the properties of the NV$^-$.
%Having understood the static properties of the charge environment,

%% Our work can be extended in several new directions.
%% It has been demonstrated that one can selectively ionize different kinds of defects using laser excitations at various wavelengths.
%% Such a technique may allow controlling the charge environment of

%% (different laser color to control charge.)
%% (our novel method to MW can be used for other defects).

% Applications
%  - necessary for sensing at low-field (strain, e-field)
%  - suppressing magnetic noise

We gratefully acknowledge the insights of and discussions
with  A.~Blezynski-Jayich, B.~Hausmann, J.~Moore, P.~Maurer, P.~Kehayias, J.~Choi, E.~Demler, and M.~Lukin.
This work was supported as part of the Center for Novel Pathways to Quantum Coherence in Materials, an Energy Frontier Research Center funded by the U.S. Department of Energy, Office of Science, Basic Energy Sciences under Award \#DE-AC02-05CH11231.
SH acknowledges support by the National Science Foundation Graduate Research Fellowship under Grant No. DGE 1752814. AJ acknowledges support from the Army Research Laboratory under Cooperative Agreement No. W911NF-18-2-0037. SC acknowledges the Miller Institute for Basic Research in Science. DB acknowledges support by the EU FET-OPEN Flagship Project ASTERIQS (action \#820394), the German Federal Ministry of Education and Research (BMBF) within the Quantumtechnologien program (FKZ 13N14439), and the DFG through the DIP program (FO 703/2-1).

\bibliography{2018-charge-v1.5.bib}

%merlin.mbs apsrev4-1.bst 2010-07-25 4.21a (PWD, AO, DPC) hacked
%Control: key (0)
%Control: author (8) initials jnrlst
%Control: editor formatted (1) identically to author
%Control: production of article title (-1) disabled
%Control: page (0) single
%Control: year (1) truncated
%Control: production of eprint (0) enabled
\begin{thebibliography}{68}%
\makeatletter
\providecommand \@ifxundefined [1]{%
 \@ifx{#1\undefined}
}%
\providecommand \@ifnum [1]{%
 \ifnum #1\expandafter \@firstoftwo
 \else \expandafter \@secondoftwo
 \fi
}%
\providecommand \@ifx [1]{%
 \ifx #1\expandafter \@firstoftwo
 \else \expandafter \@secondoftwo
 \fi
}%
\providecommand \natexlab [1]{#1}%
\providecommand \enquote  [1]{``#1''}%
\providecommand \bibnamefont  [1]{#1}%
\providecommand \bibfnamefont [1]{#1}%
\providecommand \citenamefont [1]{#1}%
\providecommand \href@noop [0]{\@secondoftwo}%
\providecommand \href [0]{\begingroup \@sanitize@url \@href}%
\providecommand \@href[1]{\@@startlink{#1}\@@href}%
\providecommand \@@href[1]{\endgroup#1\@@endlink}%
\providecommand \@sanitize@url [0]{\catcode `\\12\catcode `\$12\catcode
  `\&12\catcode `\#12\catcode `\^12\catcode `\_12\catcode `\%12\relax}%
\providecommand \@@startlink[1]{}%
\providecommand \@@endlink[0]{}%
\providecommand \url  [0]{\begingroup\@sanitize@url \@url }%
\providecommand \@url [1]{\endgroup\@href {#1}{\urlprefix }}%
\providecommand \urlprefix  [0]{URL }%
\providecommand \Eprint [0]{\href }%
\providecommand \doibase [0]{http://dx.doi.org/}%
\providecommand \selectlanguage [0]{\@gobble}%
\providecommand \bibinfo  [0]{\@secondoftwo}%
\providecommand \bibfield  [0]{\@secondoftwo}%
\providecommand \translation [1]{[#1]}%
\providecommand \BibitemOpen [0]{}%
\providecommand \bibitemStop [0]{}%
\providecommand \bibitemNoStop [0]{.\EOS\space}%
\providecommand \EOS [0]{\spacefactor3000\relax}%
\providecommand \BibitemShut  [1]{\csname bibitem#1\endcsname}%
\let\auto@bib@innerbib\@empty
%</preamble>
\bibitem [{\citenamefont {Aharonovich}\ \emph {et~al.}(2016)\citenamefont
  {Aharonovich}, \citenamefont {Englund},\ and\ \citenamefont
  {Toth}}]{Aharonovich:2016hm}%
  \BibitemOpen
  \bibfield  {author} {\bibinfo {author} {\bibfnamefont {I.}~\bibnamefont
  {Aharonovich}}, \bibinfo {author} {\bibfnamefont {D.}~\bibnamefont
  {Englund}}, \ and\ \bibinfo {author} {\bibfnamefont {M.}~\bibnamefont
  {Toth}},\ }\href@noop {} {\bibfield  {journal} {\bibinfo  {journal} {Nature
  Photonics}\ }\textbf {\bibinfo {volume} {10}},\ \bibinfo {pages} {631}
  (\bibinfo {year} {2016})}\BibitemShut {NoStop}%
\bibitem [{\citenamefont {Schirhagl}\ \emph {et~al.}(2014)\citenamefont
  {Schirhagl}, \citenamefont {Chang}, \citenamefont {Loretz},\ and\
  \citenamefont {Degen}}]{Schirhagl:2014}%
  \BibitemOpen
  \bibfield  {author} {\bibinfo {author} {\bibfnamefont {R.}~\bibnamefont
  {Schirhagl}}, \bibinfo {author} {\bibfnamefont {K.}~\bibnamefont {Chang}},
  \bibinfo {author} {\bibfnamefont {M.}~\bibnamefont {Loretz}}, \ and\ \bibinfo
  {author} {\bibfnamefont {C.~L.}\ \bibnamefont {Degen}},\ }\href {\doibase
  10.1146/annurev-physchem-040513-103659} {\bibfield  {journal} {\bibinfo
  {journal} {Annual Review of Physical Chemistry}\ }\textbf {\bibinfo {volume}
  {65}},\ \bibinfo {pages} {83} (\bibinfo {year} {2014})},\ \bibinfo {note}
  {pMID: 24274702},\ \Eprint
  {http://arxiv.org/abs/https://doi.org/10.1146/annurev-physchem-040513-103659}
  {https://doi.org/10.1146/annurev-physchem-040513-103659} \BibitemShut
  {NoStop}%
\bibitem [{\citenamefont {Doherty}\ \emph {et~al.}(2013)\citenamefont
  {Doherty}, \citenamefont {Manson}, \citenamefont {Delaney}, \citenamefont
  {Jelezko}, \citenamefont {Wrachtrup},\ and\ \citenamefont
  {Hollenberg}}]{doherty2013nitrogen}%
  \BibitemOpen
  \bibfield  {author} {\bibinfo {author} {\bibfnamefont {M.~W.}\ \bibnamefont
  {Doherty}}, \bibinfo {author} {\bibfnamefont {N.~B.}\ \bibnamefont {Manson}},
  \bibinfo {author} {\bibfnamefont {P.}~\bibnamefont {Delaney}}, \bibinfo
  {author} {\bibfnamefont {F.}~\bibnamefont {Jelezko}}, \bibinfo {author}
  {\bibfnamefont {J.}~\bibnamefont {Wrachtrup}}, \ and\ \bibinfo {author}
  {\bibfnamefont {L.~C.}\ \bibnamefont {Hollenberg}},\ }\href@noop {}
  {\bibfield  {journal} {\bibinfo  {journal} {Physics Reports}\ }\textbf
  {\bibinfo {volume} {528}},\ \bibinfo {pages} {1} (\bibinfo {year}
  {2013})}\BibitemShut {NoStop}%
\bibitem [{\citenamefont {Maze}\ \emph {et~al.}(2008)\citenamefont {Maze},
  \citenamefont {Stanwix}, \citenamefont {Hodges}, \citenamefont {Hong},
  \citenamefont {Taylor}, \citenamefont {Cappellaro}, \citenamefont {Jiang},
  \citenamefont {Dutt}, \citenamefont {Togan}, \citenamefont {Zibrov} \emph
  {et~al.}}]{maze2008nanoscale}%
  \BibitemOpen
  \bibfield  {author} {\bibinfo {author} {\bibfnamefont {J.}~\bibnamefont
  {Maze}}, \bibinfo {author} {\bibfnamefont {P.}~\bibnamefont {Stanwix}},
  \bibinfo {author} {\bibfnamefont {J.}~\bibnamefont {Hodges}}, \bibinfo
  {author} {\bibfnamefont {S.}~\bibnamefont {Hong}}, \bibinfo {author}
  {\bibfnamefont {J.}~\bibnamefont {Taylor}}, \bibinfo {author} {\bibfnamefont
  {P.}~\bibnamefont {Cappellaro}}, \bibinfo {author} {\bibfnamefont
  {L.}~\bibnamefont {Jiang}}, \bibinfo {author} {\bibfnamefont {M.~G.}\
  \bibnamefont {Dutt}}, \bibinfo {author} {\bibfnamefont {E.}~\bibnamefont
  {Togan}}, \bibinfo {author} {\bibfnamefont {A.}~\bibnamefont {Zibrov}},
  \emph {et~al.},\ }\href@noop {} {\bibfield  {journal} {\bibinfo  {journal}
  {Nature}\ }\textbf {\bibinfo {volume} {455}},\ \bibinfo {pages} {644}
  (\bibinfo {year} {2008})}\BibitemShut {NoStop}%
\bibitem [{\citenamefont {Mamin}\ \emph {et~al.}(2013)\citenamefont {Mamin},
  \citenamefont {Kim}, \citenamefont {Sherwood}, \citenamefont {Rettner},
  \citenamefont {Ohno}, \citenamefont {Awschalom},\ and\ \citenamefont
  {Rugar}}]{mamin2013nanoscale}%
  \BibitemOpen
  \bibfield  {author} {\bibinfo {author} {\bibfnamefont {H.}~\bibnamefont
  {Mamin}}, \bibinfo {author} {\bibfnamefont {M.}~\bibnamefont {Kim}}, \bibinfo
  {author} {\bibfnamefont {M.}~\bibnamefont {Sherwood}}, \bibinfo {author}
  {\bibfnamefont {C.}~\bibnamefont {Rettner}}, \bibinfo {author} {\bibfnamefont
  {K.}~\bibnamefont {Ohno}}, \bibinfo {author} {\bibfnamefont {D.}~\bibnamefont
  {Awschalom}}, \ and\ \bibinfo {author} {\bibfnamefont {D.}~\bibnamefont
  {Rugar}},\ }\href@noop {} {\bibfield  {journal} {\bibinfo  {journal}
  {Science}\ }\textbf {\bibinfo {volume} {339}},\ \bibinfo {pages} {557}
  (\bibinfo {year} {2013})}\BibitemShut {NoStop}%
\bibitem [{\citenamefont {Toyli}\ \emph {et~al.}(2012)\citenamefont {Toyli},
  \citenamefont {Christle}, \citenamefont {Alkauskas}, \citenamefont {Buckley},
  \citenamefont {Van~de Walle},\ and\ \citenamefont
  {Awschalom}}]{toyli2012measurement}%
  \BibitemOpen
  \bibfield  {author} {\bibinfo {author} {\bibfnamefont {D.}~\bibnamefont
  {Toyli}}, \bibinfo {author} {\bibfnamefont {D.}~\bibnamefont {Christle}},
  \bibinfo {author} {\bibfnamefont {A.}~\bibnamefont {Alkauskas}}, \bibinfo
  {author} {\bibfnamefont {B.}~\bibnamefont {Buckley}}, \bibinfo {author}
  {\bibfnamefont {C.}~\bibnamefont {Van~de Walle}}, \ and\ \bibinfo {author}
  {\bibfnamefont {D.}~\bibnamefont {Awschalom}},\ }\href@noop {} {\bibfield
  {journal} {\bibinfo  {journal} {Physical Review X}\ }\textbf {\bibinfo
  {volume} {2}},\ \bibinfo {pages} {031001} (\bibinfo {year}
  {2012})}\BibitemShut {NoStop}%
\bibitem [{\citenamefont {Acosta}\ \emph
  {et~al.}(2010{\natexlab{a}})\citenamefont {Acosta}, \citenamefont {Bauch},
  \citenamefont {Jarmola}, \citenamefont {Zipp}, \citenamefont {Ledbetter},\
  and\ \citenamefont {Budker}}]{Acosta:2010go}%
  \BibitemOpen
  \bibfield  {author} {\bibinfo {author} {\bibfnamefont {V.~M.}\ \bibnamefont
  {Acosta}}, \bibinfo {author} {\bibfnamefont {E.}~\bibnamefont {Bauch}},
  \bibinfo {author} {\bibfnamefont {A.}~\bibnamefont {Jarmola}}, \bibinfo
  {author} {\bibfnamefont {L.~J.}\ \bibnamefont {Zipp}}, \bibinfo {author}
  {\bibfnamefont {M.~P.}\ \bibnamefont {Ledbetter}}, \ and\ \bibinfo {author}
  {\bibfnamefont {D.}~\bibnamefont {Budker}},\ }\href@noop {} {\bibfield
  {journal} {\bibinfo  {journal} {Applied Physics Letters}\ }\textbf {\bibinfo
  {volume} {97}},\ \bibinfo {pages} {174104} (\bibinfo {year}
  {2010}{\natexlab{a}})}\BibitemShut {NoStop}%
\bibitem [{\citenamefont {Epstein}\ \emph {et~al.}(2005)\citenamefont
  {Epstein}, \citenamefont {Mendoza}, \citenamefont {Kato},\ and\ \citenamefont
  {Awschalom}}]{epstein2005anisotropic}%
  \BibitemOpen
  \bibfield  {author} {\bibinfo {author} {\bibfnamefont {R.}~\bibnamefont
  {Epstein}}, \bibinfo {author} {\bibfnamefont {F.}~\bibnamefont {Mendoza}},
  \bibinfo {author} {\bibfnamefont {Y.}~\bibnamefont {Kato}}, \ and\ \bibinfo
  {author} {\bibfnamefont {D.}~\bibnamefont {Awschalom}},\ }\href@noop {}
  {\bibfield  {journal} {\bibinfo  {journal} {Nature physics}\ }\textbf
  {\bibinfo {volume} {1}},\ \bibinfo {pages} {94} (\bibinfo {year}
  {2005})}\BibitemShut {NoStop}%
\bibitem [{\citenamefont {Dolde}\ \emph {et~al.}(2011)\citenamefont {Dolde},
  \citenamefont {Fedder}, \citenamefont {Doherty}, \citenamefont {N{\"o}bauer},
  \citenamefont {Rempp}, \citenamefont {Balasubramanian}, \citenamefont {Wolf},
  \citenamefont {Reinhard}, \citenamefont {Hollenberg}, \citenamefont
  {Jelezko},\ and\ \citenamefont {Wrachtrup}}]{Dolde:2011}%
  \BibitemOpen
  \bibfield  {author} {\bibinfo {author} {\bibfnamefont {F.}~\bibnamefont
  {Dolde}}, \bibinfo {author} {\bibfnamefont {H.}~\bibnamefont {Fedder}},
  \bibinfo {author} {\bibfnamefont {M.~W.}\ \bibnamefont {Doherty}}, \bibinfo
  {author} {\bibfnamefont {T.}~\bibnamefont {N{\"o}bauer}}, \bibinfo {author}
  {\bibfnamefont {F.}~\bibnamefont {Rempp}}, \bibinfo {author} {\bibfnamefont
  {G.}~\bibnamefont {Balasubramanian}}, \bibinfo {author} {\bibfnamefont
  {T.}~\bibnamefont {Wolf}}, \bibinfo {author} {\bibfnamefont {F.}~\bibnamefont
  {Reinhard}}, \bibinfo {author} {\bibfnamefont {L.~C.~L.}\ \bibnamefont
  {Hollenberg}}, \bibinfo {author} {\bibfnamefont {F.}~\bibnamefont {Jelezko}},
  \ and\ \bibinfo {author} {\bibfnamefont {J.}~\bibnamefont {Wrachtrup}},\
  }\href@noop {} {\bibfield  {journal} {\bibinfo  {journal} {Nature Physics}\
  }\textbf {\bibinfo {volume} {7}},\ \bibinfo {pages} {459} (\bibinfo {year}
  {2011})}\BibitemShut {NoStop}%
\bibitem [{\citenamefont {Dolde}\ \emph {et~al.}(2014)\citenamefont {Dolde},
  \citenamefont {Doherty}, \citenamefont {Michl}, \citenamefont {Jakobi},
  \citenamefont {Naydenov}, \citenamefont {Pezzagna}, \citenamefont {Meijer},
  \citenamefont {Neumann}, \citenamefont {Jelezko}, \citenamefont {Manson},\
  and\ \citenamefont {Wrachtrup}}]{Dolde:2014hc}%
  \BibitemOpen
  \bibfield  {author} {\bibinfo {author} {\bibfnamefont {F.}~\bibnamefont
  {Dolde}}, \bibinfo {author} {\bibfnamefont {M.~W.}\ \bibnamefont {Doherty}},
  \bibinfo {author} {\bibfnamefont {J.}~\bibnamefont {Michl}}, \bibinfo
  {author} {\bibfnamefont {I.}~\bibnamefont {Jakobi}}, \bibinfo {author}
  {\bibfnamefont {B.}~\bibnamefont {Naydenov}}, \bibinfo {author}
  {\bibfnamefont {S.}~\bibnamefont {Pezzagna}}, \bibinfo {author}
  {\bibfnamefont {J.}~\bibnamefont {Meijer}}, \bibinfo {author} {\bibfnamefont
  {P.}~\bibnamefont {Neumann}}, \bibinfo {author} {\bibfnamefont
  {F.}~\bibnamefont {Jelezko}}, \bibinfo {author} {\bibfnamefont {N.~B.}\
  \bibnamefont {Manson}}, \ and\ \bibinfo {author} {\bibfnamefont
  {J.}~\bibnamefont {Wrachtrup}},\ }\href@noop {} {\bibfield  {journal}
  {\bibinfo  {journal} {Physical Review Letters}\ }\textbf {\bibinfo {volume}
  {112}},\ \bibinfo {pages} {097603} (\bibinfo {year} {2014})}\BibitemShut
  {NoStop}%
\bibitem [{\citenamefont {Doherty}\ \emph {et~al.}(2014)\citenamefont
  {Doherty}, \citenamefont {Struzhkin}, \citenamefont {Simpson}, \citenamefont
  {McGuinness}, \citenamefont {Meng}, \citenamefont {Stacey}, \citenamefont
  {Karle}, \citenamefont {Hemley}, \citenamefont {Manson}, \citenamefont
  {Hollenberg} \emph {et~al.}}]{doherty2014electronic}%
  \BibitemOpen
  \bibfield  {author} {\bibinfo {author} {\bibfnamefont {M.~W.}\ \bibnamefont
  {Doherty}}, \bibinfo {author} {\bibfnamefont {V.~V.}\ \bibnamefont
  {Struzhkin}}, \bibinfo {author} {\bibfnamefont {D.~A.}\ \bibnamefont
  {Simpson}}, \bibinfo {author} {\bibfnamefont {L.~P.}\ \bibnamefont
  {McGuinness}}, \bibinfo {author} {\bibfnamefont {Y.}~\bibnamefont {Meng}},
  \bibinfo {author} {\bibfnamefont {A.}~\bibnamefont {Stacey}}, \bibinfo
  {author} {\bibfnamefont {T.~J.}\ \bibnamefont {Karle}}, \bibinfo {author}
  {\bibfnamefont {R.~J.}\ \bibnamefont {Hemley}}, \bibinfo {author}
  {\bibfnamefont {N.~B.}\ \bibnamefont {Manson}}, \bibinfo {author}
  {\bibfnamefont {L.~C.}\ \bibnamefont {Hollenberg}},  \emph {et~al.},\
  }\href@noop {} {\bibfield  {journal} {\bibinfo  {journal} {Physical review
  letters}\ }\textbf {\bibinfo {volume} {112}},\ \bibinfo {pages} {047601}
  (\bibinfo {year} {2014})}\BibitemShut {NoStop}%
\bibitem [{\citenamefont {Ledbetter}\ \emph {et~al.}(2012)\citenamefont
  {Ledbetter}, \citenamefont {Jensen}, \citenamefont {Fischer}, \citenamefont
  {Jarmola},\ and\ \citenamefont {Budker}}]{Ledbetter:2012kh}%
  \BibitemOpen
  \bibfield  {author} {\bibinfo {author} {\bibfnamefont {M.~P.}\ \bibnamefont
  {Ledbetter}}, \bibinfo {author} {\bibfnamefont {K.}~\bibnamefont {Jensen}},
  \bibinfo {author} {\bibfnamefont {R.}~\bibnamefont {Fischer}}, \bibinfo
  {author} {\bibfnamefont {A.}~\bibnamefont {Jarmola}}, \ and\ \bibinfo
  {author} {\bibfnamefont {D.}~\bibnamefont {Budker}},\ }\href@noop {}
  {\bibfield  {journal} {\bibinfo  {journal} {Physical Review A}\ }\textbf
  {\bibinfo {volume} {86}},\ \bibinfo {pages} {052116} (\bibinfo {year}
  {2012})}\BibitemShut {NoStop}%
\bibitem [{\citenamefont {Ajoy}\ and\ \citenamefont
  {Cappellaro}(2012)}]{Ajoy:2012is}%
  \BibitemOpen
  \bibfield  {author} {\bibinfo {author} {\bibfnamefont {A.}~\bibnamefont
  {Ajoy}}\ and\ \bibinfo {author} {\bibfnamefont {P.}~\bibnamefont
  {Cappellaro}},\ }\href@noop {} {\bibfield  {journal} {\bibinfo  {journal}
  {Physical Review A}\ }\textbf {\bibinfo {volume} {86}},\ \bibinfo {pages}
  {062104} (\bibinfo {year} {2012})}\BibitemShut {NoStop}%
\bibitem [{\citenamefont {Le~Sage}\ \emph {et~al.}(2013)\citenamefont
  {Le~Sage}, \citenamefont {Arai}, \citenamefont {Glenn}, \citenamefont
  {DeVience}, \citenamefont {Pham}, \citenamefont {Rahn-Lee}, \citenamefont
  {Lukin}, \citenamefont {Yacoby}, \citenamefont {Komeili},\ and\ \citenamefont
  {Walsworth}}]{le2013optical}%
  \BibitemOpen
  \bibfield  {author} {\bibinfo {author} {\bibfnamefont {D.}~\bibnamefont
  {Le~Sage}}, \bibinfo {author} {\bibfnamefont {K.}~\bibnamefont {Arai}},
  \bibinfo {author} {\bibfnamefont {D.}~\bibnamefont {Glenn}}, \bibinfo
  {author} {\bibfnamefont {S.}~\bibnamefont {DeVience}}, \bibinfo {author}
  {\bibfnamefont {L.}~\bibnamefont {Pham}}, \bibinfo {author} {\bibfnamefont
  {L.}~\bibnamefont {Rahn-Lee}}, \bibinfo {author} {\bibfnamefont
  {M.}~\bibnamefont {Lukin}}, \bibinfo {author} {\bibfnamefont
  {A.}~\bibnamefont {Yacoby}}, \bibinfo {author} {\bibfnamefont
  {A.}~\bibnamefont {Komeili}}, \ and\ \bibinfo {author} {\bibfnamefont
  {R.}~\bibnamefont {Walsworth}},\ }\href@noop {} {\bibfield  {journal}
  {\bibinfo  {journal} {Nature}\ }\textbf {\bibinfo {volume} {496}},\ \bibinfo
  {pages} {486} (\bibinfo {year} {2013})}\BibitemShut {NoStop}%
\bibitem [{\citenamefont {Kucsko}\ \emph {et~al.}(2013)\citenamefont {Kucsko},
  \citenamefont {Maurer}, \citenamefont {Yao}, \citenamefont {Kubo},
  \citenamefont {Noh}, \citenamefont {Lo}, \citenamefont {Park},\ and\
  \citenamefont {Lukin}}]{Kucsko:2013}%
  \BibitemOpen
  \bibfield  {author} {\bibinfo {author} {\bibfnamefont {G.}~\bibnamefont
  {Kucsko}}, \bibinfo {author} {\bibfnamefont {P.~C.}\ \bibnamefont {Maurer}},
  \bibinfo {author} {\bibfnamefont {N.~Y.}\ \bibnamefont {Yao}}, \bibinfo
  {author} {\bibfnamefont {M.}~\bibnamefont {Kubo}}, \bibinfo {author}
  {\bibfnamefont {H.~J.}\ \bibnamefont {Noh}}, \bibinfo {author} {\bibfnamefont
  {P.~K.}\ \bibnamefont {Lo}}, \bibinfo {author} {\bibfnamefont
  {H.}~\bibnamefont {Park}}, \ and\ \bibinfo {author} {\bibfnamefont {M.~D.}\
  \bibnamefont {Lukin}},\ }\href@noop {} {\bibfield  {journal} {\bibinfo
  {journal} {Nature}\ }\textbf {\bibinfo {volume} {500}},\ \bibinfo {pages}
  {54} (\bibinfo {year} {2013})}\BibitemShut {NoStop}%
\bibitem [{\citenamefont {Toyli}\ \emph {et~al.}(2013)\citenamefont {Toyli},
  \citenamefont {Charles}, \citenamefont {Christle}, \citenamefont
  {Dobrovitski},\ and\ \citenamefont {Awschalom}}]{toyli2013fluorescence}%
  \BibitemOpen
  \bibfield  {author} {\bibinfo {author} {\bibfnamefont {D.~M.}\ \bibnamefont
  {Toyli}}, \bibinfo {author} {\bibfnamefont {F.}~\bibnamefont {Charles}},
  \bibinfo {author} {\bibfnamefont {D.~J.}\ \bibnamefont {Christle}}, \bibinfo
  {author} {\bibfnamefont {V.~V.}\ \bibnamefont {Dobrovitski}}, \ and\ \bibinfo
  {author} {\bibfnamefont {D.~D.}\ \bibnamefont {Awschalom}},\ }\href@noop {}
  {\bibfield  {journal} {\bibinfo  {journal} {Proceedings of the National
  Academy of Sciences}\ }\textbf {\bibinfo {volume} {110}},\ \bibinfo {pages}
  {8417} (\bibinfo {year} {2013})}\BibitemShut {NoStop}%
\bibitem [{\citenamefont {McGuinness}\ \emph {et~al.}(2011)\citenamefont
  {McGuinness}, \citenamefont {Yan}, \citenamefont {Stacey}, \citenamefont
  {Simpson}, \citenamefont {Hall}, \citenamefont {Maclaurin}, \citenamefont
  {Prawer}, \citenamefont {Mulvaney}, \citenamefont {Wrachtrup}, \citenamefont
  {Caruso}, \citenamefont {Scholten},\ and\ \citenamefont
  {Hollenberg}}]{McGuinness2011}%
  \BibitemOpen
  \bibfield  {author} {\bibinfo {author} {\bibfnamefont {L.~P.}\ \bibnamefont
  {McGuinness}}, \bibinfo {author} {\bibfnamefont {Y.}~\bibnamefont {Yan}},
  \bibinfo {author} {\bibfnamefont {A.}~\bibnamefont {Stacey}}, \bibinfo
  {author} {\bibfnamefont {D.~A.}\ \bibnamefont {Simpson}}, \bibinfo {author}
  {\bibfnamefont {L.~T.}\ \bibnamefont {Hall}}, \bibinfo {author}
  {\bibfnamefont {D.}~\bibnamefont {Maclaurin}}, \bibinfo {author}
  {\bibfnamefont {S.}~\bibnamefont {Prawer}}, \bibinfo {author} {\bibfnamefont
  {P.}~\bibnamefont {Mulvaney}}, \bibinfo {author} {\bibfnamefont
  {J.}~\bibnamefont {Wrachtrup}}, \bibinfo {author} {\bibfnamefont
  {F.}~\bibnamefont {Caruso}}, \bibinfo {author} {\bibfnamefont {R.~E.}\
  \bibnamefont {Scholten}}, \ and\ \bibinfo {author} {\bibfnamefont {L.~C.~L.}\
  \bibnamefont {Hollenberg}},\ }\href {http://dx.doi.org/10.1038/nnano.2011.64}
  {\bibfield  {journal} {\bibinfo  {journal} {Nature Nanotechnology}\ }\textbf
  {\bibinfo {volume} {6}},\ \bibinfo {pages} {358 EP } (\bibinfo {year}
  {2011})}\BibitemShut {NoStop}%
\bibitem [{\citenamefont {Laraoui}\ \emph {et~al.}(2015)\citenamefont
  {Laraoui}, \citenamefont {Aycock-Rizzo}, \citenamefont {Gao}, \citenamefont
  {Lu}, \citenamefont {Riedo},\ and\ \citenamefont {Meriles}}]{Laraoui:2015ii}%
  \BibitemOpen
  \bibfield  {author} {\bibinfo {author} {\bibfnamefont {A.}~\bibnamefont
  {Laraoui}}, \bibinfo {author} {\bibfnamefont {H.}~\bibnamefont
  {Aycock-Rizzo}}, \bibinfo {author} {\bibfnamefont {Y.}~\bibnamefont {Gao}},
  \bibinfo {author} {\bibfnamefont {X.}~\bibnamefont {Lu}}, \bibinfo {author}
  {\bibfnamefont {E.}~\bibnamefont {Riedo}}, \ and\ \bibinfo {author}
  {\bibfnamefont {C.~A.}\ \bibnamefont {Meriles}},\ }\href@noop {} {\bibfield
  {journal} {\bibinfo  {journal} {Nature Communications}\ }\textbf {\bibinfo
  {volume} {6}},\ \bibinfo {pages} {8954} (\bibinfo {year} {2015})}\BibitemShut
  {NoStop}%
\bibitem [{\citenamefont {Pelliccione}\ \emph {et~al.}(2016)\citenamefont
  {Pelliccione}, \citenamefont {Jenkins}, \citenamefont {Ovartchaiyapong},
  \citenamefont {Reetz}, \citenamefont {Emmanouilidou}, \citenamefont {Ni},\
  and\ \citenamefont {Jayich}}]{pelliccione2016scanned}%
  \BibitemOpen
  \bibfield  {author} {\bibinfo {author} {\bibfnamefont {M.}~\bibnamefont
  {Pelliccione}}, \bibinfo {author} {\bibfnamefont {A.}~\bibnamefont
  {Jenkins}}, \bibinfo {author} {\bibfnamefont {P.}~\bibnamefont
  {Ovartchaiyapong}}, \bibinfo {author} {\bibfnamefont {C.}~\bibnamefont
  {Reetz}}, \bibinfo {author} {\bibfnamefont {E.}~\bibnamefont
  {Emmanouilidou}}, \bibinfo {author} {\bibfnamefont {N.}~\bibnamefont {Ni}}, \
  and\ \bibinfo {author} {\bibfnamefont {A.~C.~B.}\ \bibnamefont {Jayich}},\
  }\href@noop {} {\bibfield  {journal} {\bibinfo  {journal} {Nature
  nanotechnology}\ }\textbf {\bibinfo {volume} {11}},\ \bibinfo {pages} {700}
  (\bibinfo {year} {2016})}\BibitemShut {NoStop}%
\bibitem [{\citenamefont {Du}\ \emph {et~al.}(2017)\citenamefont {Du},
  \citenamefont {Van~der Sar}, \citenamefont {Zhou}, \citenamefont {Upadhyaya},
  \citenamefont {Casola}, \citenamefont {Zhang}, \citenamefont {Onbasli},
  \citenamefont {Ross}, \citenamefont {Walsworth}, \citenamefont {Tserkovnyak}
  \emph {et~al.}}]{du2017control}%
  \BibitemOpen
  \bibfield  {author} {\bibinfo {author} {\bibfnamefont {C.}~\bibnamefont
  {Du}}, \bibinfo {author} {\bibfnamefont {T.}~\bibnamefont {Van~der Sar}},
  \bibinfo {author} {\bibfnamefont {T.~X.}\ \bibnamefont {Zhou}}, \bibinfo
  {author} {\bibfnamefont {P.}~\bibnamefont {Upadhyaya}}, \bibinfo {author}
  {\bibfnamefont {F.}~\bibnamefont {Casola}}, \bibinfo {author} {\bibfnamefont
  {H.}~\bibnamefont {Zhang}}, \bibinfo {author} {\bibfnamefont {M.~C.}\
  \bibnamefont {Onbasli}}, \bibinfo {author} {\bibfnamefont {C.~A.}\
  \bibnamefont {Ross}}, \bibinfo {author} {\bibfnamefont {R.~L.}\ \bibnamefont
  {Walsworth}}, \bibinfo {author} {\bibfnamefont {Y.}~\bibnamefont
  {Tserkovnyak}},  \emph {et~al.},\ }\href@noop {} {\bibfield  {journal}
  {\bibinfo  {journal} {Science}\ }\textbf {\bibinfo {volume} {357}},\ \bibinfo
  {pages} {195} (\bibinfo {year} {2017})}\BibitemShut {NoStop}%
\bibitem [{\citenamefont {Dovzhenko}\ \emph {et~al.}(2016)\citenamefont
  {Dovzhenko}, \citenamefont {Casola}, \citenamefont {Schlotter}, \citenamefont
  {Zhou}, \citenamefont {B{\"u}ttner}, \citenamefont {Walsworth}, \citenamefont
  {Beach},\ and\ \citenamefont {Yacoby}}]{dovzhenko2016imaging}%
  \BibitemOpen
  \bibfield  {author} {\bibinfo {author} {\bibfnamefont {Y.}~\bibnamefont
  {Dovzhenko}}, \bibinfo {author} {\bibfnamefont {F.}~\bibnamefont {Casola}},
  \bibinfo {author} {\bibfnamefont {S.}~\bibnamefont {Schlotter}}, \bibinfo
  {author} {\bibfnamefont {T.~X.}\ \bibnamefont {Zhou}}, \bibinfo {author}
  {\bibfnamefont {F.}~\bibnamefont {B{\"u}ttner}}, \bibinfo {author}
  {\bibfnamefont {R.~L.}\ \bibnamefont {Walsworth}}, \bibinfo {author}
  {\bibfnamefont {G.~S.}\ \bibnamefont {Beach}}, \ and\ \bibinfo {author}
  {\bibfnamefont {A.}~\bibnamefont {Yacoby}},\ }\href@noop {} {\bibfield
  {journal} {\bibinfo  {journal} {arXiv preprint arXiv:1611.00673}\ } (\bibinfo
  {year} {2016})}\BibitemShut {NoStop}%
\bibitem [{\citenamefont {Gross}\ \emph {et~al.}(2017)\citenamefont {Gross},
  \citenamefont {Akhtar}, \citenamefont {Garcia}, \citenamefont
  {Mart{\'\i}nez}, \citenamefont {Chouaieb}, \citenamefont {Garcia},
  \citenamefont {Carr{\'e}t{\'e}ro}, \citenamefont {Barth{\'e}l{\'e}my},
  \citenamefont {Appel}, \citenamefont {Maletinsky} \emph
  {et~al.}}]{gross2017real}%
  \BibitemOpen
  \bibfield  {author} {\bibinfo {author} {\bibfnamefont {I.}~\bibnamefont
  {Gross}}, \bibinfo {author} {\bibfnamefont {W.}~\bibnamefont {Akhtar}},
  \bibinfo {author} {\bibfnamefont {V.}~\bibnamefont {Garcia}}, \bibinfo
  {author} {\bibfnamefont {L.}~\bibnamefont {Mart{\'\i}nez}}, \bibinfo {author}
  {\bibfnamefont {S.}~\bibnamefont {Chouaieb}}, \bibinfo {author}
  {\bibfnamefont {K.}~\bibnamefont {Garcia}}, \bibinfo {author} {\bibfnamefont
  {C.}~\bibnamefont {Carr{\'e}t{\'e}ro}}, \bibinfo {author} {\bibfnamefont
  {A.}~\bibnamefont {Barth{\'e}l{\'e}my}}, \bibinfo {author} {\bibfnamefont
  {P.}~\bibnamefont {Appel}}, \bibinfo {author} {\bibfnamefont
  {P.}~\bibnamefont {Maletinsky}},  \emph {et~al.},\ }\href@noop {} {\bibfield
  {journal} {\bibinfo  {journal} {Nature}\ }\textbf {\bibinfo {volume} {549}},\
  \bibinfo {pages} {252} (\bibinfo {year} {2017})}\BibitemShut {NoStop}%
\bibitem [{\citenamefont {Waldherr}\ \emph {et~al.}(2011)\citenamefont
  {Waldherr}, \citenamefont {Neumann}, \citenamefont {Huelga}, \citenamefont
  {Jelezko},\ and\ \citenamefont {Wrachtrup}}]{waldherr2011violation}%
  \BibitemOpen
  \bibfield  {author} {\bibinfo {author} {\bibfnamefont {G.}~\bibnamefont
  {Waldherr}}, \bibinfo {author} {\bibfnamefont {P.}~\bibnamefont {Neumann}},
  \bibinfo {author} {\bibfnamefont {S.}~\bibnamefont {Huelga}}, \bibinfo
  {author} {\bibfnamefont {F.}~\bibnamefont {Jelezko}}, \ and\ \bibinfo
  {author} {\bibfnamefont {J.}~\bibnamefont {Wrachtrup}},\ }\href@noop {}
  {\bibfield  {journal} {\bibinfo  {journal} {Physical Review Letters}\
  }\textbf {\bibinfo {volume} {107}},\ \bibinfo {pages} {090401} (\bibinfo
  {year} {2011})}\BibitemShut {NoStop}%
\bibitem [{\citenamefont {Bernien}\ \emph {et~al.}(2013)\citenamefont
  {Bernien}, \citenamefont {Hensen}, \citenamefont {Pfaff}, \citenamefont
  {Koolstra}, \citenamefont {Blok}, \citenamefont {Robledo}, \citenamefont
  {Taminiau}, \citenamefont {Markham}, \citenamefont {Twitchen}, \citenamefont
  {Childress} \emph {et~al.}}]{bernien2013heralded}%
  \BibitemOpen
  \bibfield  {author} {\bibinfo {author} {\bibfnamefont {H.}~\bibnamefont
  {Bernien}}, \bibinfo {author} {\bibfnamefont {B.}~\bibnamefont {Hensen}},
  \bibinfo {author} {\bibfnamefont {W.}~\bibnamefont {Pfaff}}, \bibinfo
  {author} {\bibfnamefont {G.}~\bibnamefont {Koolstra}}, \bibinfo {author}
  {\bibfnamefont {M.}~\bibnamefont {Blok}}, \bibinfo {author} {\bibfnamefont
  {L.}~\bibnamefont {Robledo}}, \bibinfo {author} {\bibfnamefont
  {T.}~\bibnamefont {Taminiau}}, \bibinfo {author} {\bibfnamefont
  {M.}~\bibnamefont {Markham}}, \bibinfo {author} {\bibfnamefont
  {D.}~\bibnamefont {Twitchen}}, \bibinfo {author} {\bibfnamefont
  {L.}~\bibnamefont {Childress}},  \emph {et~al.},\ }\href@noop {} {\bibfield
  {journal} {\bibinfo  {journal} {Nature}\ }\textbf {\bibinfo {volume} {497}},\
  \bibinfo {pages} {86} (\bibinfo {year} {2013})}\BibitemShut {NoStop}%
\bibitem [{\citenamefont {Hensen}\ \emph {et~al.}(2015)\citenamefont {Hensen},
  \citenamefont {Bernien}, \citenamefont {Dr{\'e}au}, \citenamefont {Reiserer},
  \citenamefont {Kalb}, \citenamefont {Blok}, \citenamefont {Ruitenberg},
  \citenamefont {Vermeulen}, \citenamefont {Schouten}, \citenamefont
  {Abell{\'a}n}, \citenamefont {Amaya}, \citenamefont {Pruneri}, \citenamefont
  {Mitchell}, \citenamefont {Markham}, \citenamefont {Twitchen}, \citenamefont
  {Elkouss}, \citenamefont {Wehner}, \citenamefont {Taminiau},\ and\
  \citenamefont {Hanson}}]{Hensen:2015dw}%
  \BibitemOpen
  \bibfield  {author} {\bibinfo {author} {\bibfnamefont {B.}~\bibnamefont
  {Hensen}}, \bibinfo {author} {\bibfnamefont {H.}~\bibnamefont {Bernien}},
  \bibinfo {author} {\bibfnamefont {A.~E.}\ \bibnamefont {Dr{\'e}au}}, \bibinfo
  {author} {\bibfnamefont {A.}~\bibnamefont {Reiserer}}, \bibinfo {author}
  {\bibfnamefont {N.}~\bibnamefont {Kalb}}, \bibinfo {author} {\bibfnamefont
  {M.~S.}\ \bibnamefont {Blok}}, \bibinfo {author} {\bibfnamefont
  {J.}~\bibnamefont {Ruitenberg}}, \bibinfo {author} {\bibfnamefont {R.~F.~L.}\
  \bibnamefont {Vermeulen}}, \bibinfo {author} {\bibfnamefont {R.~N.}\
  \bibnamefont {Schouten}}, \bibinfo {author} {\bibfnamefont {C.}~\bibnamefont
  {Abell{\'a}n}}, \bibinfo {author} {\bibfnamefont {W.}~\bibnamefont {Amaya}},
  \bibinfo {author} {\bibfnamefont {V.}~\bibnamefont {Pruneri}}, \bibinfo
  {author} {\bibfnamefont {M.~W.}\ \bibnamefont {Mitchell}}, \bibinfo {author}
  {\bibfnamefont {M.}~\bibnamefont {Markham}}, \bibinfo {author} {\bibfnamefont
  {D.~J.}\ \bibnamefont {Twitchen}}, \bibinfo {author} {\bibfnamefont
  {D.}~\bibnamefont {Elkouss}}, \bibinfo {author} {\bibfnamefont
  {S.}~\bibnamefont {Wehner}}, \bibinfo {author} {\bibfnamefont {T.~H.}\
  \bibnamefont {Taminiau}}, \ and\ \bibinfo {author} {\bibfnamefont
  {R.}~\bibnamefont {Hanson}},\ }\href@noop {} {\bibfield  {journal} {\bibinfo
  {journal} {Nature}\ }\textbf {\bibinfo {volume} {526}},\ \bibinfo {pages}
  {682} (\bibinfo {year} {2015})}\BibitemShut {NoStop}%
\bibitem [{\citenamefont {Wasilewski}\ \emph {et~al.}(2010)\citenamefont
  {Wasilewski}, \citenamefont {Jensen}, \citenamefont {Krauter}, \citenamefont
  {Renema}, \citenamefont {Balabas},\ and\ \citenamefont
  {Polzik}}]{wasilewski2010quantum}%
  \BibitemOpen
  \bibfield  {author} {\bibinfo {author} {\bibfnamefont {W.}~\bibnamefont
  {Wasilewski}}, \bibinfo {author} {\bibfnamefont {K.}~\bibnamefont {Jensen}},
  \bibinfo {author} {\bibfnamefont {H.}~\bibnamefont {Krauter}}, \bibinfo
  {author} {\bibfnamefont {J.~J.}\ \bibnamefont {Renema}}, \bibinfo {author}
  {\bibfnamefont {M.}~\bibnamefont {Balabas}}, \ and\ \bibinfo {author}
  {\bibfnamefont {E.~S.}\ \bibnamefont {Polzik}},\ }\href@noop {} {\bibfield
  {journal} {\bibinfo  {journal} {Physical Review Letters}\ }\textbf {\bibinfo
  {volume} {104}},\ \bibinfo {pages} {133601} (\bibinfo {year}
  {2010})}\BibitemShut {NoStop}%
\bibitem [{\citenamefont {Simmons}\ \emph {et~al.}(2010)\citenamefont
  {Simmons}, \citenamefont {Jones}, \citenamefont {Karlen}, \citenamefont
  {Ardavan},\ and\ \citenamefont {Morton}}]{simmons2010magnetic}%
  \BibitemOpen
  \bibfield  {author} {\bibinfo {author} {\bibfnamefont {S.}~\bibnamefont
  {Simmons}}, \bibinfo {author} {\bibfnamefont {J.~A.}\ \bibnamefont {Jones}},
  \bibinfo {author} {\bibfnamefont {S.~D.}\ \bibnamefont {Karlen}}, \bibinfo
  {author} {\bibfnamefont {A.}~\bibnamefont {Ardavan}}, \ and\ \bibinfo
  {author} {\bibfnamefont {J.~J.}\ \bibnamefont {Morton}},\ }\href@noop {}
  {\bibfield  {journal} {\bibinfo  {journal} {Physical Review A}\ }\textbf
  {\bibinfo {volume} {82}},\ \bibinfo {pages} {022330} (\bibinfo {year}
  {2010})}\BibitemShut {NoStop}%
\bibitem [{\citenamefont {Jones}\ \emph {et~al.}(2009)\citenamefont {Jones},
  \citenamefont {Karlen}, \citenamefont {Fitzsimons}, \citenamefont {Ardavan},
  \citenamefont {Benjamin}, \citenamefont {Briggs},\ and\ \citenamefont
  {Morton}}]{jones2009magnetic}%
  \BibitemOpen
  \bibfield  {author} {\bibinfo {author} {\bibfnamefont {J.~A.}\ \bibnamefont
  {Jones}}, \bibinfo {author} {\bibfnamefont {S.~D.}\ \bibnamefont {Karlen}},
  \bibinfo {author} {\bibfnamefont {J.}~\bibnamefont {Fitzsimons}}, \bibinfo
  {author} {\bibfnamefont {A.}~\bibnamefont {Ardavan}}, \bibinfo {author}
  {\bibfnamefont {S.~C.}\ \bibnamefont {Benjamin}}, \bibinfo {author}
  {\bibfnamefont {G.~A.~D.}\ \bibnamefont {Briggs}}, \ and\ \bibinfo {author}
  {\bibfnamefont {J.~J.}\ \bibnamefont {Morton}},\ }\href@noop {} {\bibfield
  {journal} {\bibinfo  {journal} {Science}\ }\textbf {\bibinfo {volume}
  {324}},\ \bibinfo {pages} {1166} (\bibinfo {year} {2009})}\BibitemShut
  {NoStop}%
\bibitem [{\citenamefont {Cappellaro}\ and\ \citenamefont
  {Lukin}(2009)}]{cappellaro2009quantum}%
  \BibitemOpen
  \bibfield  {author} {\bibinfo {author} {\bibfnamefont {P.}~\bibnamefont
  {Cappellaro}}\ and\ \bibinfo {author} {\bibfnamefont {M.~D.}\ \bibnamefont
  {Lukin}},\ }\href@noop {} {\bibfield  {journal} {\bibinfo  {journal}
  {Physical Review A}\ }\textbf {\bibinfo {volume} {80}},\ \bibinfo {pages}
  {032311} (\bibinfo {year} {2009})}\BibitemShut {NoStop}%
\bibitem [{\citenamefont {{Choi}}\ \emph {et~al.}(2018)\citenamefont {{Choi}},
  \citenamefont {{Yao}},\ and\ \citenamefont {{Lukin}}}]{2018arXiv180100042C}%
  \BibitemOpen
  \bibfield  {author} {\bibinfo {author} {\bibfnamefont {S.}~\bibnamefont
  {{Choi}}}, \bibinfo {author} {\bibfnamefont {N.~Y.}\ \bibnamefont {{Yao}}}, \
  and\ \bibinfo {author} {\bibfnamefont {M.~D.}\ \bibnamefont {{Lukin}}},\
  }\href@noop {} {\bibfield  {journal} {\bibinfo  {journal} {ArXiv e-prints}\ }
  (\bibinfo {year} {2018})},\ \Eprint {http://arxiv.org/abs/1801.00042}
  {arXiv:1801.00042 [quant-ph]} \BibitemShut {NoStop}%
\bibitem [{\citenamefont {Acosta}\ \emph {et~al.}(2009)\citenamefont {Acosta},
  \citenamefont {Bauch}, \citenamefont {Ledbetter}, \citenamefont {Santori},
  \citenamefont {Fu}, \citenamefont {Barclay}, \citenamefont {Beausoleil},
  \citenamefont {Linget}, \citenamefont {Roch}, \citenamefont {Treussart},
  \citenamefont {Chemerisov}, \citenamefont {Gawlik},\ and\ \citenamefont
  {Budker}}]{Acosta:2009gu}%
  \BibitemOpen
  \bibfield  {author} {\bibinfo {author} {\bibfnamefont {V.~M.}\ \bibnamefont
  {Acosta}}, \bibinfo {author} {\bibfnamefont {E.}~\bibnamefont {Bauch}},
  \bibinfo {author} {\bibfnamefont {M.~P.}\ \bibnamefont {Ledbetter}}, \bibinfo
  {author} {\bibfnamefont {C.}~\bibnamefont {Santori}}, \bibinfo {author}
  {\bibfnamefont {K.~M.~C.}\ \bibnamefont {Fu}}, \bibinfo {author}
  {\bibfnamefont {P.~E.}\ \bibnamefont {Barclay}}, \bibinfo {author}
  {\bibfnamefont {R.~G.}\ \bibnamefont {Beausoleil}}, \bibinfo {author}
  {\bibfnamefont {H.}~\bibnamefont {Linget}}, \bibinfo {author} {\bibfnamefont
  {J.~F.}\ \bibnamefont {Roch}}, \bibinfo {author} {\bibfnamefont
  {F.}~\bibnamefont {Treussart}}, \bibinfo {author} {\bibfnamefont
  {S.}~\bibnamefont {Chemerisov}}, \bibinfo {author} {\bibfnamefont
  {W.}~\bibnamefont {Gawlik}}, \ and\ \bibinfo {author} {\bibfnamefont
  {D.}~\bibnamefont {Budker}},\ }\href@noop {} {\bibfield  {journal} {\bibinfo
  {journal} {Physical Review B}\ }\textbf {\bibinfo {volume} {80}},\ \bibinfo
  {pages} {115202} (\bibinfo {year} {2009})}\BibitemShut {NoStop}%
\bibitem [{\citenamefont {Steinert}\ \emph {et~al.}(2010)\citenamefont
  {Steinert}, \citenamefont {Dolde}, \citenamefont {Neumann}, \citenamefont
  {Aird}, \citenamefont {Naydenov}, \citenamefont {Balasubramanian},
  \citenamefont {Jelezko},\ and\ \citenamefont {Wrachtrup}}]{Steinert:2010kk}%
  \BibitemOpen
  \bibfield  {author} {\bibinfo {author} {\bibfnamefont {S.}~\bibnamefont
  {Steinert}}, \bibinfo {author} {\bibfnamefont {F.}~\bibnamefont {Dolde}},
  \bibinfo {author} {\bibfnamefont {P.}~\bibnamefont {Neumann}}, \bibinfo
  {author} {\bibfnamefont {A.}~\bibnamefont {Aird}}, \bibinfo {author}
  {\bibfnamefont {B.}~\bibnamefont {Naydenov}}, \bibinfo {author}
  {\bibfnamefont {G.}~\bibnamefont {Balasubramanian}}, \bibinfo {author}
  {\bibfnamefont {F.}~\bibnamefont {Jelezko}}, \ and\ \bibinfo {author}
  {\bibfnamefont {J.}~\bibnamefont {Wrachtrup}},\ }\href@noop {} {\bibfield
  {journal} {\bibinfo  {journal} {Review of Scientific Instruments}\ }\textbf
  {\bibinfo {volume} {81}},\ \bibinfo {pages} {043705} (\bibinfo {year}
  {2010})}\BibitemShut {NoStop}%
\bibitem [{\citenamefont {Maertz}\ \emph {et~al.}(2010)\citenamefont {Maertz},
  \citenamefont {Wijnheijmer}, \citenamefont {Fuchs}, \citenamefont
  {Nowakowski},\ and\ \citenamefont {Awschalom}}]{maertz2010vector}%
  \BibitemOpen
  \bibfield  {author} {\bibinfo {author} {\bibfnamefont {B.}~\bibnamefont
  {Maertz}}, \bibinfo {author} {\bibfnamefont {A.}~\bibnamefont {Wijnheijmer}},
  \bibinfo {author} {\bibfnamefont {G.}~\bibnamefont {Fuchs}}, \bibinfo
  {author} {\bibfnamefont {M.}~\bibnamefont {Nowakowski}}, \ and\ \bibinfo
  {author} {\bibfnamefont {D.}~\bibnamefont {Awschalom}},\ }\href@noop {}
  {\bibfield  {journal} {\bibinfo  {journal} {Applied Physics Letters}\
  }\textbf {\bibinfo {volume} {96}},\ \bibinfo {pages} {092504} (\bibinfo
  {year} {2010})}\BibitemShut {NoStop}%
\bibitem [{\citenamefont {Stanwix}\ \emph {et~al.}(2010)\citenamefont
  {Stanwix}, \citenamefont {Pham}, \citenamefont {Maze}, \citenamefont
  {Le~Sage}, \citenamefont {Yeung}, \citenamefont {Cappellaro}, \citenamefont
  {Hemmer}, \citenamefont {Yacoby}, \citenamefont {Lukin},\ and\ \citenamefont
  {Walsworth}}]{Stanwix:2010ko}%
  \BibitemOpen
  \bibfield  {author} {\bibinfo {author} {\bibfnamefont {P.~L.}\ \bibnamefont
  {Stanwix}}, \bibinfo {author} {\bibfnamefont {L.~M.}\ \bibnamefont {Pham}},
  \bibinfo {author} {\bibfnamefont {J.~R.}\ \bibnamefont {Maze}}, \bibinfo
  {author} {\bibfnamefont {D.}~\bibnamefont {Le~Sage}}, \bibinfo {author}
  {\bibfnamefont {T.~K.}\ \bibnamefont {Yeung}}, \bibinfo {author}
  {\bibfnamefont {P.}~\bibnamefont {Cappellaro}}, \bibinfo {author}
  {\bibfnamefont {P.~R.}\ \bibnamefont {Hemmer}}, \bibinfo {author}
  {\bibfnamefont {A.}~\bibnamefont {Yacoby}}, \bibinfo {author} {\bibfnamefont
  {M.~D.}\ \bibnamefont {Lukin}}, \ and\ \bibinfo {author} {\bibfnamefont
  {R.~L.}\ \bibnamefont {Walsworth}},\ }\href@noop {} {\bibfield  {journal}
  {\bibinfo  {journal} {Physical Review B}\ }\textbf {\bibinfo {volume} {82}},\
  \bibinfo {pages} {201201} (\bibinfo {year} {2010})}\BibitemShut {NoStop}%
\bibitem [{\citenamefont {Pham}\ \emph {et~al.}(2011)\citenamefont {Pham},
  \citenamefont {Le~Sage}, \citenamefont {Stanwix}, \citenamefont {Yeung},
  \citenamefont {Glenn}, \citenamefont {Trifonov}, \citenamefont {Cappellaro},
  \citenamefont {Hemmer}, \citenamefont {Lukin}, \citenamefont {Park},
  \citenamefont {Yacoby},\ and\ \citenamefont {Walsworth}}]{Pham:2011dc}%
  \BibitemOpen
  \bibfield  {author} {\bibinfo {author} {\bibfnamefont {L.~M.}\ \bibnamefont
  {Pham}}, \bibinfo {author} {\bibfnamefont {D.}~\bibnamefont {Le~Sage}},
  \bibinfo {author} {\bibfnamefont {P.~L.}\ \bibnamefont {Stanwix}}, \bibinfo
  {author} {\bibfnamefont {T.~K.}\ \bibnamefont {Yeung}}, \bibinfo {author}
  {\bibfnamefont {D.}~\bibnamefont {Glenn}}, \bibinfo {author} {\bibfnamefont
  {A.}~\bibnamefont {Trifonov}}, \bibinfo {author} {\bibfnamefont
  {P.}~\bibnamefont {Cappellaro}}, \bibinfo {author} {\bibfnamefont {P.~R.}\
  \bibnamefont {Hemmer}}, \bibinfo {author} {\bibfnamefont {M.~D.}\
  \bibnamefont {Lukin}}, \bibinfo {author} {\bibfnamefont {H.}~\bibnamefont
  {Park}}, \bibinfo {author} {\bibfnamefont {A.}~\bibnamefont {Yacoby}}, \ and\
  \bibinfo {author} {\bibfnamefont {R.~L.}\ \bibnamefont {Walsworth}},\
  }\href@noop {} {\bibfield  {journal} {\bibinfo  {journal} {New Journal of
  Physics}\ }\textbf {\bibinfo {volume} {13}},\ \bibinfo {pages} {045021}
  (\bibinfo {year} {2011})}\BibitemShut {NoStop}%
\bibitem [{\citenamefont {Jarmola}\ \emph {et~al.}(2012)\citenamefont
  {Jarmola}, \citenamefont {Acosta}, \citenamefont {Jensen}, \citenamefont
  {Chemerisov},\ and\ \citenamefont {Budker}}]{Jarmola:2012co}%
  \BibitemOpen
  \bibfield  {author} {\bibinfo {author} {\bibfnamefont {A.}~\bibnamefont
  {Jarmola}}, \bibinfo {author} {\bibfnamefont {V.~M.}\ \bibnamefont {Acosta}},
  \bibinfo {author} {\bibfnamefont {K.}~\bibnamefont {Jensen}}, \bibinfo
  {author} {\bibfnamefont {S.}~\bibnamefont {Chemerisov}}, \ and\ \bibinfo
  {author} {\bibfnamefont {D.}~\bibnamefont {Budker}},\ }\href@noop {}
  {\bibfield  {journal} {\bibinfo  {journal} {Physical Review Letters}\
  }\textbf {\bibinfo {volume} {108}},\ \bibinfo {pages} {197601} (\bibinfo
  {year} {2012})}\BibitemShut {NoStop}%
\bibitem [{\citenamefont {Bar-Gill}\ \emph {et~al.}(2013)\citenamefont
  {Bar-Gill}, \citenamefont {Pham}, \citenamefont {Jarmola}, \citenamefont
  {Budker},\ and\ \citenamefont {Walsworth}}]{BarGill:2013dq}%
  \BibitemOpen
  \bibfield  {author} {\bibinfo {author} {\bibfnamefont {N.}~\bibnamefont
  {Bar-Gill}}, \bibinfo {author} {\bibfnamefont {L.~M.}\ \bibnamefont {Pham}},
  \bibinfo {author} {\bibfnamefont {A.}~\bibnamefont {Jarmola}}, \bibinfo
  {author} {\bibfnamefont {D.}~\bibnamefont {Budker}}, \ and\ \bibinfo {author}
  {\bibfnamefont {R.~L.}\ \bibnamefont {Walsworth}},\ }\href@noop {} {\bibfield
   {journal} {\bibinfo  {journal} {Nature Communications}\ }\textbf {\bibinfo
  {volume} {4}},\ \bibinfo {pages} {1743} (\bibinfo {year} {2013})}\BibitemShut
  {NoStop}%
\bibitem [{\citenamefont {Jarmola}\ \emph {et~al.}(2015)\citenamefont
  {Jarmola}, \citenamefont {Berzins}, \citenamefont {Smits}, \citenamefont
  {Smits}, \citenamefont {Prikulis}, \citenamefont {Gahbauer}, \citenamefont
  {Ferber}, \citenamefont {Erts}, \citenamefont {Auzinsh},\ and\ \citenamefont
  {Budker}}]{Jarmola:2015gc}%
  \BibitemOpen
  \bibfield  {author} {\bibinfo {author} {\bibfnamefont {A.}~\bibnamefont
  {Jarmola}}, \bibinfo {author} {\bibfnamefont {A.}~\bibnamefont {Berzins}},
  \bibinfo {author} {\bibfnamefont {J.}~\bibnamefont {Smits}}, \bibinfo
  {author} {\bibfnamefont {K.}~\bibnamefont {Smits}}, \bibinfo {author}
  {\bibfnamefont {J.}~\bibnamefont {Prikulis}}, \bibinfo {author}
  {\bibfnamefont {F.}~\bibnamefont {Gahbauer}}, \bibinfo {author}
  {\bibfnamefont {R.}~\bibnamefont {Ferber}}, \bibinfo {author} {\bibfnamefont
  {D.}~\bibnamefont {Erts}}, \bibinfo {author} {\bibfnamefont {M.}~\bibnamefont
  {Auzinsh}}, \ and\ \bibinfo {author} {\bibfnamefont {D.}~\bibnamefont
  {Budker}},\ }\href@noop {} {\bibfield  {journal} {\bibinfo  {journal}
  {Applied Physics Letters}\ }\textbf {\bibinfo {volume} {107}},\ \bibinfo
  {pages} {242403} (\bibinfo {year} {2015})}\BibitemShut {NoStop}%
\bibitem [{\citenamefont {Weitekamp}\ \emph {et~al.}(1983)\citenamefont
  {Weitekamp}, \citenamefont {Bielecki}, \citenamefont {Zax}, \citenamefont
  {Zilm},\ and\ \citenamefont {Pines}}]{weitekamp1983zero}%
  \BibitemOpen
  \bibfield  {author} {\bibinfo {author} {\bibfnamefont {D.}~\bibnamefont
  {Weitekamp}}, \bibinfo {author} {\bibfnamefont {A.}~\bibnamefont {Bielecki}},
  \bibinfo {author} {\bibfnamefont {D.}~\bibnamefont {Zax}}, \bibinfo {author}
  {\bibfnamefont {K.}~\bibnamefont {Zilm}}, \ and\ \bibinfo {author}
  {\bibfnamefont {A.}~\bibnamefont {Pines}},\ }\href@noop {} {\bibfield
  {journal} {\bibinfo  {journal} {Physical review letters}\ }\textbf {\bibinfo
  {volume} {50}},\ \bibinfo {pages} {1807} (\bibinfo {year}
  {1983})}\BibitemShut {NoStop}%
\bibitem [{\citenamefont {Thayer}\ and\ \citenamefont
  {Pines}(2002)}]{Thayer:2002hc}%
  \BibitemOpen
  \bibfield  {author} {\bibinfo {author} {\bibfnamefont {A.~M.}\ \bibnamefont
  {Thayer}}\ and\ \bibinfo {author} {\bibfnamefont {A.}~\bibnamefont {Pines}},\
  }\href@noop {} {\bibfield  {journal} {\bibinfo  {journal} {Accounts of
  Chemical Research}\ }\textbf {\bibinfo {volume} {20}},\ \bibinfo {pages} {47}
  (\bibinfo {year} {2002})}\BibitemShut {NoStop}%
\bibitem [{\citenamefont {Gruber}\ \emph {et~al.}(1997)\citenamefont {Gruber},
  \citenamefont {Dr{\"a}benstedt}, \citenamefont {Tietz}, \citenamefont
  {Fleury}, \citenamefont {Wrachtrup},\ and\ \citenamefont {von
  Borczyskowski}}]{Gruber:1997}%
  \BibitemOpen
  \bibfield  {author} {\bibinfo {author} {\bibfnamefont {A.}~\bibnamefont
  {Gruber}}, \bibinfo {author} {\bibfnamefont {A.}~\bibnamefont
  {Dr{\"a}benstedt}}, \bibinfo {author} {\bibfnamefont {C.}~\bibnamefont
  {Tietz}}, \bibinfo {author} {\bibfnamefont {L.}~\bibnamefont {Fleury}},
  \bibinfo {author} {\bibfnamefont {J.}~\bibnamefont {Wrachtrup}}, \ and\
  \bibinfo {author} {\bibfnamefont {C.}~\bibnamefont {von Borczyskowski}},\
  }\href@noop {} {\bibfield  {journal} {\bibinfo  {journal} {Science}\ }\textbf
  {\bibinfo {volume} {276}},\ \bibinfo {pages} {2012} (\bibinfo {year}
  {1997})}\BibitemShut {NoStop}%
\bibitem [{\citenamefont {Barson}\ \emph {et~al.}(2017)\citenamefont {Barson},
  \citenamefont {Peddibhotla}, \citenamefont {Ovartchaiyapong}, \citenamefont
  {Ganesan}, \citenamefont {Taylor}, \citenamefont {Gebert}, \citenamefont
  {Mielens}, \citenamefont {Koslowski}, \citenamefont {Simpson}, \citenamefont
  {McGuinness}, \citenamefont {McCallum}, \citenamefont {Prawer}, \citenamefont
  {Onoda}, \citenamefont {Ohshima}, \citenamefont {Jayich}, \citenamefont
  {Jelezko}, \citenamefont {Manson},\ and\ \citenamefont
  {Doherty}}]{Barson:2017}%
  \BibitemOpen
  \bibfield  {author} {\bibinfo {author} {\bibfnamefont {M.~S.~J.}\
  \bibnamefont {Barson}}, \bibinfo {author} {\bibfnamefont {P.}~\bibnamefont
  {Peddibhotla}}, \bibinfo {author} {\bibfnamefont {P.}~\bibnamefont
  {Ovartchaiyapong}}, \bibinfo {author} {\bibfnamefont {K.}~\bibnamefont
  {Ganesan}}, \bibinfo {author} {\bibfnamefont {R.~L.}\ \bibnamefont {Taylor}},
  \bibinfo {author} {\bibfnamefont {M.}~\bibnamefont {Gebert}}, \bibinfo
  {author} {\bibfnamefont {Z.}~\bibnamefont {Mielens}}, \bibinfo {author}
  {\bibfnamefont {B.}~\bibnamefont {Koslowski}}, \bibinfo {author}
  {\bibfnamefont {D.~A.}\ \bibnamefont {Simpson}}, \bibinfo {author}
  {\bibfnamefont {L.~P.}\ \bibnamefont {McGuinness}}, \bibinfo {author}
  {\bibfnamefont {J.}~\bibnamefont {McCallum}}, \bibinfo {author}
  {\bibfnamefont {S.}~\bibnamefont {Prawer}}, \bibinfo {author} {\bibfnamefont
  {S.}~\bibnamefont {Onoda}}, \bibinfo {author} {\bibfnamefont
  {T.}~\bibnamefont {Ohshima}}, \bibinfo {author} {\bibfnamefont {A.~C.~B.}\
  \bibnamefont {Jayich}}, \bibinfo {author} {\bibfnamefont {F.}~\bibnamefont
  {Jelezko}}, \bibinfo {author} {\bibfnamefont {N.~B.}\ \bibnamefont {Manson}},
  \ and\ \bibinfo {author} {\bibfnamefont {M.~W.}\ \bibnamefont {Doherty}},\
  }\href@noop {} {\bibfield  {journal} {\bibinfo  {journal} {Nano Letters}\
  }\textbf {\bibinfo {volume} {17}},\ \bibinfo {pages} {1496} (\bibinfo {year}
  {2017})}\BibitemShut {NoStop}%
\bibitem [{\citenamefont {Igarashi}\ \emph {et~al.}(2012)\citenamefont
  {Igarashi}, \citenamefont {Yoshinari}, \citenamefont {Yokota}, \citenamefont
  {Sugi}, \citenamefont {Sugihara}, \citenamefont {Ikeda}, \citenamefont
  {Sumiya}, \citenamefont {Tsuji}, \citenamefont {Mori}, \citenamefont
  {Tochio}, \citenamefont {Harada},\ and\ \citenamefont
  {Shirakawa}}]{Igarashi:2012}%
  \BibitemOpen
  \bibfield  {author} {\bibinfo {author} {\bibfnamefont {R.}~\bibnamefont
  {Igarashi}}, \bibinfo {author} {\bibfnamefont {Y.}~\bibnamefont {Yoshinari}},
  \bibinfo {author} {\bibfnamefont {H.}~\bibnamefont {Yokota}}, \bibinfo
  {author} {\bibfnamefont {T.}~\bibnamefont {Sugi}}, \bibinfo {author}
  {\bibfnamefont {F.}~\bibnamefont {Sugihara}}, \bibinfo {author}
  {\bibfnamefont {K.}~\bibnamefont {Ikeda}}, \bibinfo {author} {\bibfnamefont
  {H.}~\bibnamefont {Sumiya}}, \bibinfo {author} {\bibfnamefont
  {S.}~\bibnamefont {Tsuji}}, \bibinfo {author} {\bibfnamefont
  {I.}~\bibnamefont {Mori}}, \bibinfo {author} {\bibfnamefont {H.}~\bibnamefont
  {Tochio}}, \bibinfo {author} {\bibfnamefont {Y.}~\bibnamefont {Harada}}, \
  and\ \bibinfo {author} {\bibfnamefont {M.}~\bibnamefont {Shirakawa}},\
  }\href@noop {} {\bibfield  {journal} {\bibinfo  {journal} {Nano Letters}\
  }\textbf {\bibinfo {volume} {12}},\ \bibinfo {pages} {5726} (\bibinfo {year}
  {2012})}\BibitemShut {NoStop}%
\bibitem [{\citenamefont {{Forneris}}\ \emph {et~al.}(2017)\citenamefont
  {{Forneris}}, \citenamefont {{Ditalia Tchernij}}, \citenamefont {{Traina}},
  \citenamefont {{Moreva}}, \citenamefont {{Skukan}}, \citenamefont {{Jak{\v
  s}i{\'c}}}, \citenamefont {{Grilj}}, \citenamefont {{Croin}}, \citenamefont
  {{Amato}}, \citenamefont {{Degiovanni}}, \citenamefont {{Naydenov}},
  \citenamefont {{Jelezko}}, \citenamefont {{Genovese}},\ and\ \citenamefont
  {{Olivero}}}]{2017arXiv170607935F}%
  \BibitemOpen
  \bibfield  {author} {\bibinfo {author} {\bibfnamefont {J.}~\bibnamefont
  {{Forneris}}}, \bibinfo {author} {\bibfnamefont {S.}~\bibnamefont {{Ditalia
  Tchernij}}}, \bibinfo {author} {\bibfnamefont {P.}~\bibnamefont {{Traina}}},
  \bibinfo {author} {\bibfnamefont {E.}~\bibnamefont {{Moreva}}}, \bibinfo
  {author} {\bibfnamefont {N.}~\bibnamefont {{Skukan}}}, \bibinfo {author}
  {\bibfnamefont {M.}~\bibnamefont {{Jak{\v s}i{\'c}}}}, \bibinfo {author}
  {\bibfnamefont {V.}~\bibnamefont {{Grilj}}}, \bibinfo {author} {\bibfnamefont
  {L.}~\bibnamefont {{Croin}}}, \bibinfo {author} {\bibfnamefont
  {G.}~\bibnamefont {{Amato}}}, \bibinfo {author} {\bibfnamefont {I.~P.}\
  \bibnamefont {{Degiovanni}}}, \bibinfo {author} {\bibfnamefont
  {B.}~\bibnamefont {{Naydenov}}}, \bibinfo {author} {\bibfnamefont
  {F.}~\bibnamefont {{Jelezko}}}, \bibinfo {author} {\bibfnamefont
  {M.}~\bibnamefont {{Genovese}}}, \ and\ \bibinfo {author} {\bibfnamefont
  {P.}~\bibnamefont {{Olivero}}},\ }\href@noop {} {\bibfield  {journal}
  {\bibinfo  {journal} {ArXiv e-prints}\ } (\bibinfo {year} {2017})},\ \Eprint
  {http://arxiv.org/abs/1706.07935} {arXiv:1706.07935 [cond-mat.mtrl-sci]}
  \BibitemShut {NoStop}%
\bibitem [{\citenamefont {Zhu}\ \emph {et~al.}(2014)\citenamefont {Zhu},
  \citenamefont {Matsuzaki}, \citenamefont {Ams{\"u}ss}, \citenamefont
  {Kakuyanagi}, \citenamefont {Shimo-Oka}, \citenamefont {Mizuochi},
  \citenamefont {Nemoto}, \citenamefont {Semba}, \citenamefont {Munro},\ and\
  \citenamefont {Saito}}]{Zhu:2014}%
  \BibitemOpen
  \bibfield  {author} {\bibinfo {author} {\bibfnamefont {X.}~\bibnamefont
  {Zhu}}, \bibinfo {author} {\bibfnamefont {Y.}~\bibnamefont {Matsuzaki}},
  \bibinfo {author} {\bibfnamefont {R.}~\bibnamefont {Ams{\"u}ss}}, \bibinfo
  {author} {\bibfnamefont {K.}~\bibnamefont {Kakuyanagi}}, \bibinfo {author}
  {\bibfnamefont {T.}~\bibnamefont {Shimo-Oka}}, \bibinfo {author}
  {\bibfnamefont {N.}~\bibnamefont {Mizuochi}}, \bibinfo {author}
  {\bibfnamefont {K.}~\bibnamefont {Nemoto}}, \bibinfo {author} {\bibfnamefont
  {K.}~\bibnamefont {Semba}}, \bibinfo {author} {\bibfnamefont {W.~J.}\
  \bibnamefont {Munro}}, \ and\ \bibinfo {author} {\bibfnamefont
  {S.}~\bibnamefont {Saito}},\ }\href@noop {} {\bibfield  {journal} {\bibinfo
  {journal} {Nature Communications}\ }\textbf {\bibinfo {volume} {5}},\
  \bibinfo {pages} {3424} (\bibinfo {year} {2014})}\BibitemShut {NoStop}%
\bibitem [{\citenamefont {Simanovskaia}\ \emph
  {et~al.}(2013{\natexlab{a}})\citenamefont {Simanovskaia}, \citenamefont
  {Jensen}, \citenamefont {Jarmola}, \citenamefont {Aulenbacher}, \citenamefont
  {Manson},\ and\ \citenamefont {Budker}}]{PhysRevB.87.224106}%
  \BibitemOpen
  \bibfield  {author} {\bibinfo {author} {\bibfnamefont {M.}~\bibnamefont
  {Simanovskaia}}, \bibinfo {author} {\bibfnamefont {K.}~\bibnamefont
  {Jensen}}, \bibinfo {author} {\bibfnamefont {A.}~\bibnamefont {Jarmola}},
  \bibinfo {author} {\bibfnamefont {K.}~\bibnamefont {Aulenbacher}}, \bibinfo
  {author} {\bibfnamefont {N.}~\bibnamefont {Manson}}, \ and\ \bibinfo {author}
  {\bibfnamefont {D.}~\bibnamefont {Budker}},\ }\href {\doibase
  10.1103/PhysRevB.87.224106} {\bibfield  {journal} {\bibinfo  {journal} {Phys.
  Rev. B}\ }\textbf {\bibinfo {volume} {87}},\ \bibinfo {pages} {224106}
  (\bibinfo {year} {2013}{\natexlab{a}})}\BibitemShut {NoStop}%
\bibitem [{\citenamefont {Acosta}\ \emph
  {et~al.}(2010{\natexlab{b}})\citenamefont {Acosta}, \citenamefont {Bauch},
  \citenamefont {Ledbetter}, \citenamefont {Waxman}, \citenamefont {Bouchard},\
  and\ \citenamefont {Budker}}]{PhysRevLett.104.070801}%
  \BibitemOpen
  \bibfield  {author} {\bibinfo {author} {\bibfnamefont {V.~M.}\ \bibnamefont
  {Acosta}}, \bibinfo {author} {\bibfnamefont {E.}~\bibnamefont {Bauch}},
  \bibinfo {author} {\bibfnamefont {M.~P.}\ \bibnamefont {Ledbetter}}, \bibinfo
  {author} {\bibfnamefont {A.}~\bibnamefont {Waxman}}, \bibinfo {author}
  {\bibfnamefont {L.-S.}\ \bibnamefont {Bouchard}}, \ and\ \bibinfo {author}
  {\bibfnamefont {D.}~\bibnamefont {Budker}},\ }\href {\doibase
  10.1103/PhysRevLett.104.070801} {\bibfield  {journal} {\bibinfo  {journal}
  {Phys. Rev. Lett.}\ }\textbf {\bibinfo {volume} {104}},\ \bibinfo {pages}
  {070801} (\bibinfo {year} {2010}{\natexlab{b}})}\BibitemShut {NoStop}%
\bibitem [{\citenamefont {Kubo}\ \emph {et~al.}(2010)\citenamefont {Kubo},
  \citenamefont {Ong}, \citenamefont {Bertet}, \citenamefont {Vion},
  \citenamefont {Jacques}, \citenamefont {Zheng}, \citenamefont {Dr{\'e}au},
  \citenamefont {Roch}, \citenamefont {Auff{\`e}ves}, \citenamefont {Jelezko},
  \citenamefont {Wrachtrup}, \citenamefont {Barthe}, \citenamefont {Bergonzo},\
  and\ \citenamefont {Esteve}}]{Kubo:2010}%
  \BibitemOpen
  \bibfield  {author} {\bibinfo {author} {\bibfnamefont {Y.}~\bibnamefont
  {Kubo}}, \bibinfo {author} {\bibfnamefont {F.~R.}\ \bibnamefont {Ong}},
  \bibinfo {author} {\bibfnamefont {P.}~\bibnamefont {Bertet}}, \bibinfo
  {author} {\bibfnamefont {D.}~\bibnamefont {Vion}}, \bibinfo {author}
  {\bibfnamefont {V.}~\bibnamefont {Jacques}}, \bibinfo {author} {\bibfnamefont
  {D.}~\bibnamefont {Zheng}}, \bibinfo {author} {\bibfnamefont
  {A.}~\bibnamefont {Dr{\'e}au}}, \bibinfo {author} {\bibfnamefont {J.~F.}\
  \bibnamefont {Roch}}, \bibinfo {author} {\bibfnamefont {A.}~\bibnamefont
  {Auff{\`e}ves}}, \bibinfo {author} {\bibfnamefont {F.}~\bibnamefont
  {Jelezko}}, \bibinfo {author} {\bibfnamefont {J.}~\bibnamefont {Wrachtrup}},
  \bibinfo {author} {\bibfnamefont {M.~F.}\ \bibnamefont {Barthe}}, \bibinfo
  {author} {\bibfnamefont {P.}~\bibnamefont {Bergonzo}}, \ and\ \bibinfo
  {author} {\bibfnamefont {D.}~\bibnamefont {Esteve}},\ }\href@noop {}
  {\bibfield  {journal} {\bibinfo  {journal} {Physical Review Letters}\
  }\textbf {\bibinfo {volume} {105}},\ \bibinfo {pages} {140502} (\bibinfo
  {year} {2010})}\BibitemShut {NoStop}%
\bibitem [{\citenamefont {Lai}\ \emph {et~al.}(2009)\citenamefont {Lai},
  \citenamefont {Zheng}, \citenamefont {Jelezko}, \citenamefont {Treussart},\
  and\ \citenamefont {Roch}}]{Lai:2009}%
  \BibitemOpen
  \bibfield  {author} {\bibinfo {author} {\bibfnamefont {N.~D.}\ \bibnamefont
  {Lai}}, \bibinfo {author} {\bibfnamefont {D.}~\bibnamefont {Zheng}}, \bibinfo
  {author} {\bibfnamefont {F.}~\bibnamefont {Jelezko}}, \bibinfo {author}
  {\bibfnamefont {F.}~\bibnamefont {Treussart}}, \ and\ \bibinfo {author}
  {\bibfnamefont {J.-F.}\ \bibnamefont {Roch}},\ }\href {\doibase
  10.1063/1.3238467} {\bibfield  {journal} {\bibinfo  {journal} {Applied
  Physics Letters}\ }\textbf {\bibinfo {volume} {95}},\ \bibinfo {pages}
  {133101} (\bibinfo {year} {2009})},\ \Eprint
  {http://arxiv.org/abs/https://doi.org/10.1063/1.3238467}
  {https://doi.org/10.1063/1.3238467} \BibitemShut {NoStop}%
\bibitem [{\citenamefont {Bourgeois}\ \emph {et~al.}(2015)\citenamefont
  {Bourgeois}, \citenamefont {Jarmola}, \citenamefont {Siyushev}, \citenamefont
  {Gulka}, \citenamefont {Hruby}, \citenamefont {Jelezko}, \citenamefont
  {Budker},\ and\ \citenamefont {Nesladek}}]{Bourgeois:2015ke}%
  \BibitemOpen
  \bibfield  {author} {\bibinfo {author} {\bibfnamefont {E.}~\bibnamefont
  {Bourgeois}}, \bibinfo {author} {\bibfnamefont {A.}~\bibnamefont {Jarmola}},
  \bibinfo {author} {\bibfnamefont {P.}~\bibnamefont {Siyushev}}, \bibinfo
  {author} {\bibfnamefont {M.}~\bibnamefont {Gulka}}, \bibinfo {author}
  {\bibfnamefont {J.}~\bibnamefont {Hruby}}, \bibinfo {author} {\bibfnamefont
  {F.}~\bibnamefont {Jelezko}}, \bibinfo {author} {\bibfnamefont
  {D.}~\bibnamefont {Budker}}, \ and\ \bibinfo {author} {\bibfnamefont
  {M.}~\bibnamefont {Nesladek}},\ }\href@noop {} {\bibfield  {journal}
  {\bibinfo  {journal} {Nature Communications}\ }\textbf {\bibinfo {volume}
  {6}},\ \bibinfo {pages} {8577} (\bibinfo {year} {2015})}\BibitemShut
  {NoStop}%
\bibitem [{\citenamefont {Rondin}\ \emph {et~al.}(2014)\citenamefont {Rondin},
  \citenamefont {Tetienne}, \citenamefont {Hingant}, \citenamefont {Roch},
  \citenamefont {Maletinsky},\ and\ \citenamefont {Jacques}}]{Rondin:2014kd}%
  \BibitemOpen
  \bibfield  {author} {\bibinfo {author} {\bibfnamefont {L.}~\bibnamefont
  {Rondin}}, \bibinfo {author} {\bibfnamefont {J.-P.}\ \bibnamefont
  {Tetienne}}, \bibinfo {author} {\bibfnamefont {T.}~\bibnamefont {Hingant}},
  \bibinfo {author} {\bibfnamefont {J.~F.}\ \bibnamefont {Roch}}, \bibinfo
  {author} {\bibfnamefont {P.}~\bibnamefont {Maletinsky}}, \ and\ \bibinfo
  {author} {\bibfnamefont {V.}~\bibnamefont {Jacques}},\ }\href@noop {}
  {\bibfield  {journal} {\bibinfo  {journal} {Reports on Progress in Physics}\
  }\textbf {\bibinfo {volume} {77}},\ \bibinfo {pages} {056503} (\bibinfo
  {year} {2014})}\BibitemShut {NoStop}%
\bibitem [{\citenamefont {Jamonneau}\ \emph {et~al.}(2016)\citenamefont
  {Jamonneau}, \citenamefont {Lesik}, \citenamefont {Tetienne}, \citenamefont
  {Alvizu}, \citenamefont {Mayer}, \citenamefont {Dr\'eau}, \citenamefont
  {Kosen}, \citenamefont {Roch}, \citenamefont {Pezzagna}, \citenamefont
  {Meijer}, \citenamefont {Teraji}, \citenamefont {Kubo}, \citenamefont
  {Bertet}, \citenamefont {Maze},\ and\ \citenamefont
  {Jacques}}]{PhysRevB.93.024305}%
  \BibitemOpen
  \bibfield  {author} {\bibinfo {author} {\bibfnamefont {P.}~\bibnamefont
  {Jamonneau}}, \bibinfo {author} {\bibfnamefont {M.}~\bibnamefont {Lesik}},
  \bibinfo {author} {\bibfnamefont {J.~P.}\ \bibnamefont {Tetienne}}, \bibinfo
  {author} {\bibfnamefont {I.}~\bibnamefont {Alvizu}}, \bibinfo {author}
  {\bibfnamefont {L.}~\bibnamefont {Mayer}}, \bibinfo {author} {\bibfnamefont
  {A.}~\bibnamefont {Dr\'eau}}, \bibinfo {author} {\bibfnamefont
  {S.}~\bibnamefont {Kosen}}, \bibinfo {author} {\bibfnamefont {J.-F.}\
  \bibnamefont {Roch}}, \bibinfo {author} {\bibfnamefont {S.}~\bibnamefont
  {Pezzagna}}, \bibinfo {author} {\bibfnamefont {J.}~\bibnamefont {Meijer}},
  \bibinfo {author} {\bibfnamefont {T.}~\bibnamefont {Teraji}}, \bibinfo
  {author} {\bibfnamefont {Y.}~\bibnamefont {Kubo}}, \bibinfo {author}
  {\bibfnamefont {P.}~\bibnamefont {Bertet}}, \bibinfo {author} {\bibfnamefont
  {J.~R.}\ \bibnamefont {Maze}}, \ and\ \bibinfo {author} {\bibfnamefont
  {V.}~\bibnamefont {Jacques}},\ }\href {\doibase 10.1103/PhysRevB.93.024305}
  {\bibfield  {journal} {\bibinfo  {journal} {Phys. Rev. B}\ }\textbf {\bibinfo
  {volume} {93}},\ \bibinfo {pages} {024305} (\bibinfo {year}
  {2016})}\BibitemShut {NoStop}%
\bibitem [{\citenamefont {Simanovskaia}\ \emph
  {et~al.}(2013{\natexlab{b}})\citenamefont {Simanovskaia}, \citenamefont
  {Jensen}, \citenamefont {Jarmola}, \citenamefont {Aulenbacher}, \citenamefont
  {Manson},\ and\ \citenamefont {Budker}}]{Simanovskaia:2013}%
  \BibitemOpen
  \bibfield  {author} {\bibinfo {author} {\bibfnamefont {M.}~\bibnamefont
  {Simanovskaia}}, \bibinfo {author} {\bibfnamefont {K.}~\bibnamefont
  {Jensen}}, \bibinfo {author} {\bibfnamefont {A.}~\bibnamefont {Jarmola}},
  \bibinfo {author} {\bibfnamefont {K.}~\bibnamefont {Aulenbacher}}, \bibinfo
  {author} {\bibfnamefont {N.}~\bibnamefont {Manson}}, \ and\ \bibinfo {author}
  {\bibfnamefont {D.}~\bibnamefont {Budker}},\ }\href {\doibase
  10.1103/PhysRevB.87.224106} {\bibfield  {journal} {\bibinfo  {journal} {Phys.
  Rev. B}\ }\textbf {\bibinfo {volume} {87}},\ \bibinfo {pages} {224106}
  (\bibinfo {year} {2013}{\natexlab{b}})}\BibitemShut {NoStop}%
\bibitem [{\citenamefont {Matsuzaki}\ \emph {et~al.}(2016)\citenamefont
  {Matsuzaki}, \citenamefont {Morishita}, \citenamefont {Shimooka},
  \citenamefont {Tashima}, \citenamefont {Kakuyanagi}, \citenamefont {Semba},
  \citenamefont {Munro}, \citenamefont {Yamaguchi}, \citenamefont {Mizuochi},\
  and\ \citenamefont {Saito}}]{Matsuzaki:2016}%
  \BibitemOpen
  \bibfield  {author} {\bibinfo {author} {\bibfnamefont {Y.}~\bibnamefont
  {Matsuzaki}}, \bibinfo {author} {\bibfnamefont {H.}~\bibnamefont
  {Morishita}}, \bibinfo {author} {\bibfnamefont {T.}~\bibnamefont {Shimooka}},
  \bibinfo {author} {\bibfnamefont {T.}~\bibnamefont {Tashima}}, \bibinfo
  {author} {\bibfnamefont {K.}~\bibnamefont {Kakuyanagi}}, \bibinfo {author}
  {\bibfnamefont {K.}~\bibnamefont {Semba}}, \bibinfo {author} {\bibfnamefont
  {W.~J.}\ \bibnamefont {Munro}}, \bibinfo {author} {\bibfnamefont
  {H.}~\bibnamefont {Yamaguchi}}, \bibinfo {author} {\bibfnamefont
  {N.}~\bibnamefont {Mizuochi}}, \ and\ \bibinfo {author} {\bibfnamefont
  {S.}~\bibnamefont {Saito}},\ }\href@noop {} {\bibfield  {journal} {\bibinfo
  {journal} {Journal of Physics: Condensed Matter}\ }\textbf {\bibinfo {volume}
  {28}},\ \bibinfo {pages} {275302} (\bibinfo {year} {2016})}\BibitemShut
  {NoStop}%
\bibitem [{\citenamefont {Chen}\ \emph {et~al.}(2017)\citenamefont {Chen},
  \citenamefont {Clevenson}, \citenamefont {Johnson}, \citenamefont {Pham},
  \citenamefont {Englund}, \citenamefont {Hemmer},\ and\ \citenamefont
  {Braje}}]{PhysRevA.95.053417}%
  \BibitemOpen
  \bibfield  {author} {\bibinfo {author} {\bibfnamefont {E.~H.}\ \bibnamefont
  {Chen}}, \bibinfo {author} {\bibfnamefont {H.~A.}\ \bibnamefont {Clevenson}},
  \bibinfo {author} {\bibfnamefont {K.~A.}\ \bibnamefont {Johnson}}, \bibinfo
  {author} {\bibfnamefont {L.~M.}\ \bibnamefont {Pham}}, \bibinfo {author}
  {\bibfnamefont {D.~R.}\ \bibnamefont {Englund}}, \bibinfo {author}
  {\bibfnamefont {P.~R.}\ \bibnamefont {Hemmer}}, \ and\ \bibinfo {author}
  {\bibfnamefont {D.~A.}\ \bibnamefont {Braje}},\ }\href {\doibase
  10.1103/PhysRevA.95.053417} {\bibfield  {journal} {\bibinfo  {journal} {Phys.
  Rev. A}\ }\textbf {\bibinfo {volume} {95}},\ \bibinfo {pages} {053417}
  (\bibinfo {year} {2017})}\BibitemShut {NoStop}%
\bibitem [{\citenamefont {Levchenko}\ \emph {et~al.}(2015)\citenamefont
  {Levchenko}, \citenamefont {Vasil'ev}, \citenamefont {Zibrov}, \citenamefont
  {Zibrov}, \citenamefont {Sivak},\ and\ \citenamefont
  {Fedotov}}]{Levchenko:2015}%
  \BibitemOpen
  \bibfield  {author} {\bibinfo {author} {\bibfnamefont {A.~O.}\ \bibnamefont
  {Levchenko}}, \bibinfo {author} {\bibfnamefont {V.~V.}\ \bibnamefont
  {Vasil'ev}}, \bibinfo {author} {\bibfnamefont {S.~A.}\ \bibnamefont
  {Zibrov}}, \bibinfo {author} {\bibfnamefont {A.~S.}\ \bibnamefont {Zibrov}},
  \bibinfo {author} {\bibfnamefont {A.~V.}\ \bibnamefont {Sivak}}, \ and\
  \bibinfo {author} {\bibfnamefont {I.~V.}\ \bibnamefont {Fedotov}},\ }\href
  {\doibase 10.1063/1.4913428} {\bibfield  {journal} {\bibinfo  {journal}
  {Applied Physics Letters}\ }\textbf {\bibinfo {volume} {106}},\ \bibinfo
  {pages} {102402} (\bibinfo {year} {2015})},\ \Eprint
  {http://arxiv.org/abs/https://doi.org/10.1063/1.4913428}
  {https://doi.org/10.1063/1.4913428} \BibitemShut {NoStop}%
\bibitem [{\citenamefont {Steele}\ \emph {et~al.}(2017)\citenamefont {Steele},
  \citenamefont {Lawson}, \citenamefont {Onyszczak}, \citenamefont {Bush},
  \citenamefont {Mei}, \citenamefont {Dioguardi}, \citenamefont {King},
  \citenamefont {Parker}, \citenamefont {Pines}, \citenamefont {Weir},
  \citenamefont {Evans}, \citenamefont {Visbeck}, \citenamefont {Vohra},\ and\
  \citenamefont {Curro}}]{Steele:2017cm}%
  \BibitemOpen
  \bibfield  {author} {\bibinfo {author} {\bibfnamefont {L.~G.}\ \bibnamefont
  {Steele}}, \bibinfo {author} {\bibfnamefont {M.}~\bibnamefont {Lawson}},
  \bibinfo {author} {\bibfnamefont {M.}~\bibnamefont {Onyszczak}}, \bibinfo
  {author} {\bibfnamefont {B.~T.}\ \bibnamefont {Bush}}, \bibinfo {author}
  {\bibfnamefont {Z.}~\bibnamefont {Mei}}, \bibinfo {author} {\bibfnamefont
  {A.~P.}\ \bibnamefont {Dioguardi}}, \bibinfo {author} {\bibfnamefont
  {J.}~\bibnamefont {King}}, \bibinfo {author} {\bibfnamefont {A.}~\bibnamefont
  {Parker}}, \bibinfo {author} {\bibfnamefont {A.}~\bibnamefont {Pines}},
  \bibinfo {author} {\bibfnamefont {S.~T.}\ \bibnamefont {Weir}}, \bibinfo
  {author} {\bibfnamefont {W.}~\bibnamefont {Evans}}, \bibinfo {author}
  {\bibfnamefont {K.}~\bibnamefont {Visbeck}}, \bibinfo {author} {\bibfnamefont
  {Y.~K.}\ \bibnamefont {Vohra}}, \ and\ \bibinfo {author} {\bibfnamefont
  {N.~J.}\ \bibnamefont {Curro}},\ }\href@noop {} {\bibfield  {journal}
  {\bibinfo  {journal} {Applied Physics Letters}\ }\textbf {\bibinfo {volume}
  {111}},\ \bibinfo {pages} {221903} (\bibinfo {year} {2017})}\BibitemShut
  {NoStop}%
\bibitem [{\citenamefont {Maze}\ \emph {et~al.}(2011)\citenamefont {Maze},
  \citenamefont {Gali}, \citenamefont {Togan}, \citenamefont {Chu},
  \citenamefont {Trifonov}, \citenamefont {Kaxiras},\ and\ \citenamefont
  {Lukin}}]{Maze:2011gw}%
  \BibitemOpen
  \bibfield  {author} {\bibinfo {author} {\bibfnamefont {J.~R.}\ \bibnamefont
  {Maze}}, \bibinfo {author} {\bibfnamefont {A.}~\bibnamefont {Gali}}, \bibinfo
  {author} {\bibfnamefont {E.}~\bibnamefont {Togan}}, \bibinfo {author}
  {\bibfnamefont {Y.}~\bibnamefont {Chu}}, \bibinfo {author} {\bibfnamefont
  {A.}~\bibnamefont {Trifonov}}, \bibinfo {author} {\bibfnamefont
  {E.}~\bibnamefont {Kaxiras}}, \ and\ \bibinfo {author} {\bibfnamefont
  {M.~D.}\ \bibnamefont {Lukin}},\ }\href@noop {} {\bibfield  {journal}
  {\bibinfo  {journal} {New Journal of Physics}\ }\textbf {\bibinfo {volume}
  {13}},\ \bibinfo {pages} {025025} (\bibinfo {year} {2011})}\BibitemShut
  {NoStop}%
\bibitem [{\citenamefont {Smeltzer}\ \emph {et~al.}(2011)\citenamefont
  {Smeltzer}, \citenamefont {Childress},\ and\ \citenamefont
  {Gali}}]{Smeltzer:2011fb}%
  \BibitemOpen
  \bibfield  {author} {\bibinfo {author} {\bibfnamefont {B.}~\bibnamefont
  {Smeltzer}}, \bibinfo {author} {\bibfnamefont {L.}~\bibnamefont {Childress}},
  \ and\ \bibinfo {author} {\bibfnamefont {A.}~\bibnamefont {Gali}},\
  }\href@noop {} {\bibfield  {journal} {\bibinfo  {journal} {New Journal of
  Physics}\ }\textbf {\bibinfo {volume} {13}},\ \bibinfo {pages} {025021}
  (\bibinfo {year} {2011})}\BibitemShut {NoStop}%
\bibitem [{\citenamefont {Oort}\ and\ \citenamefont
  {Glasbeek}(1990)}]{VANOORT1990}%
  \BibitemOpen
  \bibfield  {author} {\bibinfo {author} {\bibfnamefont {E.~V.}\ \bibnamefont
  {Oort}}\ and\ \bibinfo {author} {\bibfnamefont {M.}~\bibnamefont
  {Glasbeek}},\ }\href {\doibase https://doi.org/10.1016/0009-2614(90)85665-Y}
  {\bibfield  {journal} {\bibinfo  {journal} {Chemical Physics Letters}\
  }\textbf {\bibinfo {volume} {168}},\ \bibinfo {pages} {529 } (\bibinfo {year}
  {1990})}\BibitemShut {NoStop}%
\bibitem [{Kob()}]{Kobrin_18xx}%
  \BibitemOpen
  \href@noop {} {}\bibinfo {note} {Kobrin et al., in preparation
  (2018).}\BibitemShut {Stop}%
\bibitem [{Note1()}]{Note1}%
  \BibitemOpen
  \bibinfo {note} {We assume that the charges are independently positioned in
  three dimensions}\BibitemShut {NoStop}%
\bibitem [{Note2()}]{Note2}%
  \BibitemOpen
  \bibinfo {note} {We note that the the hyperfine interaction in the
  Hamiltonian is obtained under the secular approximation.}\BibitemShut {Stop}%
\bibitem [{sup()}]{supp}%
  \BibitemOpen
  \href@noop {} {}\bibinfo {note} {See Supplementary Material for detailed
  information.}\BibitemShut {Stop}%
\bibitem [{\citenamefont {Davies}(1977)}]{DAVIES:1977kt}%
  \BibitemOpen
  \bibfield  {author} {\bibinfo {author} {\bibfnamefont {G.}~\bibnamefont
  {Davies}},\ }\href@noop {} {\bibfield  {journal} {\bibinfo  {journal}
  {Nature}\ }\textbf {\bibinfo {volume} {269}},\ \bibinfo {pages} {498}
  (\bibinfo {year} {1977})}\BibitemShut {NoStop}%
\bibitem [{Note3()}]{Note3}%
  \BibitemOpen
  \bibinfo {note} {We measure the ODMR spectra of 68 single NV centers in an
  untreated Type-Ib sample, and find four that exhibit a significant
  electric-field-induced splitting with amplitude difference at zero magnetic
  field \cite {supp}.}\BibitemShut {Stop}%
\bibitem [{\citenamefont {Chu}\ \emph {et~al.}(2014)\citenamefont {Chu},
  \citenamefont {de~Leon}, \citenamefont {Shields}, \citenamefont {Hausmann},
  \citenamefont {Evans}, \citenamefont {Togan}, \citenamefont {Burek},
  \citenamefont {Markham}, \citenamefont {Stacey}, \citenamefont {Zibrov},
  \citenamefont {Yacoby}, \citenamefont {Twitchen}, \citenamefont {Loncar},
  \citenamefont {Park}, \citenamefont {Maletinsky},\ and\ \citenamefont
  {Lukin}}]{Chu:2014es}%
  \BibitemOpen
  \bibfield  {author} {\bibinfo {author} {\bibfnamefont {Y.}~\bibnamefont
  {Chu}}, \bibinfo {author} {\bibfnamefont {N.~P.}\ \bibnamefont {de~Leon}},
  \bibinfo {author} {\bibfnamefont {B.~J.}\ \bibnamefont {Shields}}, \bibinfo
  {author} {\bibfnamefont {B.}~\bibnamefont {Hausmann}}, \bibinfo {author}
  {\bibfnamefont {R.}~\bibnamefont {Evans}}, \bibinfo {author} {\bibfnamefont
  {E.}~\bibnamefont {Togan}}, \bibinfo {author} {\bibfnamefont {M.~J.}\
  \bibnamefont {Burek}}, \bibinfo {author} {\bibfnamefont {M.}~\bibnamefont
  {Markham}}, \bibinfo {author} {\bibfnamefont {A.}~\bibnamefont {Stacey}},
  \bibinfo {author} {\bibfnamefont {A.~S.}\ \bibnamefont {Zibrov}}, \bibinfo
  {author} {\bibfnamefont {A.}~\bibnamefont {Yacoby}}, \bibinfo {author}
  {\bibfnamefont {D.~J.}\ \bibnamefont {Twitchen}}, \bibinfo {author}
  {\bibfnamefont {M.}~\bibnamefont {Loncar}}, \bibinfo {author} {\bibfnamefont
  {H.}~\bibnamefont {Park}}, \bibinfo {author} {\bibfnamefont {P.}~\bibnamefont
  {Maletinsky}}, \ and\ \bibinfo {author} {\bibfnamefont {M.~D.}\ \bibnamefont
  {Lukin}},\ }\href@noop {} {\bibfield  {journal} {\bibinfo  {journal} {Nano
  Letters}\ }\textbf {\bibinfo {volume} {14}},\ \bibinfo {pages} {1982}
  (\bibinfo {year} {2014})}\BibitemShut {NoStop}%
\bibitem [{\citenamefont {Jelezko}\ \emph {et~al.}(2002)\citenamefont
  {Jelezko}, \citenamefont {Popa}, \citenamefont {Gruber}, \citenamefont
  {Tietz}, \citenamefont {Wrachtrup}, \citenamefont {Nizovtsev},\ and\
  \citenamefont {Kilin}}]{Jelezko:2002kx}%
  \BibitemOpen
  \bibfield  {author} {\bibinfo {author} {\bibfnamefont {F.}~\bibnamefont
  {Jelezko}}, \bibinfo {author} {\bibfnamefont {I.}~\bibnamefont {Popa}},
  \bibinfo {author} {\bibfnamefont {A.}~\bibnamefont {Gruber}}, \bibinfo
  {author} {\bibfnamefont {C.}~\bibnamefont {Tietz}}, \bibinfo {author}
  {\bibfnamefont {J.}~\bibnamefont {Wrachtrup}}, \bibinfo {author}
  {\bibfnamefont {A.}~\bibnamefont {Nizovtsev}}, \ and\ \bibinfo {author}
  {\bibfnamefont {S.}~\bibnamefont {Kilin}},\ }\href@noop {} {\bibfield
  {journal} {\bibinfo  {journal} {Applied Physics Letters}\ }\textbf {\bibinfo
  {volume} {81}},\ \bibinfo {pages} {2160} (\bibinfo {year}
  {2002})}\BibitemShut {NoStop}%
\end{thebibliography}%


%merlin.mbs apsrev4-1.bst 2010-07-25 4.21a (PWD, AO, DPC) hacked
%Control: key (0)
%Control: author (8) initials jnrlst
%Control: editor formatted (1) identically to author
%Control: production of article title (-1) disabled
%Control: page (0) single
%Control: year (1) truncated
%Control: production of eprint (0) enabled
\begin{thebibliography}{3}%
\makeatletter
\providecommand \@ifxundefined [1]{%
 \@ifx{#1\undefined}
}%
\providecommand \@ifnum [1]{%
 \ifnum #1\expandafter \@firstoftwo
 \else \expandafter \@secondoftwo
 \fi
}%
\providecommand \@ifx [1]{%
 \ifx #1\expandafter \@firstoftwo
 \else \expandafter \@secondoftwo
 \fi
}%
\providecommand \natexlab [1]{#1}%
\providecommand \enquote  [1]{``#1''}%
\providecommand \bibnamefont  [1]{#1}%
\providecommand \bibfnamefont [1]{#1}%
\providecommand \citenamefont [1]{#1}%
\providecommand \href@noop [0]{\@secondoftwo}%
\providecommand \href [0]{\begingroup \@sanitize@url \@href}%
\providecommand \@href[1]{\@@startlink{#1}\@@href}%
\providecommand \@@href[1]{\endgroup#1\@@endlink}%
\providecommand \@sanitize@url [0]{\catcode `\\12\catcode `\$12\catcode
  `\&12\catcode `\#12\catcode `\^12\catcode `\_12\catcode `\%12\relax}%
\providecommand \@@startlink[1]{}%
\providecommand \@@endlink[0]{}%
\providecommand \url  [0]{\begingroup\@sanitize@url \@url }%
\providecommand \@url [1]{\endgroup\@href {#1}{\urlprefix }}%
\providecommand \urlprefix  [0]{URL }%
\providecommand \Eprint [0]{\href }%
\providecommand \doibase [0]{http://dx.doi.org/}%
\providecommand \selectlanguage [0]{\@gobble}%
\providecommand \bibinfo  [0]{\@secondoftwo}%
\providecommand \bibfield  [0]{\@secondoftwo}%
\providecommand \translation [1]{[#1]}%
\providecommand \BibitemOpen [0]{}%
\providecommand \bibitemStop [0]{}%
\providecommand \bibitemNoStop [0]{.\EOS\space}%
\providecommand \EOS [0]{\spacefactor3000\relax}%
\providecommand \BibitemShut  [1]{\csname bibitem#1\endcsname}%
\let\auto@bib@innerbib\@empty
%</preamble>
\bibitem [{\citenamefont {Whitehead}\ and\ \citenamefont
  {Hackett}(1939)}]{Whitehead:1939df}%
  \BibitemOpen
  \bibfield  {author} {\bibinfo {author} {\bibfnamefont {S.}~\bibnamefont
  {Whitehead}}\ and\ \bibinfo {author} {\bibfnamefont {W.}~\bibnamefont
  {Hackett}},\ }\href@noop {} {\bibfield  {journal} {\bibinfo  {journal}
  {Proceedings of the Physical Society}\ }\textbf {\bibinfo {volume} {51}},\
  \bibinfo {pages} {173} (\bibinfo {year} {1939})}\BibitemShut {NoStop}%
\bibitem [{\citenamefont {Smeltzer}\ \emph {et~al.}(2011)\citenamefont
  {Smeltzer}, \citenamefont {Childress},\ and\ \citenamefont
  {Gali}}]{C13Hyper}%
  \BibitemOpen
  \bibfield  {author} {\bibinfo {author} {\bibfnamefont {B.}~\bibnamefont
  {Smeltzer}}, \bibinfo {author} {\bibfnamefont {L.}~\bibnamefont {Childress}},
  \ and\ \bibinfo {author} {\bibfnamefont {A.}~\bibnamefont {Gali}},\ }\href
  {http://stacks.iop.org/1367-2630/13/i=2/a=025021} {\bibfield  {journal}
  {\bibinfo  {journal} {New Journal of Physics}\ }\textbf {\bibinfo {volume}
  {13}},\ \bibinfo {pages} {025021} (\bibinfo {year} {2011})}\BibitemShut
  {NoStop}%
\bibitem [{\citenamefont {Oort}\ and\ \citenamefont
  {Glasbeek}(1990)}]{VANOORT1990}%
  \BibitemOpen
  \bibfield  {author} {\bibinfo {author} {\bibfnamefont {E.~V.}\ \bibnamefont
  {Oort}}\ and\ \bibinfo {author} {\bibfnamefont {M.}~\bibnamefont
  {Glasbeek}},\ }\href {\doibase https://doi.org/10.1016/0009-2614(90)85665-Y}
  {\bibfield  {journal} {\bibinfo  {journal} {Chemical Physics Letters}\
  }\textbf {\bibinfo {volume} {168}},\ \bibinfo {pages} {529 } (\bibinfo {year}
  {1990})}\BibitemShut {NoStop}%
\end{thebibliography}%

\end{document}

\title{Supplementary Material: Imaging the local charge environment of nitrogen-vacancy centers in diamond}

\author{T.~Mittiga}
\thanks{These authors contributed equally to this work}
\author{S.~Hsieh}
\thanks{These authors contributed equally to this work}
\author{C.~Zu}
\thanks{These authors contributed equally to this work}
\author{B.~Kobrin}
\author{F.~Machado}
\author{P.~Bhattacharyya}
\author{N.~Rui}
\author{A.~Jarmola}
\author{S.~Choi}
\author{D.~Budker}
\author{N.~Y.~Yao}
\date{\today}

\maketitle
\tableofcontents

\section{Materials and Methods}

\subsection{Sample details}
The six diamond samples used in this work are all sourced from Element Six. Three of them have been treated (electron irradiation at Prism Gem and vacuum annealing) to increase NV density. The details are listed in Table I.
\begin{table}[H]
\centering
\begin{tabular}{l|c|c|c|c|c|c}\hline\hline
Sample name & Synthesis & \makecell{{[}N{]} \\ (ppm)}  & \makecell{Electron \\ irradiation dose} & \makecell{Energy\\ (MeV)} & \makecell{Anneal temperature \\($^{\circ}$C) } & \makecell{Spectrum}\\\hline

Ib treated (S1)        & HPHT      & $\lesssim$200 & 2$\times$10$^{18}$ cm$^{-2}$   & 2            & 800   & Fig.~\ref{ensemble1}a, main text Fig. 1a            \\
Ib treated (S2)        & HPHT      & $\lesssim$200 & 1$\times$10$^{17}$ cm$^{-2}$     & 14           & 400; 800; 1200     & Fig~\ref{ensemble1}b    \\
IIa treated (S3)         & CVD       & $\lesssim$ 1   & 1$\times$10$^{17}$ cm$^{-2}$     & 2            & 700; 875    & Fig.~\ref{ensemble1}c, main text Fig. 2a           \\ \hline
Ib untreated (S4)      & HPHT      & $\lesssim$200 & n/a                            &    n/a           & n/a  & Fig.~\ref{ensemble2}a  \\
Ib untreated (S5)       & HPHT      & $\lesssim$200 & n/a                            &    n/a           &    n/a   & Fig.~\ref{ensemble2}b, main text Fig. 2b   \\  
IIa untreated (S6)        & CVD       & $\lesssim$1   & n/a                            &   n/a            &     n/a      & Fig.~\ref{ensemble2}c             \\
\hline\hline
\end{tabular}
\caption{Details of all samples shown in main and supplementary text. All samples are sourced from Element Six. [N] is specified by the manufacturer. }
\end{table}
% \todo[inline]{Add column specifying which figure a given sample is shown}
\subsection{Experimental apparatus}

\begin{figure}
\centering
\includegraphics[width=1\textwidth]{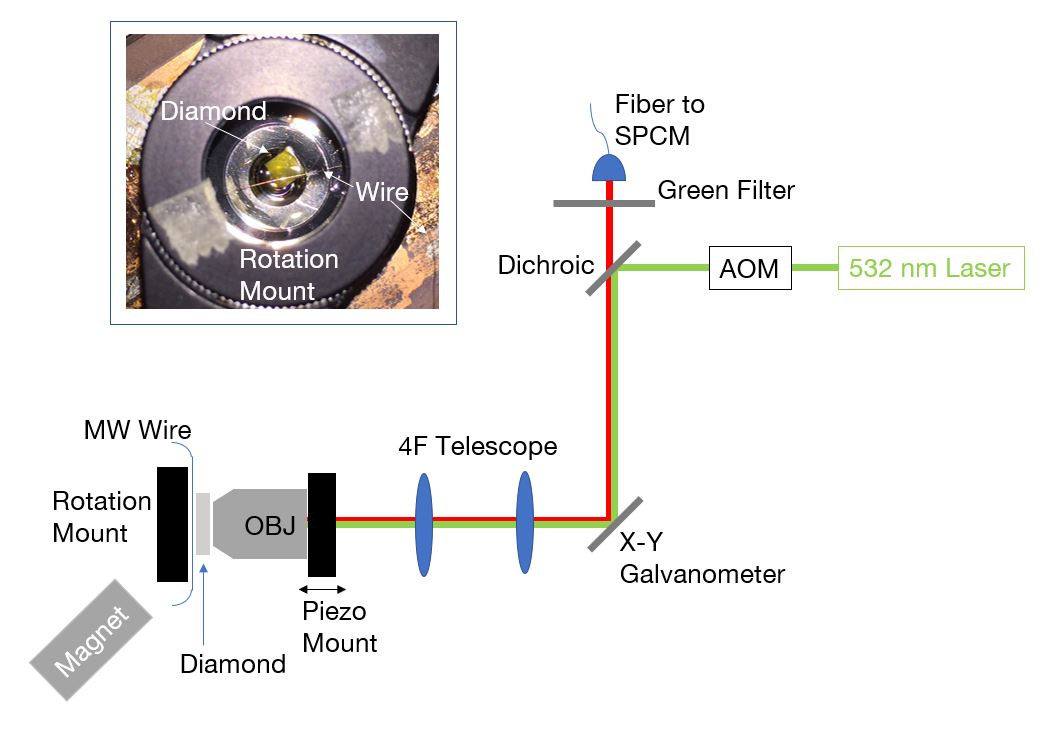}
\caption{Experimental Apparatus: A 532~nm laser shuttered by an AOM light switch excites the NVs, both for state preparation and read-out. A 4$f$ telescope permits the galvonometer to scan the surface of the diamond and a piezo-mounted objective controls the depth of the focal plane. The objective lens focuses the excitation beam and collects fluorescence. Microwave fields are delivered by a magnet wire (as pictured) or a coplanar waveguide. Inset: Magnet wire stretched onto an optical rotation mount hovers over the surface of the diamond }
\label{setup image}
\end{figure}

We conduct single and ensemble NV measurements in a scanning confocal microscope equipped with controllable magnetic field and microwave delivery (Fig.~\ref{setup image}). %To probe single NVs in diamond, we constructed a scanning confocal microscope, equipped with magnetic field control and microwave delivery.
A 532 nm laser beam (Coherent Compass 315M) shuttered by an acousto-optic modulator (AOM, Gooch $\&$ Housego AOMO 3110-120) is used for both ground state preparation and spin state detection. 
An objective lens focuses the beam to a diffraction limited spot size. 
We use an oil immersion objective lens (Nikon Plan Fluor 100x, NA 1.49) for resolving single NV centers or an air objective lens (Olympus LUCPLFLN, NA 0.6) for ensemble measurements.
The combined action of an X-Y galvanometer (Thorlabs GVS212) and a 4$f$ telescope provides the ability to scan the sample at the focal plane of the objective lens. 
A piezo mount for the objective lens serves to move the scanning plane in the longitudinal direction for depth scans.
%NV detection relies upon optically exciting the NV and collecting its fluorescence. 
% A 532 nm Coherent Compass green laser excites the vibronic sideband of the NV optical transition. The excitation beam is shuttered by an acousto-optical modulator (Gooch $\&$ Housego AOMO 3110-120), and focused to a diffraction-limited spot-size by either an oil immersion lens (Nikon Plan Fluor 100x, NA 1.49) for resolving single NVs or an air objective lens (Olympus LUCPLFLN, NA 0.6) for ensemble measurements.
%We scan focal planes parallel to the surface of the diamond with an open-loop X-Y galvanometer (Thorlabs GVS212). A 4F telescope images the galvanometer’s mirror angles onto the input of the objective lens without varying the position of the beam on the input. %The objective lens maps each incidence angle at its input to a position on the focal plane, permitting scanning the plane with minimal alteration of transmission power. Such surface scans are limited to a 200x200 $\mu$m region. 
%The depth of a single focal plane is controlled by a closed-loop piezo objective lens mount (PI). %The focal plane is always a constant distance from the objective lens, so varying the objective lens displacement from the diamond varies what depth the focal plane lies below the diamond surface.
%The NV fluoresces in a spherical uniform radiation pattern. The objective lens collects and collimates a solid angle of the radiation and sends it counter-propagating along same optical path used by the excitation beam.

The fluorescence photons collected by the objective lens are separated from the excitation beam path by a dichroic mirror (Semrock FF552-Di02). 
The coupling of the fluorescence beam to a single mode fiber serves as an effective pinhole for the confocal microscope. 
The fiber shuttles the fluorescence photons to a single photon counting module (SPCM, Excelitas SPCM-AQRH-64-FC) or avalanche photodiode (Thorlabs APD410A). 
We use a Data Aquisition card (DAQ) for fluorescence measurements and subsequent data processing (National Instruments USB 6343).
%The fluorescence photons diverge from the excitation beam path at the dichroic mirror (Semrock), which transmits red, but reflects green. After passing a filter (Semrock) to remove errant excitation photons, the fluorescence photons are coupled into a single-mode fiber, serving as the pinhole of the confocal microscope, and delivered to a single photon counting module (Excelitas SPCM-AQRH-64-FC). Photon detection events on the SPCM are counted by a DAQ (National Instruments USB-6343).

A microwave source (Stanford Research Systems SG384) in combination with a 16W amplifier (Mini-Circuits ZHL-16W-43+) serves to generate signals for spin state manipulation. For ensemble measurements, microwave signals are delivered using a coplanar waveguide (CPW) deposited on a coverslip. For single NV experiments, a 46~AWG magnet wire taped to a rotation mount (Thorlabs RSP05) is used (Fig.~\ref{setup image} inset). The magnet wire is adjusted to sit parallel to, and approximately 550~$\mu m$ above, the focal plane of the objective lens. By rotating the wire using the mount, we effectively change the polarization of the microwaves at the site of the single NV center of interest.
The calculation of the polarization angle in the NV center frame is discussed later in the Section {\bf Microwave Angle Projection}.
%%% IS THIS REALLY TRUE?? I BELIEVE WE JUST ASSUMED IT TO BE 550 UM AWAY -- PRA --> Variations over 100um do not change the angle by much, so it is safe to guess the wire is just above the diamond surface and hte diamond is the stated 500um thickness.
%We deliver microwaves via a 40 $\mu$m diameter magnet wire mounted on a lens rotation mount (Thorlabs). %, and is long enough to prevent major stress on the wire during rotation. 
%Using a translation stage (Newport) the height and XY position of the wire is set so the wire hovers a few $\mu$m above the back surface of the diamond, and approximately 550 $\mu$m above the focal plane containing our single NVs. 
% A 16W amplifier (Mini-Circuits ZHL-16W-43+) amplifies the microwave source (Stanford Research Systems SG384). 
%This configuration was designed for simplicity, rather than efficiency. While approximating the microwave linear polarization angle at the location of the NV is simplified, the power delivery to the NV is severely limited, as will be more explicitly addressed in the discussion of Rabi frequency calibration. The wire heats due to poor transmission (Measured value), but since it is not in thermal contact with the diamond, positional drift of single NVs is undetectable.
% When needed, magnetic fields are set by a permanent magnet mounted on adjustable posts. %Since only rough alignment to the NV z-axis is required for our purposes, the magnetic fields are aligned by hand. The posts can be moved to different locations on the optical table for alignment to each of the four possible NV orientations.

\subsection{Pulse sequence for measuring magnetic resonance spectra of NVs}

\begin{figure}[h]
\centering
\includegraphics[width=0.5\textwidth]{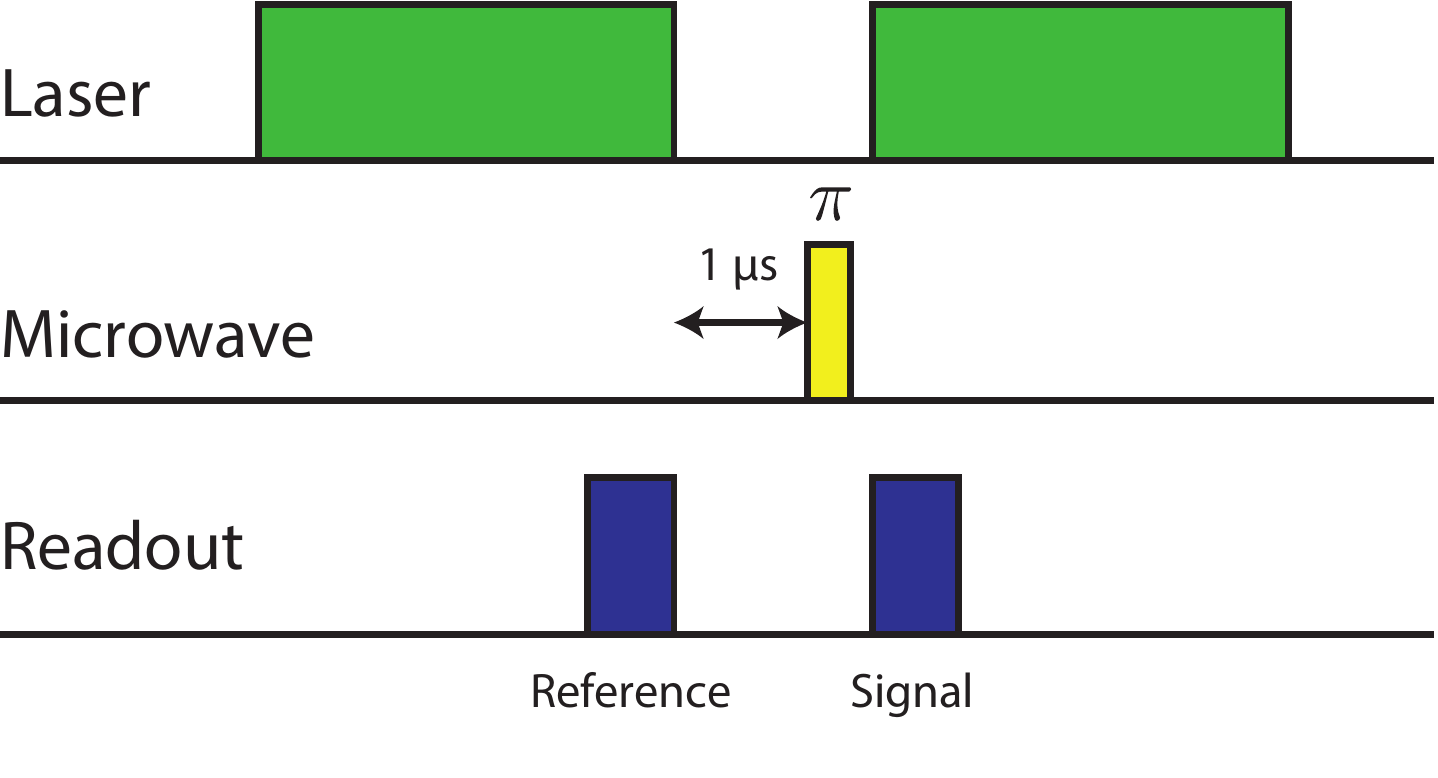}
\caption{Pulse sequence for ODMR measurement.}
\label{Pulse sequence}
\end{figure}

To measure the optically detected magnetic resonance (ODMR) spectra of NVs, we first use a $10~\mu s$ 532~nm laser pulse to initialize the spin triplet ground states to $m_s = 0$ (Fig.~\ref{Pulse sequence}).
After turning off the laser for 1~$\mu$s to allow the excited state population to decay, we apply a microwave $\pi$ pulse and sweep its frequency.
Our $\pi$ pulse length is chosen as 2-6~$\mu$s for measurements on Type-Ib diamonds and 8-10~$\mu$s for Type-IIa diamonds to avoid power broadening.
At the end, we apply another 10~$\mu$s laser pulse to detect the NV spin state by collecting the resulting fluorescence photons (Signal).
In addition, we collect the photon counts at the end of the initialization laser pulse (Reference), and normalize the measured contrast.

\subsection{Isolating single NVs}
\begin{figure}[h]
\centering
\includegraphics[width=0.5\textwidth]{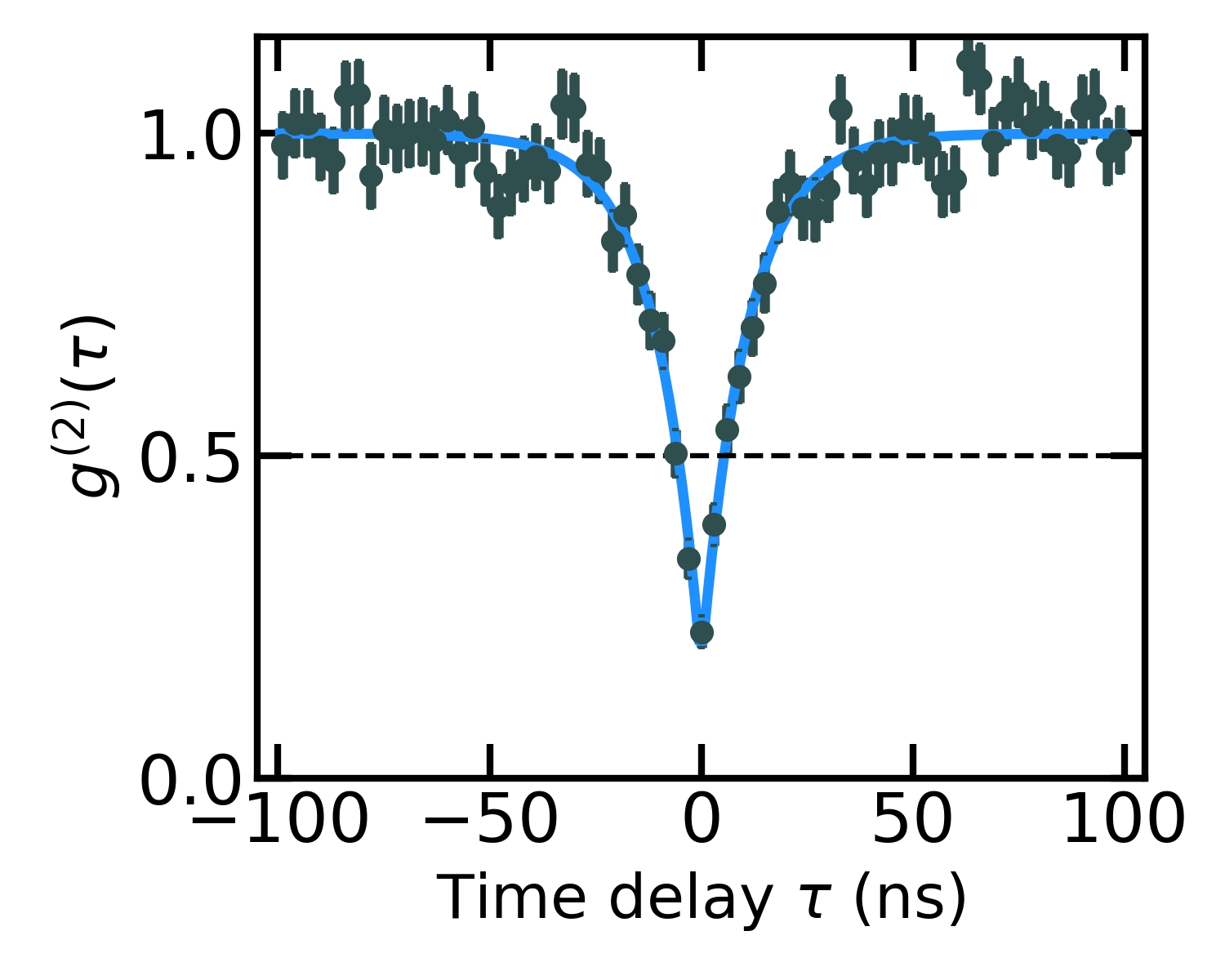}
\caption{$g^2(\tau)$ measurement on NV1: the extracted $g^2(0)= 0.17^{+0.05}_{-0.03} <0.5$ definitively confirms it is a single NV center.}
\label{g2}
\end{figure}

The diamond sample used for single NV experiments is sample S4 (untreated type Ib).
We found a region of the sample where we could isolate single NVs as confirmed by a $g^{(2)}$ measurement (Fig.~\ref{g2}).
 
\section{Charge model and ensemble spectrum}

In this section, we provide additional details regarding our charge model. This includes an analysis of the electric and magnetic field distributions, as well as an explanation of the fitting procedure of the ensemble spectra and the estimation of error bars. 

\subsection{Electric field distribution}
\label{sec:EleDist}
In our model, we consider each NV to be surrounded by an equal density, $\rho_c$ of positive and negative point-like charges. 
We simulate the positions of these charges by randomly placing a large number ($N_{\textrm{charge}}\sim 100$) of points within a spherical volume. The radius of the sphere, $R$, is determined such that the average density of the charges matches $\rho_{\textrm{c}}$; in particular, this implies 
\begin{equation}\label{radius}
R = \left(\frac 3 {4 \pi} \frac{N_\textrm{charge}}{n_0 \rho_{\textrm{c}}} \right)^{\frac 1 3}
\end{equation} 
where $n_0 = 1.76\times 10^{-4}$ (ppm$\cdot$nm$^3$)$^{-1}$ is the factor relating the number density (in ppm) to the volume density.
Based on the positions of the charges $\{\vec{r}_i\}$, we calculate the electric field at the center of the sphere (the NV's location):
\begin{equation} \label{coulomb}
\vec E = \sum_i \frac{e}{4\pi \epsilon_0 \epsilon_r}\frac{\hat r_i}{r_i^2}
\end{equation}
where $\epsilon_r = 5.7$ is the relative permittivity of diamond\cite{Whitehead:1939df}.

\begin{figure}[b] 
\centering
\includegraphics[width=0.6\textwidth]{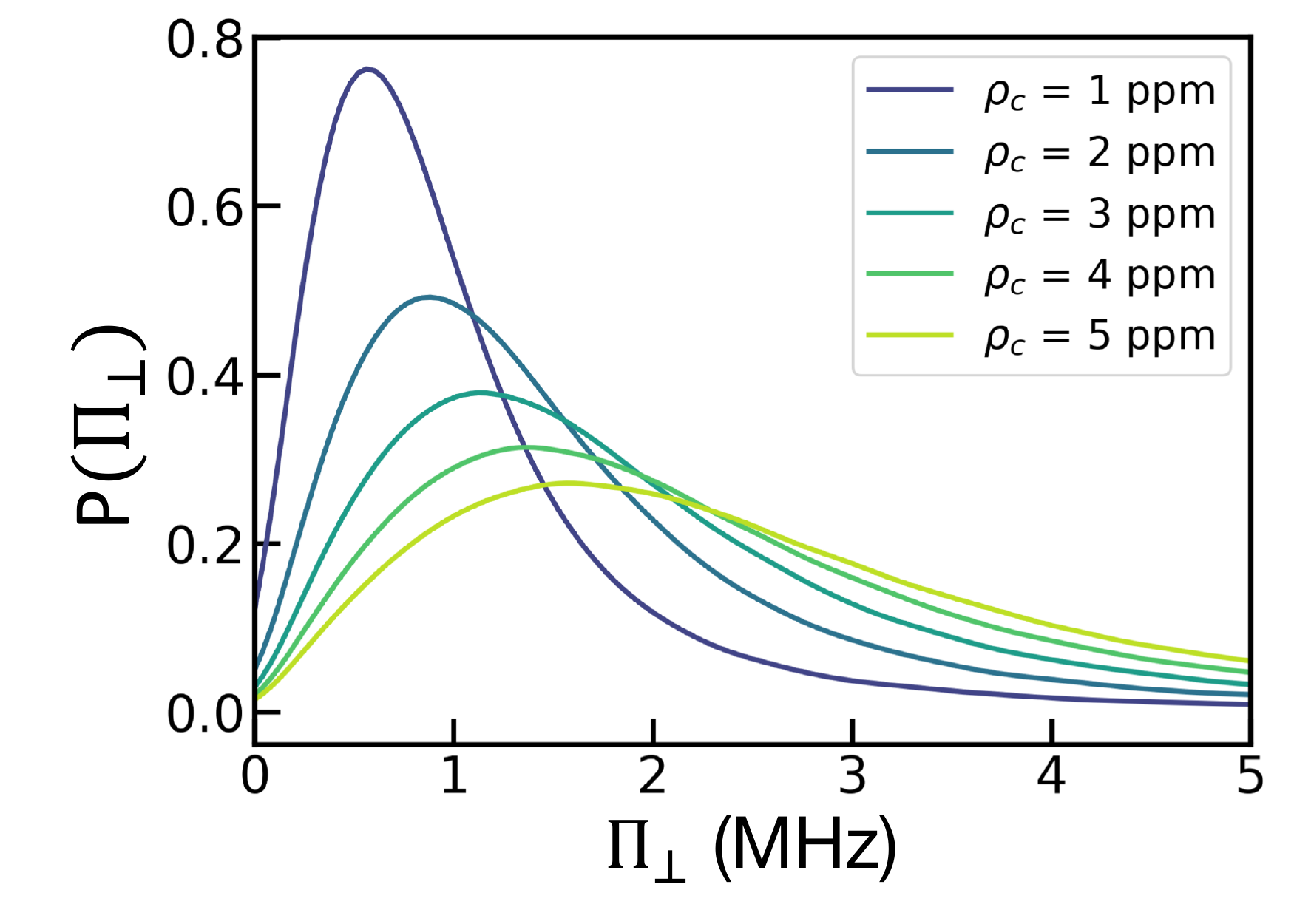}
\caption{Distributions for the transverse electric field component, $\Pi_\perp=d_\perp E_\perp$, at various charge densities. The distributions were generated by the charge sampling procedure described in the text.}
\label{edist}
\end{figure}

Sampling over $\{\vec{r}_i\}$ yields a distribution for $\vec E$. 
We are particularly interested in the transverse component, $E_\perp$, which couples $\sim 50$ times stronger to the NV, i.e. $\Pi_\perp = d_\perp E_\perp$. 
The distribution $P(\Pi_\perp)$ for various densities are shown in Fig.~\ref{edist}. 
We note that these distributions are related to each other by a simple rescaling, $\Pi_\perp \rightarrow \rho_{\textrm{c}}^{2/3} \Pi_\perp$, though we do not incorporate this rescaling explicitly in our sampling procedure.

\subsection{Magnetic field distribution}
\label{sec:MagDist}

We assume that the local magnetic environment arises from interactions with other magnetic impurities. For Type-Ib diamond, the dominant impurities are the electronic spins associated with P1 centers. For Type-IIa diamond, the leading contribution comes from the nuclear spins associated with $^{13}$C (1.1\% natural abundance). 

In both cases, we model the effect of the magnetic impurities as a dipolar interaction between the NV and a bath of electronic spins $\left(s = \frac{1}{2}\right)$ at density $\rho_{\textrm{s}}$:
\begin{equation}
H_{\textrm{dipolar}} = \sum_i -\frac{J_0}{r_i^3} \left(3(\hat S \cdot \hat r_i)(\hat P_i \cdot \hat r_i)-\hat S \cdot \hat P_i\right).
\end{equation}
Here $\{\vec r_i\}$ are the positions of the magnetic impurities, $\hat S$, $\hat P_i$ are the spin operators for the NV and impurities, respectively, and $J_0 = (2\pi)52$ MHz$\cdot$nm$^3$. Under the secular approximation, this interaction further simplifies to:
\begin{align} \label{dipolar}
H_{\textrm{dipolar}}  = \delta B_z ~S_z\quad , \quad \delta B_z  = \sum_i -\frac{J_0}{r_i^3}\left(3\hat n^z_i - 1\right)p_i ~,
\end{align}
where $\hat n_i^z = \hat z \cdot \hat r_i$, and $p_i = \pm 1/2$ are the spins of the magnetic impurities at the mean-field level. 

A few remarks are in order. 
First, the coupling strength for nuclear spins is $\sim 2600$ times weaker.
This can be effectively modeled by an electronic spin bath, whose the density is reduced by the same factor. 
All samples are then characterized with a single parameter $\rho_{\textrm{s}}$.
Second, the $^{13}$C nuclear spins give rise to an additional interaction via the Fermi contact term \cite{C13Hyper}. 
Because directly accounting for this is difficult, we approximate its effect as an extra contribution to $\rho_{\textrm{s}}$. 
The resulting spectra are in quantitative agreement with the experimental data at high field (Fig.~\ref{ensemble1},\ref{ensemble2}), validating this approximation.

Similar to the electric field distribution, we sample $\{\vec r_i\}$ for $N_{spin} \sim 100$ from a sphere whose radius is chosen to be consistent with $\rho_{\textrm{s}}$ (Eq.~\ref{radius}). In this case, we also sample a configuration of spins $\{p_i\}$ from a uniform distribution. Inserting $\{\vec r_i\}$ and $\rho_{\textrm{s}}$ into Eq.~\ref{dipolar} allows us to calculate $\delta B_z$ for each realization.

\subsection{Fitting procedure and error estimation}

\begin{figure}[p] 
\centering
\includegraphics[width=\textwidth]{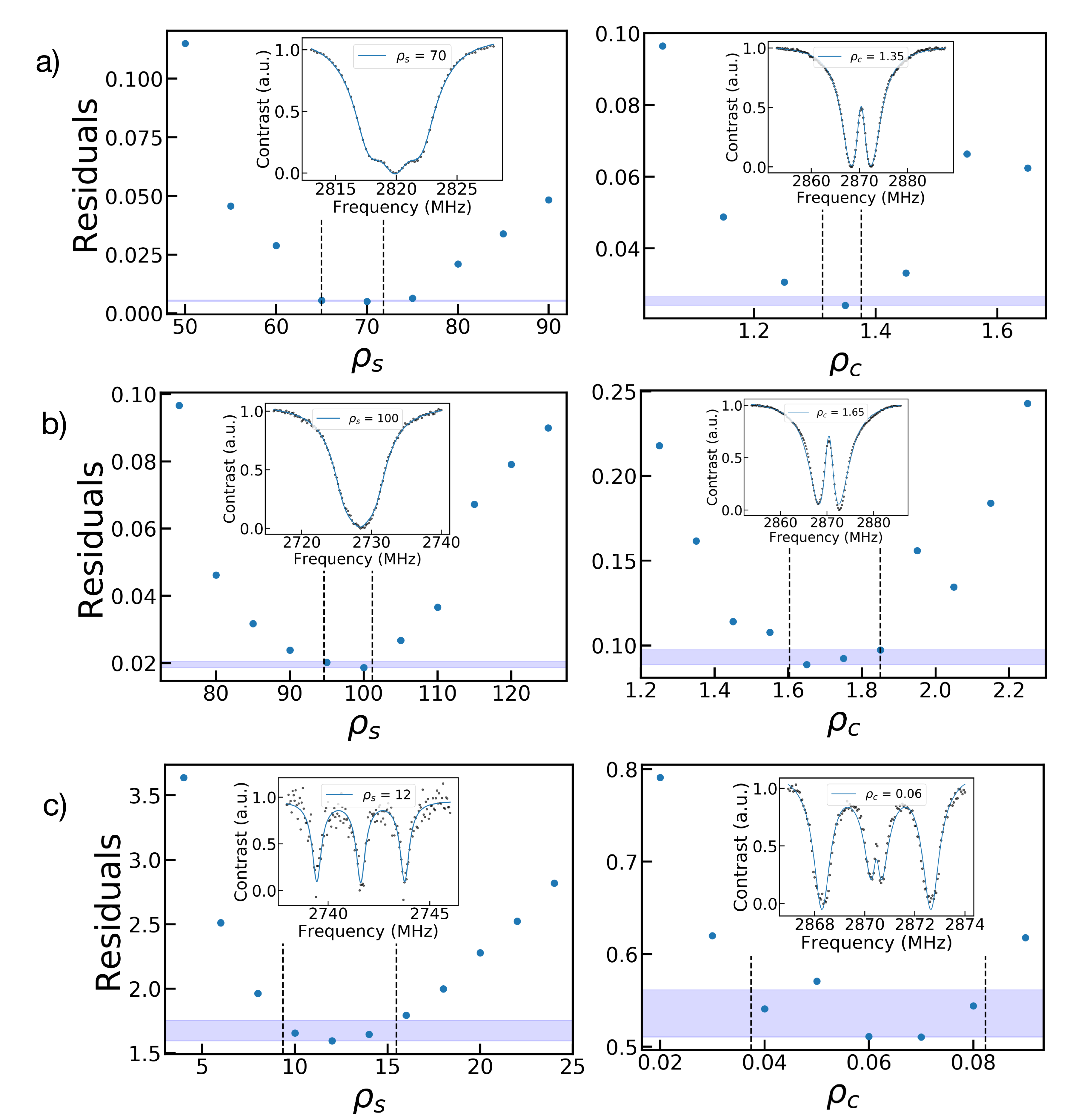}
\caption{Ensemble fitting procedure applied to the treated samples: (a) Ib treated (S1), (b) Ib treated (S2), and (c) IIa treated (S3). The main plots show the least-square residuals as a function of $\rho_\textrm{s}$ (left) and $\rho_\textrm{c}$ (right) under large ($\sim 25$-$50$~G) and zero applied field, respectively. We identify the best-fit values for $\rho_\textrm{s}$, $\rho_\textrm{c}$ based on the minimum residual, and we estimate their error from the range values whose residuals lie within 10\% of the minimum (blue shaded regions). The insets depict the best-fit spectra (blue curve), along with the experimental data (black points).}
\label{ensemble1}
\end{figure}

\begin{figure}[h!] 
\centering
\includegraphics[width=\textwidth]{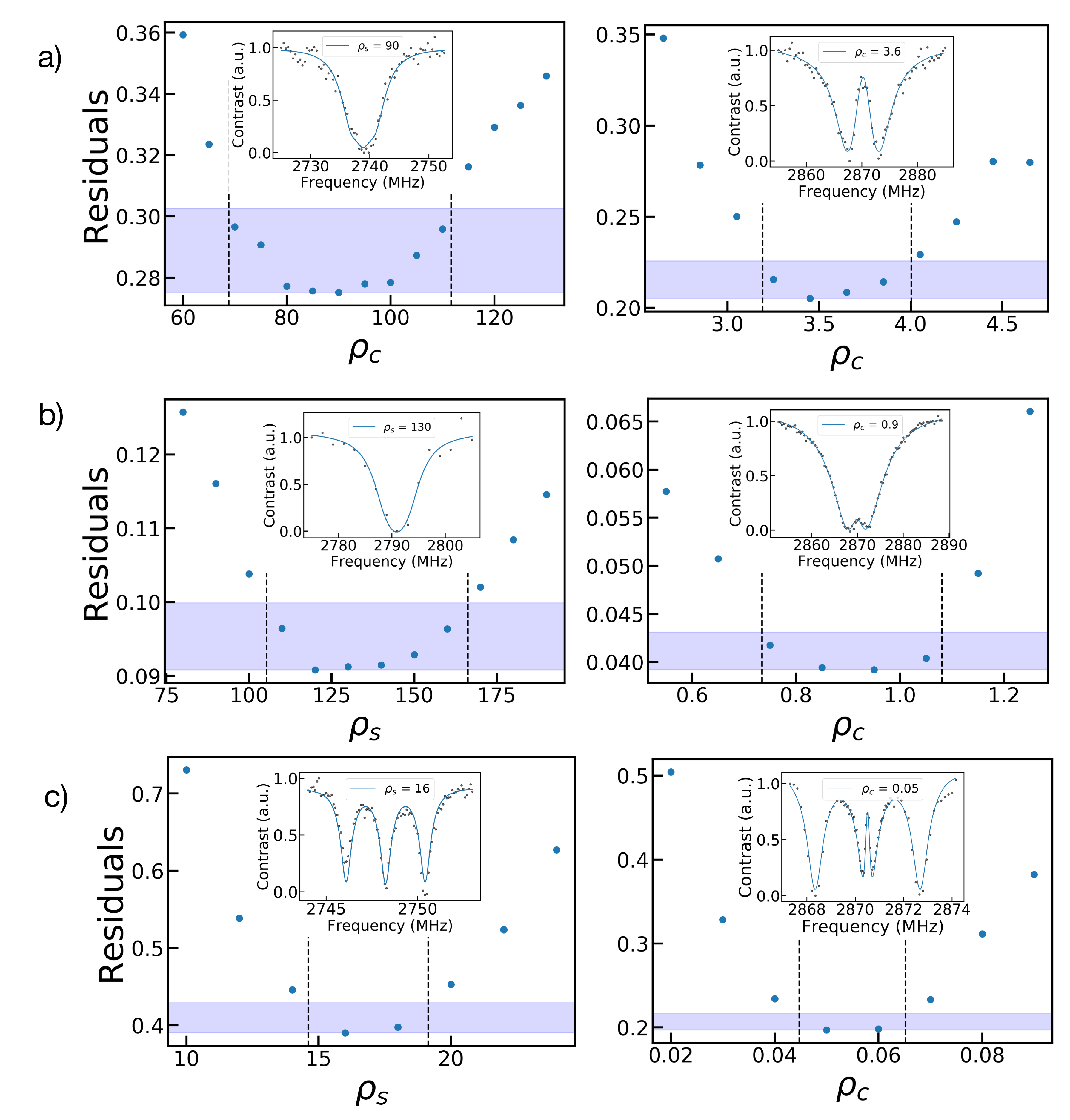}
\caption{Fitting procedure applied to the untreated samples: (a) Ib untreated (S4), (b) Ib untreated (S5), and (c) IIa treated (S6). See caption of Fig.~\ref{ensemble1} for description.}
\label{ensemble2}
\end{figure}

Our fitting procedure for each ensemble sample consists of two steps.
First, we fit a spectrum taken at high magnetic field, where the effects of electric fields are highly suppressed and the dominant broadening is due to magnetic impurities (Figs.~\ref{ensemble1} and \ref{ensemble2}, left column). 
This allows us to characterize $\rho_{\textrm{s}}$ independently.
Second, we fit a spectrum at zero applied field using the previously determined magnetic noise and an additional contribution due to electric fields (Figs.~\ref{ensemble1} and \ref{ensemble2}, right column). 
This determines the charge density $\rho_{\textrm{c}}$, as well the the natural linewidth $\Gamma$.

For the high-field spectra, we sample over the magnetic impurities configurations following the procedure outlined in the previous section. 
For each configuration, we calculate the NV's resonance frequencies using the full Hamiltonian of the system, Eq.~(1) of the main text. 
Repeating this procedure for $\sim 5000$ realizations, we obtain a histogram of resonance energies that is proportional to the experimentally observed spectra.
We generate such spectra for a range of $\rho_{\textrm{s}}$ and fit each individually to the high-field measurement, optimizing with respect to the center frequency, vertical offset, and overall amplitude. We characterize $\rho_{\textrm{s}}$ by calculating the least-square residuals between our simulated spectra and the experimental data (Figs.~\ref{ensemble1} and \ref{ensemble2}, left column). In particular, we identify $\rho_{\textrm{s}}$ that minimizes the residual as the best-fit parameter and estimate its error from the range of values whose residuals lie within 10\% of the minimum. 

The fitting procedure for the zero-field spectra follows in close analogy, except we now average over both the charge distribution and the magnetic impurity distribution. 
Specifically, we first sample the positions of the charges and calculate the electric field at the NV center ($\sim 5000$ realizations). For each charge realization, we then sample over many configurations of magnetic impurities to simulate the magnetic noise (additional $\sim 5000$ realizations). Another important difference from before is that we now incorporate a natural linewidth for each resonance.
To do so, we convolve the distribution of resonance frequencies with a Lorentzian profile characterized by a full-width-half-maximum $\Gamma$. 
This linewidth accounts for broadening that is independent from the charge environment or static magnetic fields.
For example, it would include contributions from power broadening, fluctuating fields in the environment (i.e.~$T_{2,\textrm{echo}}$), and strain-induced broadening.

To isolate the effects due to the charge environment, we fit the zero-field spectra as a function of $\rho_{\textrm{c}}$ while fixing the magnetic noise ($\rho_{\textrm{s}}$) based on the previous step.
For each value of $\rho_{\textrm{c}}$, we optimize with respect to the natural linewidth $\Gamma$, the center frequency, overall amplitude, and vertical offset. 
These results are shown in the right column of Figs.~\ref{ensemble1} and \ref{ensemble2}. As before, we estimate the error on $\rho_{\textrm{c}}$ from the 10\% interval of the residuals, while for $\Gamma$ we take the standard error estimated by the fitting routine.

All simulated spectra agree quantitatively with the experimental data, and the extracted $\rho_{\textrm{s}}$, $\rho_{\textrm{c}}$ and $\Gamma$ are listed in Table I in the main text.
We note that for one of the six samples (S5), the linewidth contribution from $\delta B_z$ is on the same order as $\Gamma$.
Since we assume $\delta B_z$ is the dominant source of noise in the high field spectra when extracting $\rho_{\textrm{s}}$, the magnetic impurity density for this sample may not be precise. 

\section{Charge localization using single NVs}

In this section, we discuss the details associated with the charge localization based on a single NV. 
We consider the derivation of the imbalance and relate it to the electric field orientation and the microwave polarization.
We note that the imbalance of the resonances is strong evidence for the presence of a nearby charge, as most other interactions would not modify the transition amplitudes differentially with respect to linearly polarized microwave fields.
%We note that the imbalance of the resonances is very strong evidence for the presence of a nearby charge, as most other interactions would leave the spectrum symmetric respect to linearly polarized microwave fields.

To extract the position of the charge, we first calculate the polarization of the microwave field in the reference frame of the NV, $\phi_{\textrm{MW}}$ (Fig. \ref{edist}a inset of the main text).
By varying $\phi_{\textrm{MW}}$, and measuring the imbalance one can directly extract the transverse orientation of the electric field $\phi_{E}$.
Combined with the observed splitting 2$\Pi_\perp$ and shifting $\Pi_z$ we can fully determine the local electric field vector and localize the corresponding charge.
These procedures are detailed below.

\subsection{Derivation of the Imbalance}

In order to quantitatively extract the orientation of the electric field $\phi_E$, we introduce the notion of imbalance as the difference in the weights of the resonances in the observed spectra.
This imbalance $\mathcal{I}$ is directly related to  $\phi_{\textrm{MW}}$ and the transverse orientation of the electric field $\phi_E$.

We begin by focusing our attention to the states with $^{14}$N nuclear spin $m_I = 0$ (two inner resonances). 
In the presence of an electric field, these states are described by the Hamiltonian:
\begin{equation}
H = (D_{gs}+\Pi_z) S_z^2+\Pi_x(S_y^2-S_x^2)+\Pi_y(S_x S_y+S_y S_x).
\end{equation}
One finds that the electric field couples only the $\ket{m_s = \pm 1}$ states, leading to the new eigenstates:
\begin{align} \label{eig0}
\ket + &= \frac 1 {\sqrt 2}\left(\ket{m_s = +1} - e^{-i \phi_E} \ket{m_s = -1}\right) \\
\ket - &= \frac 1 {\sqrt 2}\left(e^{i \phi_E} \ket{m_s = +1} + \ket{m_s =-1}\right)
\end{align}
with energy splitting $2\Pi_\perp = 2\sqrt{\Pi_x^2 + \Pi_y^2}$.

The magnetic resonance spectrum is obtained by driving transitions from the $\ket{m_s=0}$ state to the $\ket{\pm}$ states using a linearly polarized microwave field. 
The matrix elements associated with these transitions are
\begin{align}
\mathcal{M}_{\pm} &= \bra{0} S_x \cos{\phi_{\textrm{MW}}}+S_y \sin{\phi_{\textrm{MW}}} \ket{\pm} \\
&= \frac{1}{2}\left[ e^{-i \phi_{\textrm{MW}}} \mp e^{i (\phi_E + \phi_{\textrm{MW}})} \right]
\end{align}
where $\phi_{\textrm{MW}}$ is the direction of microwave polarization. This results in two resonances with amplitudes, $A_{\pm} \equiv \left|\mathcal{M}_{\pm}\right|^2$: 
\begin{equation} \label{in_weight}
A_{\pm} = \frac{1}{2} \mp \frac{1}{2}\cos(2\phi_{\textrm{MW}} + \phi_E).
\end{equation}
By defining the imbalance $\mathcal{I} \equiv \frac {A_+ - A_-}{A_+ + A_-} $, we recover Eq. (2) in the main text:
\begin{equation} \label{inner}
\mathcal{I} = - \cos(2\phi_{\textrm{MW}} + \phi_E).
\end{equation}
We note that the imbalance reverses direction for $\phi_{\textrm{MW}} \rightarrow \phi_{\textrm{MW}} + 90^\circ$ and that, for certain microwave angles, the amplitude of one resonance can fully vanish.

For completeness, we also derive the imbalance of the outer $^{14}$N hyperfine states, which correspond to $m_I = \pm 1$. The derivation follows the same logic as above, except the Hamiltonian is now
\begin{equation}
% H_{\pm} = -E_x\sigma_x+E_y\sigma_y \pm A_{zz} \sigma_z
H = (D_{gs}+\Pi_z) S_z^2+\Pi_x(S_y^2-S_x^2)+\Pi_y(S_x S_y+S_y S_x) \pm 2 A_{zz} S_z.
\end{equation}
The eigenstates $\ket{\pm}$ are split by $2 \sqrt{A_{zz}^2+\Pi_{\perp}^2}$. For $m_I = 1$, one finds
\begin{align}
\ket + &= \frac{1}{\sqrt{1+\xi^2}}\left(\ket {+1} - \xi e^{-i \phi_E} \ket {-1} \right) \\
\ket - &= \frac{1}{\sqrt{1+\xi^2}}\left(\xi e^{i \phi_E} \ket {+1} + \ket {-1} \right)
\end{align}
where
\begin{equation}
\xi = \frac {A_{zz}}{\Pi_{\perp}} \left(\sqrt{1+\left(\frac {\Pi_{\perp} }{A_{zz}}\right)^2}-1\right)
\end{equation}
An analogous expression holds for $m_I = -1$.
In both cases, the amplitudes of the $\ket{m_s = 0}$ $\leftrightarrow$ $\ket{\pm}$ resonances are
\begin{equation} \label{out_weight}
A_{\pm} = \frac{1}{\sqrt{1+\xi^2}}\left(1+\xi^2 \mp 2 \xi \cos(2\phi_{\textrm{MW}} + \phi_E) \right).
\end{equation}
This leads to an imbalance:
\begin{equation} \label{outer}
\mathcal{I} = \frac{- 2\xi\cos(2\phi_{\textrm{MW}} + \phi_E)}{1+\xi^2}.
\end{equation}
Thus, the imbalance of the outer resonances follows the same phase dependence as the inner resonances, but the maximum imbalance depends on the ratio $\Pi_{\perp} / A_{zz}$. In particular, in the limit $\Pi_{\perp} \gg A_{zz}$, $\xi \approx 1$ and a fully dark state is still possible; whereas, for $\Pi_{\perp} \ll A_{zz}$, the maximum imbalance is reduced to $\mathcal{I}_\textrm{max} \approx \Pi_{\perp} / A_{zz}$.

The resulting dependence on $\phi_{MW}$ and $\phi_{E}$ does not change if we include the interaction with a nearby $^{13}$C (within the secular approximation), since it interacts with the NV in a similar fashion to $^{14}$N hyperfine.

\subsection{Microwave Angle Projection}
\label{Microwave Angle Projection}

% DO WE WANT FIGURES IN THIS SECTION??

\begin{figure}
\centering
\includegraphics[width=1\textwidth]{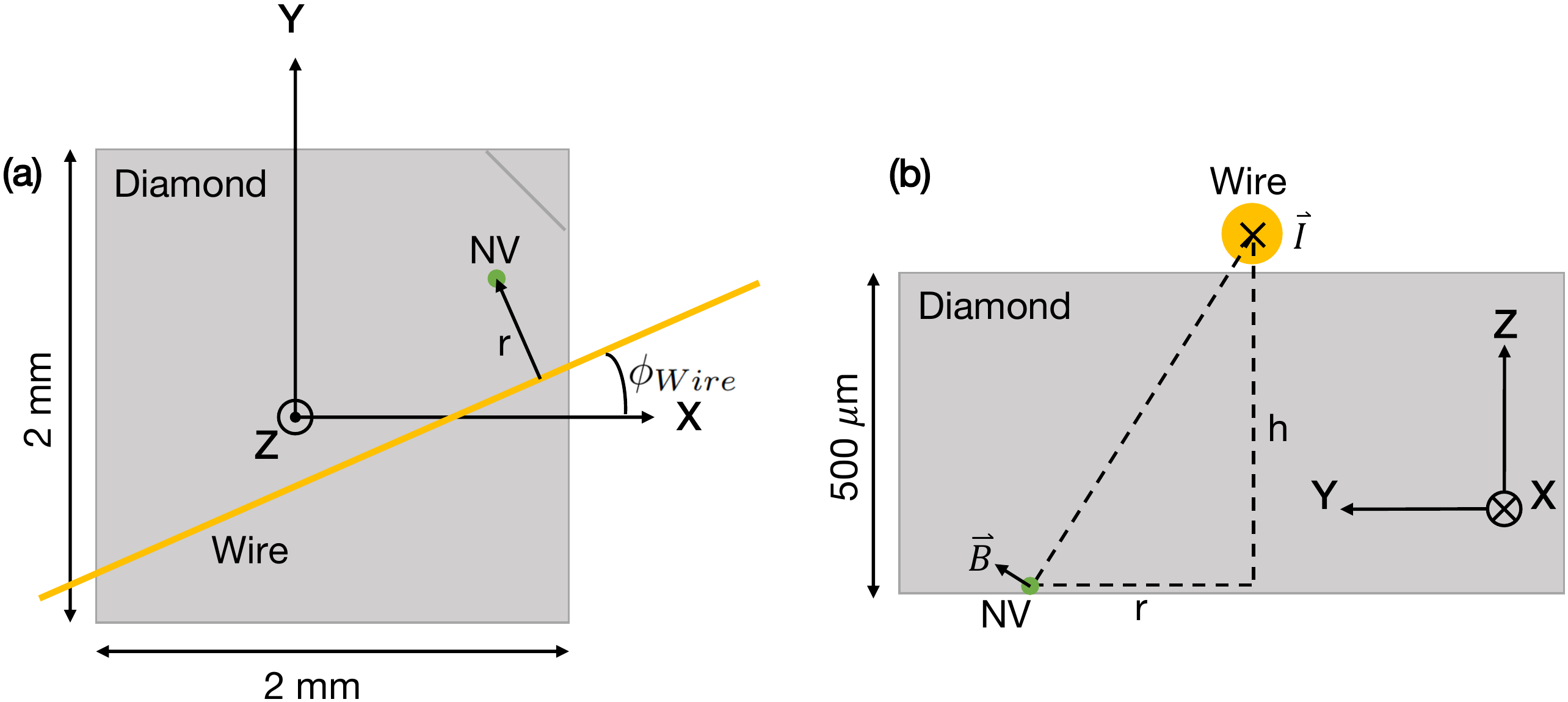}
\caption{a) Top view of lab frame, $\hat{X}$, $\hat{Y}$, and $\hat{Z}$ axes are defined as shown. Wire is displayed at an angle $\phi_{\textrm{Wire}}$ relative to X, and $r$ is the distance between the wire and the NV. b) Side view of lab frame. With $\phi_{\textrm{Wire}} = 0$, when the oscillating current $\vec{I}$ flows in the direction shown, we calculate the direction of the magnetic field vector $\vec{B}$ at a height $h$ below the wire as shown.}
\label{MW_projection}
\end{figure}

We define $(\hat{X},\hat{Y},\hat{Z})$ as our lab frame shown in Fig.~\ref{MW_projection}a and the NV frame $(\hat{x},\hat{y},\hat{z})$ as shown in Fig.~1a left inset in the main text. These two frames are related by the crystallographic axes of the sample. We approximate the microwave delivery wire to be infinitely long, with an angle $\phi_\textrm{Wire}$ with respect to $\hat{X}$, and an in-plane distance $r$ away from the NV. We extract $\phi_\textrm{Wire}$ and $r$ from an image of the sample geometry (Fig.~\ref{setup image} inset).
The height $h$ of the wire’s plane above the NV is assumed to be 550~$\pm$~100~$\mu$m given the thickness of the diamond 500~$\mu$m, the wire diameter $40~\mathrm{\mu m}$, and an intentional air gap to avoid contact to the sample ($\sim30~\mathrm{\mu m}$).
The wire carries a current which generates a linearly polarized microwave field at the location of the NV (Fig.~\ref{MW_projection}) whose transverse projection $\phi_{\textrm{MW}}$ drives the $\ket{m_s=0}\leftrightarrow \ket{\pm}$ transition. $\phi_{\textrm{MW}}$ is fully determined by the values $\{\phi_\textrm{Wire},h,r\}$. To estimate error in each realization of $\phi_{\textrm{Wire}}$, we use a Monte Carlo method assuming a $\pm 10^\circ$ tilting of the wire out of the plane.

\subsection{Single Charge Localization}

\begin{figure}[t]
\centering
\includegraphics[width=0.8\textwidth]{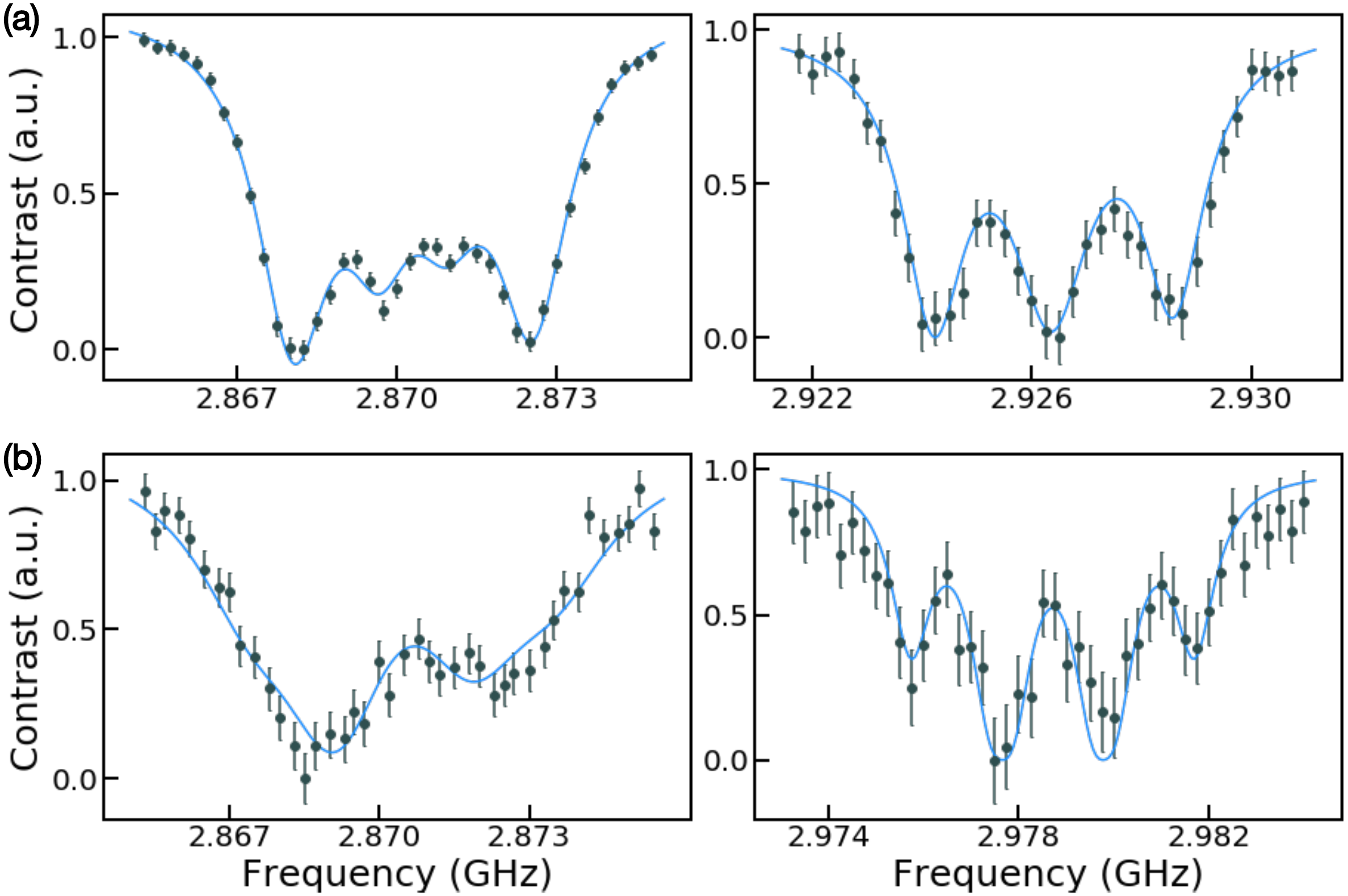}
\caption{Spectra taken with and without a magnetic field applied along the NV z-axis. a) Left: zero-field spectrum for NV1 with microscopic model fit; Right: spectrum with an applied magnetic field and a fit to 3 Lorentzians. b) Left: zero-field spectrum for NV2 with microscopic model fit; Right: spectrum with an applied magnetic field. The fit function is two sets of three Lorentzians. The Lorentzians in each set are separated by the $^{14}$N hyperfine splitting. The sets are split from each other by a fit parameter for the $^{13}$C hyperfine interaction. }
\label{SingleNV_spectra}
\end{figure}

We search through 68 single NVs and find four exhibiting a significant imbalance in the zero-field spectrum consistent with a nearby charge, from which we analyze two in this work (referred to as NV1 and NV2).
Because these spectra can also be affected by the presence of a nearby strongly-coupled $^{13}$C, we apply a bias $B_z$ field, which suppresses the effect of the electric field and identifies the source of the splitting.
The zero- and high-field spectra for these two NVs are shown in Fig.~\ref{SingleNV_spectra}.
For NV1, we find three resonances spaced $\sim$2.16 MHz apart, a signal associated exclusively with $^{14}$N hyperfine. 
In contrast, for NV2 we observe four resonances, indicating the additional presence of a strongly-coupled $^{13}$C.
We fit the spectrum of NV2 to extract the $^{13}$C hyperfine coupling strength $\approx$ 1.65(7)~MHz.
To confirm the charge origin, we then measure the full imbalance curve using dark-state spectroscopy.

\begin{figure}
\centering
\includegraphics[width=0.4\textwidth]{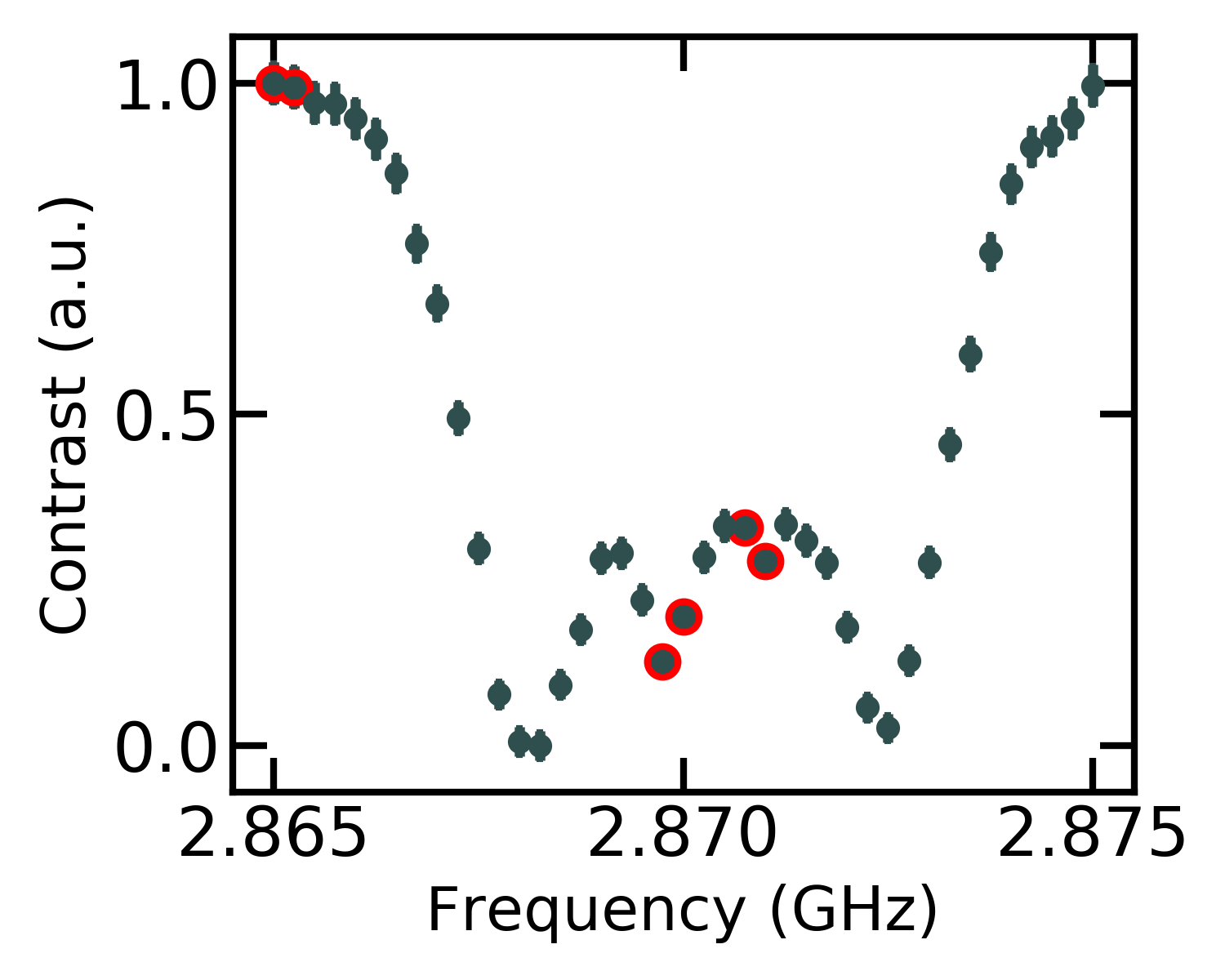}
\caption{Position of the six frequencies (red) considered when computing the imbalance.
Instead of measuring full-spectra, we take data points closely spaced at the location of each of the two inner resonances and two data points far from the resonances, so as to measure the baseline signal.
%Usually, the two data points taken far from resonance were separated, one on either side of the resonance.}
}
\label{6pointMethod}
\end{figure}

For NV1, we can clearly resolve the four resonances. 
The information about the imbalance is encoded into the amplitude of the inner two resonances.
To estimate these amplitudes we measure only six spectral data points for each $\phi_{\textrm{MW}}$ (Fig.~\ref{6pointMethod}): two data points closely spaced at the location of each of the two inner resonances and two data points far from the resonances (measurement of the baseline contrast) . 
The imbalance extracted with this method is shown in the main text Fig. 4d, from which we extract $\phi_E~=~124(5)^\circ$.
%measuring the spectra with sufficient frequency resolution required too much time. 

\begin{figure}
\centering
\includegraphics[width=0.8\textwidth]{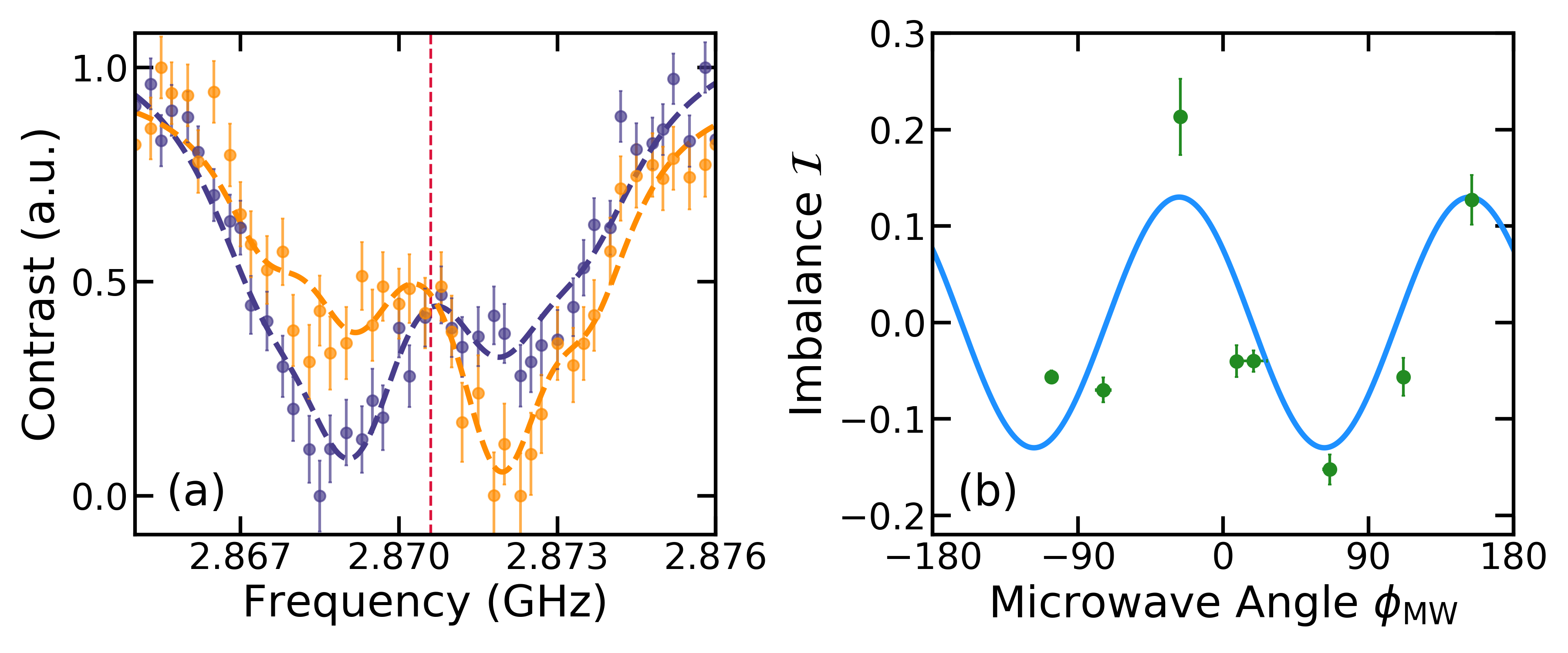}
\caption{a) Two spectra from NV2 with fit from the microscopic model at different values of $\phi_{\text{MW}}$. The dashed vertical line indicates the fit center frequency ($2.8706$ GHz). We estimate the imbalance by compare the integral on either side of the center frequency. b) Resultant imbalance sinusoid, from where we extract $\phi_E = 236(15)^\circ$.}
\label{NV7_Spectra_Sine}
\end{figure}

For NV2, since we cannot clearly resolve the four resonances due to the presence of the nearby $^{13}$C, we estimate imbalance by integrating the area on either side of the fit center frequency (Fig.~\ref{NV7_Spectra_Sine} a). 
The imbalance curve is shown in Fig.~\ref{NV7_Spectra_Sine}b, from which we extract $\phi_E~=~236(15)^\circ$. 

We note that the amplitudes of these curves are much smaller than unity.
This discrepancy from our simple theoretical model can also be explained by a few possibilities.
First, our methods do not directly probe the weight of the transitions.
Second, due to the intrinsic linewidth and power broadening, the inner and outer resonances overlap, which precludes isolating any single transition.
Third, a dynamic charge bath may generate a background spectrum that is not included in our model.

In order to localize the charge, we also need to extract the charge-induced splitting $\Pi_{\perp}$ and shifting $\Pi_z$.
In direct analogy to the treatment of ensembles, we fit the full zero-field single NV spectra using our microscopic model to extract these parameters as follows:
\begin{enumerate}
\item The spectra depend on five physical parameters: the three components of the electric field $\vec{E}$, the density of magnetic defects $\rho_{\textrm{s}}$, and the natural linewidth $\Gamma$. 
We also include a global amplitude scaling factor and background offset.% so as to match the experimental data to the theoretical prediction.
\item To account for the magnetic noise distribution, we follow a prescription similar to the previous magnetic field distribution section.
% Sec.~\ref{sec:MagDist}. 
We begin by considering the distribution of magnetic field for $\rho_{\textrm{s}}$ which yields a probability distribution for measuring a particular value of $\delta B_z$.
We then discretize over $\delta B_z$ and for each possible value, perform steps 3-5. 
Each of the resulting spectra is weighted by the probability of measuring $\delta B_z$.
\item We solve the full Hamiltonian of the system (including $^{13}$C and $^{14}$N hyperfine interactions where applicable) to find the positions of the resonances.
\item We generate a spectrum by weighting each resonance by its transition amplitude with the $\ket{m_s=0}$ state.
  We compute the wright by fixing the microwave direction in the $\hat{x}$ axis and computing $\left|\bra{0}S_x\ket{\pm}\right|^2$.
\item We broaden each resonance by a Lorentzian distribution with full-width-half-maximum of $\Gamma$.
\item We use a least-squares regression method on steps 1-5 over the seven fitting parameters, reproducing the experimental spectra.

\end{enumerate}
Note, in order to determine $\Pi_z$, we use the ensemble-averaged $D_{\textrm{gs}} = 2870.25(5)$~MHz from the adjacent region of the same diamond containing a high density of NVs as a reference value (Figure~\ref{ensemble2}a).

From the fits (see main text Figure 4a,b) we extract the shifting and splitting due to the electric field:

\begin{align}
\text{NV1: }&\quad \Pi_z = (30 \pm 50) ~\textrm{kHz} \quad, \quad\Pi_{\perp} = (650 \pm 10) ~\textrm{kHz}~\\
\text{NV2: }&\quad \Pi_z = (270 \pm 70) ~\textrm{kHz} \quad , \quad \Pi_{\perp} = (850 \pm 80) ~\textrm{kHz}~.
\end{align}

Using the susceptibilities \cite{VANOORT1990}, we extract the electric field vectors at the position of the single NVs: 
\begin{align}
\text{NV1: }&\quad(E_x, E_y, E_z) = (-2.1 \pm 0.2, \enspace3.2 \pm 0.2, \enspace9 \pm 14) ~\textrm{MV/m}\\
\text{NV2: }&\quad(E_x, E_y, E_z) = (-2.8 \pm 1.1, \enspace -4.1 \pm 0.8, \enspace 77 \pm 20) ~\textrm{MV/m}~.
\end{align}
The parameters of the electric field uniquely determine the position of the positive single fundamental charge (main text Fig.~1b and 1c).
The confidence intervals can be estimated using a Monte Carlo method.

\bibliography{Supp}